\documentclass[12pt,preprint]{aastex}
\usepackage{amsmath}
\usepackage{graphicx}
\usepackage{tabularx}

\def\ser  {{\tt p3m}}
\def\parsl{{\tt llpm-sl}}
\def\parhc{{\tt GRACOS}}

\begin{document}
  \title{\parhc: Scalable and Load Balanced P$^3$M Cosmological N-body Code}
  \author{Alexander Shirokov and Edmund Bertschinger}
  \affil{Department of Physics and Center for Space Research, MIT
  Room 37-602A, 77 Massachusetts Ave., Cambridge, MA 02139\\ {\tt
  shirokov@mit.edu, edbert@mit.edu}}

\def\dografic {{\tt GRAFIC1}\ }
\def\vol{{V_0}}
\def\mark{{\small $\blacklozenge$}\ }
\def\lfun{\mathcal{L}}
\def\gfun{\mathcal{G}}
\def\rfun{\mathcal{R}}
\def\gsl{G_{\rm sl}} 
\def\grid{{G_0}}
\def\dolsc{{L_{\rm SCH}}}
\def\ptr{{-\hspace{-2mm}>\hspace{0.5mm}}}

\newcommand{\PreserveBackslash}[1]{\let\temp=\\#1\let\\=\temp}
\let\PBS=\PreserveBackslash

\def  \donloc     {{\tt nloc}}
\def  \dosloc     {{\tt sloc}}
\def  \dodx       {dx}
\def  \don         {n}
\def  \donh        {n_h}
\def  \donc        {n_c}
\def  \dohclr     {L_h}
\def  \doeps      {\tilde{\epsilon}}
\def  \doetat     {\eta_t}
\def  \dong       {N_{\rm gr}}
\def  \donhc      {N_{\rm HC}}
\def  \dodx       {\Delta x}
\def  \dodxc      {\Delta \tilde{x}_c}
\def  \dovdxcm     {\Delta {\bf x}_c}
\def  \doK        {K}

\def  \nstep{{n}}
\def  \frob{{f}}

\def  \dorb       {r_{\rm b}}
\def  \dorn       {r_{\rm n}}
\def  \dohb       {h_{\rm b}}
\def  \dohn       {h_{\rm n}}

\def  \doheb      {h_{\rm eb}}
\def  \dohen      {h_{\rm en}}
\def  \doreb      {h_{\rm eb}}
\def  \doren      {h_{\rm en}}

\def  \dor        {r}

\def  \dormaxval  {2.78}
\def  \doh        {h}
\def  \donp       {N}
\def  \dohctoi    {{\tt hilbert\_c2i}}
\def  \dohitoc    {{\tt hilbert\_i2c}}
\def  \domhctoi   {\mathcal{H}}
\def  \domhctor   {\mathcal{H}_{\rm r}}
\def  \domhitoc   {\mathcal{H}^{-1}}
\def  \domhrtoc   {\mathcal{H}_{\rm r}^{-1}}
\def  \doVo       {V_0}
\def  \doVh       {V_{\rm h}}

\def  \donpc      {n_{\rm pr}}
\def  \dodxh      {\Delta \tilde{x}_h}
\def  \dormax     {\tilde{R}_{\max}}
\def  \dovc        {{\bf c}}
\def  \doc        {c}
\def  \dom        {m}

\def  \dopa       {{\tt pa}}
\def  \dopaf      {{\tt pa\_f}}
\def  \dopafa     {{\tt pa\_fa}}
\def  \domeas     {\mathcal{N}}

\def  \dohcavb    {{\tt hc\_avb}}
\def  \dohcstg    {{\tt hc\_stg}}

\def  \dohctol    {{\tt hc\_tol}}

  \begin{abstract}
We present a parallel implementation of the
particle-particle/particle-mesh (P$^3$M) algorithm for
distributed memory clusters.  The \parhc\ (GRAvitational
COSmology) code uses a hybrid method for both computation and
domain decomposition.  Long-range forces are computed using a
Fourier transform gravity solver on a regular mesh; the mesh is
distributed across parallel processes using a static
one-dimensional slab domain decomposition.  Short-range forces
are computed by direct summation of close pairs; particles are
distributed using a dynamic domain decomposition based on a
space-filling Hilbert curve.  A nearly-optimal method was devised
to dynamically repartition the particle distribution so as to
maintain load balance even for extremely inhomogeneous mass
distributions.  Tests using $800^3$ simulations on a 40-processor
beowulf cluster showed good load balance and scalability up to 80
processes.  We discuss the limits on scalability imposed by
communication and extreme clustering and suggest how they may be
removed by extending our algorithm to include adaptive mesh
refinement.
\end{abstract}


\section{Introduction}\label{sec_intro}

Cosmological N-body simulations are the main tool used to study
the dynamics of collisionless dark matter and its role in the
formation of cosmic structure.  They first became widely used 20
years ago after it was realized that the gravitational potentials
of galaxies are dominated by dark matter.  At the same time,
theories of the early universe were developed for dark matter
fluctuations so that galaxy formation became an initial value
problem.

Although many of the most pressing issues of galaxy formation
require simulation of gas dynamics as well as gravity, there is
still an important role for gravitational N-body simulations in
cosmology.  Dark matter halos host galaxies and therefore
gravitational N-body simulations provide the framework upon which
one adds gas dynamics and other physics. Moreover, many questions
of structure formation can be addressed with N-body simulations as
a good first approximation: the shapes and radial mass profiles of
dark matter halos, the rate of merging and its role in halo
formation, the effect of dark matter caustics on ultra-small scale
structure, etc.

In a cosmological N-body simulation, the matter is discretized
into particles that feel only the force of gravity.  A subvolume
of the universe is sampled in a rectangular (not necessarily
cubic) volume with periodic boundary conditions.  In principle,
one simply uses Newton's laws to evolve the particles from their
initial state of near-perfect Hubble expansion.  Gravity takes
care of the rest.

In practice, cosmological N-body simulation is difficult because
of the vast dynamic range required to adequately model the
physics.  Gravity knows no scales and the cosmological initial
fluctuations have power on all scales.  After numerical accuracy
and speed, dynamic range is the primary goal of the computational
cosmologist.  One would like to simulate as many particles as
possible (at least $10^{10}$ to sample galaxies well within a
supercluster-sized volume), with as great spatial resolution as
possible (at least $10^4$ per dimension), for as long as possible
($10^3$ to $10^4$ timesteps to follow the formation and evolution
of structure up to the present day).

A single computer is insufficient to achieve the maximum possible
dynamic range.  One should use many computers cooperating to solve
the problem using the technique of parallelization.  In a parallel
N-body simulation, the computation and memory are distributed
among multiple {\it processes} running on different {\it nodes}
(computers).\footnote{Because of the increasing availability of
beowulf clusters, we consider only distributed memory
parallelism.} Unfortunately, ordinary compilers cannot effectively
parallelize a cosmological N-body simulation code.  A programmer
must write special code instructing the computers how to divide up
the work and specifying the communication between processes.

A parallel code is considered successful if it produces
load-balanced and scalable simulations.  A simulation is {\it load
balanced} when the distribution of the effective workloads among
the nodes is uniform. {\it Scalability} for a given problem means
that the wall clock time spent by the computer cluster doing
simulations scales inversely with the number of nodes used.
Ideally, of course, the code should also be {\it efficient}: as
much as possible, the wall clock time should be entirely devoted
to computation.

At present, there are two main algorithms used for cosmological
N-body codes: Tree and P$^3$M (see Bertschinger 1998 for review).
The current parallel Tree code implementations include TreeSPH
\citep{ddh97}), HOT \citep{sw94}, Gadget \citep{syw01}, and
Gasoline \citep{wsq04}.  Tree codes have the advantage of
relatively easy parallelization and computing costs that scale as
$N\log N$ where $N$ is the number of particles. However, they have
relatively large memory requirements.

The P$^3$M (particle-particle/particle-mesh) method was introduced
to cosmology by \cite{ee81} and is described in detail in
\cite{edfw85} (see also Bertschinger \& Gelb 1991).  For moderate
clustering strengths, P$^3$M is faster than the Tree code but it
becomes slower when clustering is strong.  This is because P$^3$M
is a hybrid approach that splits the gravitational force field of
each particle into a long-range part computed quickly on a mesh
plus a short-range contribution computed by direct summation over
close pairs.  When clustering is weak, the computation time scales
as $\dong\log\dong$ where $\dong\ $ is the number of grid (mesh)
points, while when clustering is strong the computation time
increases in proportion to $N^2$.  The scaling can be restored to
$N\log N$ using adaptive methods \citep{c91}.

Currently there exist several parallel implementations of the
P$^3$M algorithm, including the version of \cite{fb95} for the
(now defunct) Connection Machine CM-5 and the Hydra code of
\cite{mcpp98}.  The Hydra code uses shared memory communications
for the Cray T3E. There is a need for a message-passing based
version of P$^3$M (and its adaptive extension) to run on beowulf
clusters.  This need motivates the present work.

The difficulty of parallelizing adaptive P$^3$M has led a number
of groups to use other techniques to add short-range forces to the
particle-mesh (PM) algorithm.  The Tree and PM algorithms have
been combined by \cite{bo03} and \cite{dkp04} while \cite{mpt04}
use a two-level adaptive mesh refinement of the PM force
calculation.  The FLASH code \citep{flash} has been extended to
incorporate PM forces with multi-level adaptive mesh refinement.

When the matter distribution becomes strongly clustered, parallel
codes based on PM and P$^3$M face severe challenges to remain
load-balanced.

In general, P$^3$M and PM-based parallel codes suffer
complications when the matter becomes very clustered as happens at
the late stages of structure formation.  Most of the existing
codes use a static one-dimensional slab domain decomposition,
which is to say that the simulation volume is divided into slices
and each process works on the same slice throughout, even when the
particle distribution becomes strongly inhomogeneous.  The GOTPM
code uses dynamic domain decomposition, with the slices changing
in thickness as the simulation proceeds, resulting in superior
load balancing.  However, even this code will break down at very
strong clustering because it also uses a one-dimensional slab
domain decomposition.  The FLASH code uses a more sophisticated
domain decomposition similar in some respects to the method
introduced in the current paper.

The motivation of the current work is to produce a publicly
available code that will load balance and scale effectively for
all stages of clustering on any number of nodes in a beowulf
cluster. This paper introduces a new, scalable and load-balanced
approach to the parallelization technique for the P$^3$M force
calculation.   We achieve this by using dynamic domain
decomposition based on a space-filling Hilbert curve and by
optimizing data storage and communication in ways that we
describe.

This paper is the first of two describing our parallelization of
an adaptive P$^3$M algorithm.  The current paper describes
the domain decomposition and other issues associated with parallel
P$^3$M.  The second paper will describe the adaptive refinement
method used to speed up the short-range force calculation.

The outline of this paper is as follows.  The serial P$^3$M
algorithm (based on Gelb \& Bertschinger 1994 and Ferrell \&
Bertschinger 1994) that underlies our parallelization is
summarized in \S \ref{sec_serial}.  Section \ref{sec_par}
discusses domain decomposition methods starting with the
widely-implemented static one-dimensional slab decomposition
method.  We then introduce the space-filling Hilbert curve and
describe its use to achieve a flexible three-dimensional
decomposition.  Section \ref{sec_loadbal} presents our algorithm
for dynamically changing the domain decomposition so as to achieve
load balance. Section \ref{sec_layout} presents our techniques for
organizing the particle data so as to minimize efficiency in
memory usage, cache memory access, and interprocessor
communications.  In \S \ref{sec_hcforce} we describe the
algorithms used to parallelize the PM and PP force calculations.
Section \ref{sec_test} presents code tests emphasizing load
balance and scalability.  Conclusions are presented in \S
\ref{sec_concl}.  An appendix presents an overview of the code and
frequently appearing symbols, and another appendix briefly
describe the routines used to map the Hilbert curve onto a
three-dimensional mesh and vice versa.
\section{Serial P$^3$M C-code and force calculation}\label{sec_serial}

In this section we summarize our serial cosmological \mbox{N-body}
C implementation \ser\ based on an earlier serial Fortran
implementation of P$^3$M by one of the authors. We discuss in
detail the code units and aspects of the force calculation that
are necessary for understanding the parallelization issues covered
in the later sections.

\subsection{Long and Short Range Forces and the Pairwise Force Law}
\label{sec_force}

Given the pairwise force ${\bf F_0}({\bf r}_{12})$ between two
particles of masses $m_1$ and $m_2$ and separation ${\bf
r}_{12}={\bf x}_2-{\bf x}_1$, we define the interparticle force
law profile ${\bf \Theta}_0({\bf r}) \equiv {\bf F_0}({\bf
r}_{12})/(Gm_1m_2)$. For a system of many particles, the
gravitational acceleration of particle $i$ is $\sum_{j\ne
i}Gm_j{\bf\Theta}_0({\bf r}_{ij})$.

The required interparticle force law profile depends on the shape
of the simulation particles.  For point particles one uses the
inverse square force law profile ${\bf \Theta}_0({\bf r}) = -{\bf
r}/r^3$.  The inverse square force law is not used for simulation
of dark matter particles in order to avoid the formation of
unphysical tight binaries, which happens as a result of two-body
relaxation \citep{bt94}.  For cold dark matter simulations many
authors use the \cite{p11} force law
\begin{equation}
  {\bf \Theta}_{\rm PL}({\bf r},\epsilon) \equiv
    -\frac{\bf  r}{(r^2+\epsilon^2)^{3/2}}\ ,
\end{equation}
where $\epsilon$ is the Plummer softening length.  We take the
Plummer softening length to be constant in comoving coordinates.
With Plummer softening the particles have effective size
$\epsilon$.  In a P$^3$M code, $\epsilon$ is usually set to a
fraction of the PM-mesh spacing.

In a P$^3$M code, the desired (e.g., Plummer) force law is
approximated by the sum of a long-range (particle-mesh or PM)
force evaluated using a grid and a short-range (particle-particle
or PP) force evaluated by direct summation over close pairs.  The
PM force ${\bf\Theta}_{\rm PM}({\bf x}_i,{\bf x}_j)$ varies
slightly depending on the locations of the particles relative to
the grid (see Appendix A of Ferrell \& Bertschinger 1994).  The
average PM force law $\langle{\bf\Theta}_{\rm PM}\rangle({\bf
r}_{ij})$ can be tabulated by a set of Monte-carlo PM-force
simulations each having one massive particle surrounded by
randomly placed test particles \citep{b91}. In practice, the mean
PM force differs from the inverse square law by less than 1\% for
pair separations greater than a few PM grid spacings.  For smaller
separations, a correction (the PP force) must be applied.  The
total force is given by
\begin{equation}\label{eq_p3m}
  {\bf\Theta}_{\rm P3M}({\bf r}_{ij}) \equiv {\bf \Theta}_{\rm PP}({\bf r}_{ij})
  +{\bf\Theta}_{\rm PM}({\bf x}_i,{\bf x}_j)\ .
\end{equation}
Strictly speaking, the P$^3$M force is not translationally
invariant and therefore depends on the positions of both
particles.  The P$^3$M force differs from the exact desired
interparticle force profile ${\bf \Theta_0}$ by ${\bf\Theta}_{\rm
Error}({\bf x}_i,{\bf x}_j)\equiv {\bf\Theta}_{\rm P3M}({\bf
x}_i,{\bf x}_j)-{\bf \Theta_0}({\bf r}_{ij}) = {\bf\Theta}_{\rm
PM}({\bf x}_i,{\bf x}_j)-\langle{\bf\Theta}_{\rm PM}\rangle({\bf
r}_{ij})$.

At large separations, both the PM-force and the required force
reduce to the inverse square law (modified on the scale of the
simulation volume by periodic boundary conditions). The PP-force
can therefore be set to zero at $r \ge R_{\rm max}$ for some
$R_{\rm max}$. The PP-correction is applied only for separations
$r < R_{\rm max}$. The PM-force on the other hand is mainly
contributed by remote particles.

\subsection{Dynamic Equations and Code Units}\label{sec_eom}
The equation of motion of particles in a Robertson-Walker Universe is
\begin{equation}\label{eq_motion_cosm}
  \frac{d^2\bf x}{d\tau^2}+\frac{1}{a}\frac{da}{d\tau}\frac{d \bf x}{d\tau} =
  -\nabla_{\bf x} \phi \ ,
\end{equation}
where~${\bf x} \equiv \{x^0, x^1, x^2\}$ is the comoving position
and $\tau$ is comoving (conformal) time. The potential $\phi$
satisfies the Poisson equation
\begin{equation}
  \nabla_{\bf x}^2 \phi = 4\pi G a^2\delta \rho\,({\bf
    x},\tau) \ ,
\end{equation}
where $\delta \rho$ is the excess of the proper density over the
background uniform density.

The equations take a simpler and dimensionless form in a special
set of units that we adopt.  The coordinates, energy and time in
our code are brought to this form. Let us denote by tildes
variables expressed in code units. Then for the units of time,
position, velocity and energy (or potential), we write
$d\tilde{t}=H_0dt/a^2=H_0d\tau/a$, $\tilde{x}=x/\Delta x$,
$\tilde{v} = v(a/H_0\Delta x)$ and $\tilde{E}=E(a/H_0\Delta x)^2$
or $\tilde{\phi}=\phi(a/H_0\Delta x)^2$, where~$a$ is the
expansion factor of the universe, $v$ is the proper velocity,
$H_0$ is the Hubble constant
and~$\Delta x$ is the cell spacing of the PM density mesh in our
code (see Sec.~\ref{sec_pm}) expressed in comoving {\rm Mpc}. In
these units, the equation of motion (\ref{eq_motion_cosm}) reduces
to
\begin{equation}\label{eq_motionxv}
  \frac{d\tilde{\bf x}}{d\tilde{t}}=\tilde{\bf v}\ ,\ \
  \frac{d\tilde{\bf v}}{d\tilde{t}}=\tilde{\bf g}
    \equiv\tilde m\tilde{\bf \Theta}({\bf r})=
    -\nabla_{\bf{\tilde{x}}} \tilde{\phi}\ .
\end{equation}
We choose units of mass so that the Poisson equation takes the
following form in dimensionless variables:
\begin{equation}\label{nodim_pois}
  \nabla_{\tilde{\bf x}}^2 \tilde{\phi} =
  \frac{3\Omega_m a}{2}\delta \tilde{\rho}\ ,\ \
  \delta\tilde{\rho}\equiv\delta\rho/\bar\rho_m\ ,
\end{equation}
where $\bar\rho_m=3\Omega_mH_0^2/(8\pi G)$ is the proper mean
matter density. Particle masses are made dimensionless by $\tilde
m=m/[\bar\rho_m(a\Delta x)^3]$.  The dimensionless total mass of
all the particles is ${\tilde M}_{\rm tot}=\dong$ where $\dong$ is
the total number of PM mesh points.  Periodic boundary conditions
are assumed in each dimension so that a finite volume simulation
represents a small portion of a universe that is homogeneous on
larger scales.

As a check on overall code accuracy, we monitor global energy
conservation by integrating the Layzer-Irvine equation, which in
code units takes the form
\begin{equation}\label{eq_layzer}
  \frac{d}{d \tilde{t}}\,(\tilde{E}_{k}+\tilde{E}_{g}) =
  \frac{\tilde{E}_{g}}{a} \frac{da}{d \tilde{t}}\;,
\end{equation}
where
\begin{equation}\label{eq_energy}
  \tilde{E}_g \equiv -\frac{1}{2\dong}\sum_{i,j} \tilde{m}_i \tilde{m}_j
  \int^{\infty}_{r_{ij}}{\tilde{\bf \Theta}}(r)\,dr\ \ \mbox{and}\ \
  \tilde{E}_k \equiv \frac{1}{\dong}\sum_i
  \frac{\tilde{m}_i \tilde{v}_i^2}{2}
\end{equation}
are the dimensionless gravitational and kinetic energies in the
simulation.  Note that in a Robertson-Walker background, the
Hamiltonian is time-dependent and so the energy is not conserved
\citep{b93}. However, we can integrate equation (\ref{eq_energy})
to get a quantity that should remain constant as the simulation
progresses,
\begin{equation}\label{eq_econ}
  \tilde{E}_{\rm con} \equiv \int\,d \tilde{E_k}+\int\,d \tilde{E_g}-\int\tilde{E_g}\,d\ln{a}\;.
\end{equation}

\subsection{Particle Data Structure and Layout}\label{ser_pa}
A particle is represented in both our serial and parallel codes by
a structure, defined as
\begin{eqnarray}\label{eq_part}
  &&{\tt typedef\ struct\ part\_t\ \{float\ x0,x1,x2, mass,
    g0,g1,g2;}\nonumber\\
   &&\quad\quad\quad\quad\quad\quad\quad\quad\quad\quad\quad\ \
   {\tt int\ id;\ float\ v0,v1,v2; \}\ part\_t}\ .
\end{eqnarray}
The size of the structure is 44 bytes on 32 bit machines.  The
structure contains three positions, the mass, accelerations and
velocities of the particles along the three spatial Cartesian
directions, all made dimensionless by the choice of units
described above. In addition, the integer ${\tt id}$ is used to
tag particles. This number can be arbitrary and is not used
anywhere in force calculations or particle propagation. In the
serial code, the particles are stored in memory simply as an array
with base pointer $\dopa$ and end pointer $\dopaf = \dopa+\donp,$
where $\donp$ is the total number of particles in the simulation
volume. To scan all the particles, e.g. for their imaging, we loop
over all the pointers ${\tt p}$ within the range ${\tt p \in
[\,\dopa, \dopaf)}$.  The particle masses are not required to be
equal to each other in general.

\subsection{Particle Integration and Timestep Criterion}\label{sec_leapfrog}
All the particles in the code are positioned within the simulation
box of size $L^i = n^i\dodx$, where $i\in\{0,1,2\}$ labels the
spatial dimension.  (We allow for unequal lengths with $n^i/n^j$
equalling a ratio of small integers.)  Periodic boundary
conditions are applied to bring particles that move outside back
into the simulation volume.  We currently use a Drift-Kick-Drift
(DKD) leapfrog integrator scheme \citep{r83,st92} to integrate the
equations of motion (\ref{eq_motionxv}) for the particles:
\begin{equation}\label{eq_leapfrog}
  \begin{array}{l}
    {\bf x}_{n+1/2} = {\bf x}_n+\frac{1}{2}{{\bf v}_n\, \Delta t}\\[3pt]
    {\rm Force \; calculation\;} {\bf g}_{n+1/2}\\[3pt]
    {\bf v}_{n+1} = {\bf v}_n+{\bf g}_{n+1/2}\, \Delta t\\[1pt]
    {\bf x}_{n+1} = {\bf x}_{n+1/2}+\frac{1}{2}{{\bf v}_{n+1}\, \Delta t}\ ,
  \end{array}
\end{equation}
where the subscripts denote timesteps.  \cite{he88} discuss the
accuracy and stability of this scheme. Note that the P$^3$M force
calculation is needed only once each timestep.

All integrators have advantages and limitations. For our problem,
which can be expressed as a continuum Hamiltonian time evolution,
the leapfrog integrator of equation (\ref{eq_leapfrog}) is a good
choice, since with a constant timestep it is {\it symplectic}.  A
symplectic integrator preserves the Poincar\'e integral invariants
and follows the time evolution under a discrete Hamiltonian that
is close to the continuum Hamiltonian of interest.  The difference
between the discrete and continuum Hamiltonian or the discrete
integrator error is itself a Hamiltonian.  When the error is a
Hamiltonian, and is sufficiently small, according to the KAM
theorem \citep{a78} the difference between the Hamiltonian paths
evolved by the two Hamiltonians is a set of finite measure.
Therefore most of the structure of the Hamiltonian flow evolved by
the continuum Hamiltonian will be preserved when evolved by the
discrete Hamiltonian with the symplectic integrator. Most of the
stable orbits in the continuum Hamiltonian system will remain
stable under the discrete Hamiltonian evolution and vice versa.

Higher order symplectic integrators for Hamiltonian evolution can
be constructed using the method of \cite{y90}, which requires more
force evaluations per timestep. In general, a $N$-th order
symplectic integrator requires at least~$N-1$ force evaluations
per complete timestep.

In a cosmological simulation, particles become more clustered with
time. It is not practical therefore to have a fixed value of the
timestep for the whole simulation. Currently we advance equation
(\ref{eq_leapfrog}) with the same value of timestep $\Delta t$ for
all the particles but allow it to change with time.  The choice
for $\Delta t$ is based on the current particle data and therefore
depends on the phase space variables.  Consequently, equation
(\ref{eq_leapfrog}) is no longer an exact symplectic map.
Nevertheless, it remains in practice well-behaved provided the
timestep varies sufficiently slowly.

The timestep must at least satisfy the leapfrog stability
criterion $(da/dx)_{\max}(\Delta t)^2 < 4$ given by equation
(4-42) of \cite{he88}. This stability requirement is essentially
equivalent to the constraint that the global timestep must be
small enough not to exceed the local dynamical time at any point
within the simulation box, $\Delta t \sim
\sqrt{\eta_t/(G\rho_{\max})}$ where $\eta_t$ is a dimensionless
constant. The density is somewhat expensive to obtain, but given
the particle accelerations and using the approximation $g \sim
GM/r^2 \sim G\rho r^3/r^2 \sim G\rho\epsilon$, we have now,
expressed in code units,
\begin{equation}\label{eq_timestep}
  \Delta \tilde{t} = \sqrt{\frac{\eta_t\,\tilde{\epsilon}}
    {\tilde{g}_{\max}}}\ ,
\end{equation}
where $\tilde{\epsilon}$ is the dimensionless Plummer softening
length (see \S~\ref{sec_force}), $\eta_t$ is a free parameter in
the code usually set to a small value such as $\eta_t=0.05$, and
$\tilde{g}_{\rm max}$ is the maximum acceleration of a particle in
the simulation box in code units.  This criterion is conservative
in assuring that the orbits of all particles are well sampled.

To further improve the integration technique, one may consider
adaptive integrators, with individual particle timesteps changing
according to the local dynamical time at the given position within
the simulation volume \citep{qks97}.  On the other hand, one may
consider higher order symplectic integrators, which would require
more force evaluations per timestep.  Some of the non-symplectic
higher order integrators, such as Runge-Kutta, are known not to
preserve the Hamiltonian flow structure even with fixed timesteps.
For example, it can be shown by integration of Kepler orbit that
the popular fourth order Runge-Kutta integrator yields a divergent
orbit very quickly. On the other hand, for second order
integrators, the DKD scheme shows stable orbits with errors
behaving as small perturbations as expected on the basis of KAM
theorem.

In this paper we adopt the leapfrog integrator with variable
timestep (set to the same value for all particles), leaving the
implementation of a higher order symplectic integrator and
individual particle timesteps for future work.

\subsection{Particle-Mesh Force Calculation}\label{sec_pm}
The particle--mesh (PM) force is the long-range force that can be
computed using Fast Fourier Transforms (FFT). In our code, we use
the FFTW Fourier transform implementation \citep{fftw} and the PM
algorithm of \cite{fb94}.  For large total number of particles
$\donp$ in the simulation box, the PM force computation is faster
than the direct summation method, requiring only $\propto \dong
\log {\dong}$ operations in total ($\dong \equiv n^0n^1n^2$ is the
number of PM grid points), as opposed to $O(\donp^2)$.

The rectangular PM-density mesh is allocated for the whole
simulation volume in the serial code. This grid is to be filled
with the density values interpolated from the particles nearby.
\cite{he88} discuss a number of methods for the density
interpolation with increasing smoothness, ranging from Nearest
Grid Point (NGP) to Cloud-in-Cell (CIC) to the Triangular-Shaped
Cloud (TSC) method. The highest of accuracy of these is given by
the TSC interpolation scheme and that is the scheme we have
implemented. As shown by \cite{he88}, an interpolation of the mass
value from a particle at position $\tilde{\bf x}$ to a grid point
at position $\tilde{\bf x}_{\rm gr}$ within the PM mesh and vice
versa takes place if and only if
\begin{equation}\label{eq_int}
  \max\limits_{i = \{ 0,1,2\}} \left| \tilde{x}^i-\tilde{x}^i_{\rm gr}
    \right|\le \dolsc\ ,
\end{equation}
where the absolute value is taken with the proper account for the
boundary conditions, and $\dolsc$ is the window function domain
locality length, specific to the interpolation scheme used, e.g.
$L_{\rm NGP}=\frac{1}{2}$, $L_{\rm CIC}=1$ and $L_{\rm
TSC}=\frac{3}{2}$. In our code we have $\dolsc = L_{\rm TSC}$.

\begin{figure}[t]
  \begin{center}
    \includegraphics[scale=0.17]{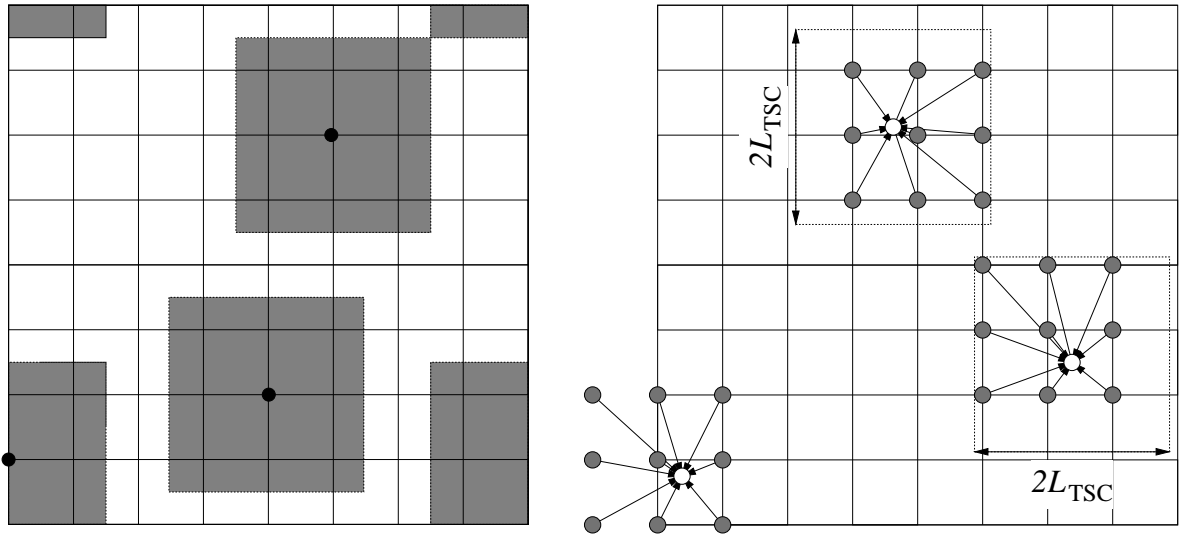}
  \end{center}
    \caption{Left: Density interpolation from particles to grid
      points {PM Step 1}.  To interpolate the density to a
      given grid point one needs to add contributions from all of
      the particles inside the shaded box of length
      $2L_{\rm TSC}$ centered on the grid point.
      Right: force interpolation from grid points to particles
      {PM Step 5}.  To get the force on a given particle
      (open circle), force values are used from all of the
      surrounding grid points.
      }\label{fg_grid}
\end{figure}

There are several steps involved for one PM force calculation:
\begin{enumerate}
  \item Density interpolation: Masses of particles are interpolated
  to a rectangular density mesh of grid points using a forward TSC
  interpolation scheme as illustrated by the left Figure \ref{fg_grid}.
  Details are given on pp. 142--146 of \cite{he88} and equation
  (A.16) of \cite{fb94}.
  \item The mesh density is Fourier transformed to the complex domain.
  \item The force is computed in the complex domain using a
  pretabulated Green's function given by equation (A.14) of \cite{fb94}.
  \item The mesh force field is inversely Fourier transformed to return
  to the real domain.
  \item Force interpolation: Forces are interpolated from the force
  mesh to particles using a backward TSC interpolation scheme, as shown
  in the right Figure \ref{fg_grid}. This step is opposite to Step 1.
\end{enumerate}

In  Step 5, information flows in exactly the opposite direction as
Step 1. Only the same grid points satisfying equation
(\ref{eq_int}) that acquired their density values from the
particles in Step 1 are used for the interpolation of the forces
to only the same particles in Step 5.  If an exchange of the
information between a grid point and a particle ever occurs, {\it
it has to be both ways}. This point will be very useful when we
discuss density and force grid messages for the parallel code in
\S \ref{sec_pm_denfor}.

The timing of the PM force evaluation scales as
\begin{equation}\label{tpm}
  t_{\rm PM}\propto A\,\donp + B\,\dong\log(\dong)\ ,
\end{equation}
where the first term is due to the density and force interpolation
and the second is due to the Fast Fourier Transform. The
coefficients $A$ and $B$ do not depend on $\donp$ and $\dong$. The
coefficient $A$ depends on the interpolation scheme used. For the
TSC interpolation scheme in $d=3$ dimensions, the density is
always interpolated from a particle to the $(2L_{\rm TSC})^d=27$
nearby grid points satisfying the condition (\ref{eq_int}). During
the force interpolation, the inverse occurs three times: once for
each of the three spatial dimensions. The factor of~$4\times27$
therefore enters into an expression for~$A$ when TSC interpolation
is used. The coefficient~$B$ is independent of~$A$ and is given by
the existing benchmarks for the FFTW implementation \citep{fftw}.

\subsection{PP Force Calculation and the Chaining Mesh}
\label{sec_pp}

In order to calculate the short range force, we must first find
all the pairs separated by less than $\dormax$. This is
accomplished using a fast linked-list sorting procedure
\citep{he88}.  At the start of a simulation the whole simulation
volume is partitioned into rectangular {\it chaining mesh} cells
whose spacings in dimension $i$ are constrained by
\begin{equation}\label{eq_dxcm}
  \dodxc^i \ge \dormax\ .
\end{equation}
Given this constraint, for any particle in any chaining mesh cell,
only the particles within the same or one of the adjacent chaining
mesh cells need be included in the short range force calculation,
since the PP force is zero for separations greater than $\dormax$.
Choosing the smallest possible value satisfying equation
(\ref{eq_dxcm}), this leads to
\begin{equation}\label{eq_dxcm_eq}
  \dodxc^i = \frac{\tilde{L}^i}{\donc^i}\ ,\ \
  \mbox{where}\ \ \donc^i = \left[\,{\tilde L}^i/\dormax\right]\ ,
\end{equation}
where the square brackets signify taking the integer part.  At the
start of the run, we sort all the particles into chaining mesh
cells occupying the 3D volume and form linked lists of particles
belonging to each cell. Each chaining mesh cell then contains the
root of the linked list to all the particles within that cell.

\begin{figure}[t]
\begin{center}
  \includegraphics[scale=1.0]{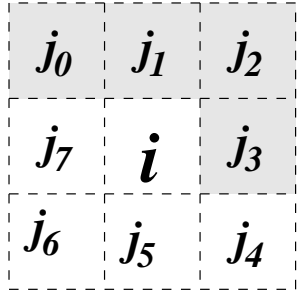}
\end{center}
  \caption{
    The cells $j_0 \ldots j_3$ constitute the 4 cells
    needed for PP force calculation in two dimensions,
    $B^{\rm PP}=\{j_0, j_1, j_2, j_3\}$.
  }\label{fg_Bpp}
\end{figure}

In order to apply a short range force correction to a particle $p$
within the simulation volume, we access particles contained within
the same cell as well as the particles within the $3^d-1 = 26$
surrounding chaining mesh cells. Since the short range correction
procedure is applied for each pair of particles within the
simulation volume, we need to traverse only half of the
surrounding cells, as illustrated in Figure \ref{fg_Bpp}.  For a
given chaining mesh cell $j$, let $N_{\rm CM}(j)$ be the number of
particles within the cell and $B^{\rm PP}(j)$ be the set of the
$(3^d-1)/2$ surrounding cells used for the short range force
calculation.  The number of floating point operations needed in
order to apply the short range force correction for every particle
within the simulation volume scales as
\begin{equation}\label{eq_tpp}
  t_{\rm PP}\propto \sum_{j} N_{\rm CM}(j)\;
  \left[ \frac{1}{2}N_{\rm CM}(j)+\sum_{j' \in B^{\rm PP}}
    N_{\rm CM}(j\,') \right] \ .
\end{equation}
The PP force calculation takes a lot of time when particles are
highly clustered because of the quadratic dependence on numbers.

%
%
\subsection{Memory requirements}\label{sec_mem}
\noindent
\begin{table}[t]
{
\scriptsize
    \begin{tabular}{|l|@{\quad} > {\PBS}p{0.3\textwidth}| @{\quad} > {\PBS}p{0.3\textwidth}|}\hline
                               & Memory size                            & Memory size, for a $32^3$ P$^3$M simulation, in bytes.\\[8pt]\hline
      Particle array           & $4\,\hbox{bytes} \times 11\donp $      & 1,441,792\\
      Particle linked list     & $4\,\hbox{bytes} \times 2\donp$       & 262,144\\
      Chaining mesh            & $4\,\hbox{bytes} \times n_c^0n_c^1n_c^2$      & 5,324\\
      Green's function         & $4\,\hbox{bytes} \times  n^0n^1(n^2/2+1)$     & 69,632\\
      Density and Force meshes & $4\,\hbox{bytes}\times 2\times n^0n^1(n^2+2)$ & 278,528\\[8pt]\hline
      Total                    & $4\,\hbox{bytes}\times(13\donp +\newline 5\don^0\don^1(\don^2/2+1)+n_c^0n_c^1n_c^2)$ & 2,057,420\\[8pt]\hline
    \end{tabular}
    \caption{\label{tb_sermem}
      Dominant memory requirements of the serial~\ser\ code
    }
}
\end{table}

The total memory requirement for the serial code consists of
several significant parts listed in Table \ref{tb_sermem}, where
the variables $n^i, n_c^i$ and~$\donp$ are defined in Table
\ref{tb_cvars} of Appendix \ref{sec_overview}. Using the serial
N-body code with an average of $p$ particles per PM gridpoint, the
total memory requirement for a P$^3$M code is
$$
M_{\rm tot} = \left(13p+\frac{5}{2}+\frac{1}{\dormax^3}\right)
  n^0n^1n^2 \times 4\ \mbox{bytes}
$$
The maximum amount of memory available for dynamic allocation for
a 32-bit machine in Unix is 2 GB. In practice the amount of memory
available for our application is about 30\% smaller. For a
simulation having one particle per density mesh cell with a cubic
grid, $M_{\rm tot} = (31/2)\, \dong^3\times 4$ bytes. The maximum
problem size for such a simulation with the upper limit on total
memory of $1.4$Gb is $\dong=296^3$.  This severe limitation on
problem size is avoided using the parallel code described in the
rest of this paper.
\section{Hilbert Curve Domain Decomposition}\label{sec_par}

In order to perform simulations with more than $296^3$ particles
and gridpoints, we distribute the computation to multiple
processors of a parallel computer.  We are using the Single
Program, Multiple Data (SPMD) model in which one program runs on
multiple processors which perform computations on different
subsets of the data.  The first decision to be made is how to
distribute the data and computation.  The computational volume is
divided into parts called domains and the memory and computation
associated with each domain is assigned to a different parallel
process.

The problem of domain decomposition is to decide how to partition
the computational volume into domains. As we will see, there are a
number of considerations that enter this decision.  This section
first describes the simplest method, one-dimensional static domain
decomposition, which is well suited for spatially homogeneous
problems but not for strongly clustered N-body simulations.  We
then introduce the Hilbert curve method of dynamic domain
decomposition used in our parallel code.

We use the word {\it process} to refer to one of the instances of
our parallel program being applied to the data in its domain.  A
process may correspond to one CPU (or one virtual CPU, in the case
of hyperthreading) or there may be multiple processes on one CPU.

\subsection{Static Slab Domain Decomposition}\label{sec_slab}

In a static slab domain decomposition, the volume is divided by
fixed planes with equal spacing.  This it the method used, for
example, in the FFTW Fast Fourier Transform \citep{fftw}.  It is
well suited for problems in which the computation is uniformly
distributed over volume.  A variation on this method is to use a
two-dimensional lattice of columns instead of a one-dimensional
lattice of slabs.

Several groups have implemented static domain decomposition in
parallel N-body codes based on PM or P$^3$M (see \S
\ref{sec_intro}). As a first step, we developed our own
implementation \parsl\ of the static slab domain decomposition
Particle-Mesh N-body code.

\begin{figure}[th]
  \begin{center}
    \includegraphics[scale=0.5]{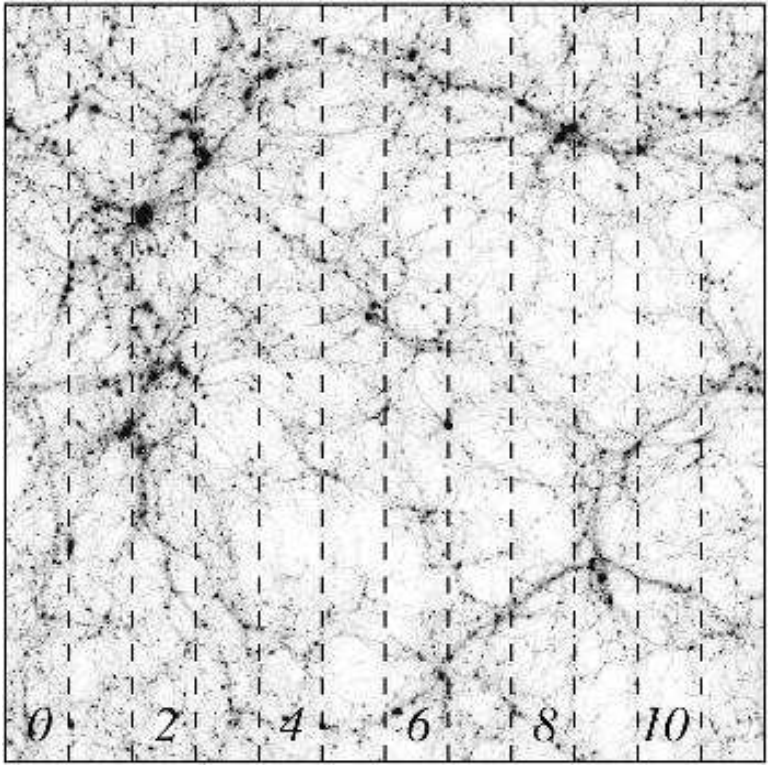}
  \end{center}
    \caption{Sample particle distribution with 12 computing processes
      and a static slab domain decomposition.  Processes 1, 2, 3, 8,
      and 9 have significantly more particles than the others.
    }\label{fg_virgo_sl}
\end{figure}

A static slab or any other static particle domain decomposition is
a good strategy when the number density distribution of particles
across the simulation box is nearly uniform and each slab contains
approximately the same number of particles to process each
timestep.  However, gravitational instability destroys the spatial
uniformity leading to serious inefficiency.  As particle clusters
grow, the memory and computational resources of the processes
containing the largest clusters (e.g. processes 1, 2, 3, 8, and 9
in Figure \ref{fg_virgo_sl}) grow quickly.  Other processes finish
their work and have to wait idle.  Worse, the heavily loaded
processes may run out of memory causing paging to disk. The
inevitable result is that the computation becomes unbalanced and
the code grinds to a halt (see the timing results in \S
\ref{sec_test} for a $512^3$ test run).  The same problem will
arise in any gravitational N-body code that uses static domain
decomposition.

Such a situation, when the performance of the cluster degrades as
a result of hugely varying workloads, is called work~{\it load
imbalance}.  In the remainder of this section we introduce an
alternative method of dynamic domain decomposition that solves the
load imbalance problem for strongly clustered systems.

\subsection{Dynamic Domain Decomposition with a Hilbert Curve}\label{sec_hc}

As we have seen from the slab domain decomposition example in the
previous section, it is important for an N-body code to load
balance.  We solve the load balancing problem by the
implementation of dynamic particle domain decomposition defined by
a Hilbert space-filling curve as suggested by \cite{pb94}.  Domain
decomposition methods based on Morton ordering (a different
space-filling curve) have been used by \cite{sw94} and
\cite{flash}.

\begin{figure}[th]
  \begin{center}
     \includegraphics[scale=0.55]{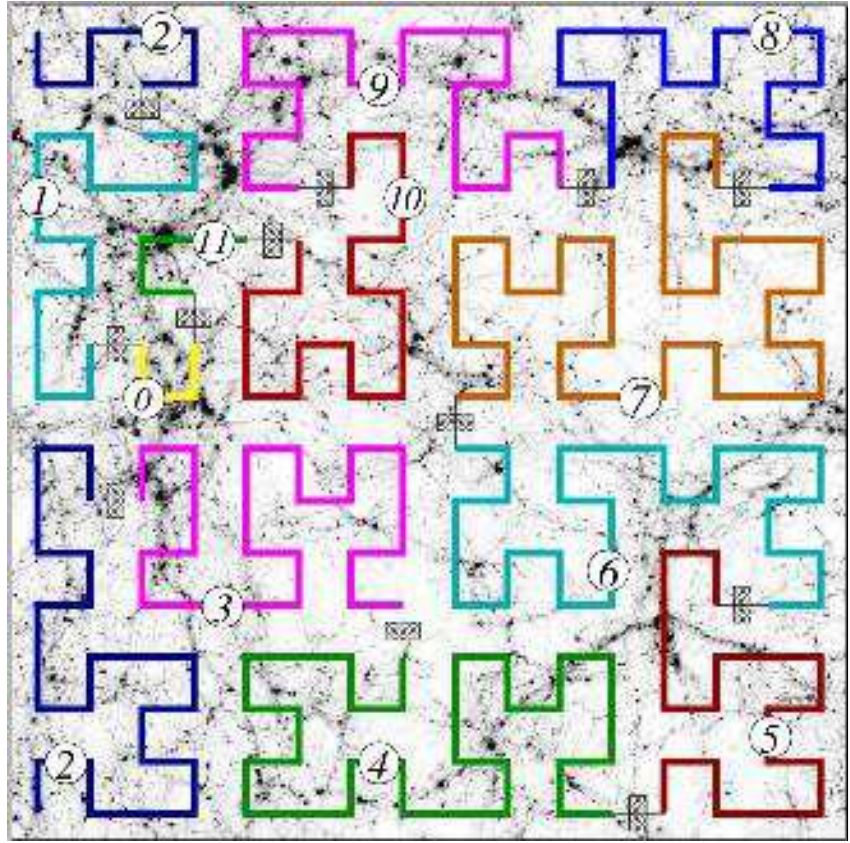}
  \end{center}
    \caption{The same matter distribution as in
      Fig.~\ref{fg_virgo_sl}, but now superimposed by a Hilbert curve.
      The Hilbert curve is divided into 12 colored segments
      separated by cross-hatched bars and labelled by the circled numbers.
      During the run the partitions will move along the Hilbert curve
      so that each process will have approximately the same amount
      of work to do.  In a real simulation the Hilbert curve is
      divided into much finer segments.}
    \label{fg_virgo}
\end{figure}

The Hilbert curve (HC) is a fractal invented by the German
mathematician in 1891 and is one of the possible space-filling
curves that completely fill a cubic rectangular volume.  A unique
HC is defined for any positive integer $m$ (the {\it HC order}),
and dimensionality $d$, for which the HC will fill each cell of a
$d$-dimensional cube of length $2^\dom$.  For $d=2$, examples are
given in Figures \ref{fg_virgo} (with $m=4$) and \ref{fg_hc8}
(with $m=3$). The HC provides a bijective (one-to-one) mapping
between the index $\doh$ along the curve (the {\it HC index}) and
the cell within the volume. In our code the mapping was provided
by the Hilbert curve implementation of \cite{doug} (see Appendix
\ref{sec_moore}). The real simulation volume and the space filling
curve we use are in fact three-dimensional but the two-dimensional
case is used in the figures throughout the paper in order to
simplify the presentation.

The main idea of Hilbert curve domain decomposition is to take a
three-dimensional volume with inhomogeneous workload and to
convert it into a one-dimensional curve that is easily partitioned
into approximately equal workloads.  The key advantage compared
with slab decomposition is that the Hilbert curve method breaks up
the problem into $2^{md}$ chunks of work with $d=3$ instead of
$d=1$.  With much finer granularity it is possible to load balance
extremely inhomogeneous problems.  In addition, the Hilbert curve
minimizes communication between processes, as we show below.

The Hilbert curve has the following properties:\newline ${\it
1)}$~{\it Compactness}: it tends to fill the space very compactly.
A~set of cells defined by a continuous section of a HC tends to be
quasi-spherical, having small surface to volume ratio.  One can
approximate the surface to volume ratio of any continuous segment
of $n$ cells along the three-dimensional Hilbert curve with
\begin{equation}\label{eq_sv}
  \mbox{S.V.R.}(n) \approx \frac{4.8}{n^{1/3}}\ ,
\end{equation}
which decreases with the increasing $n$. This approximation is
crude at small $n$. The maximum possible ratio $\mbox{S.V.R.}(1) =
26$ is reached for $n=1$, since one volume cell is surrounded by
26 adjacent surface cells.
\newline
${\it 2)}$~{\it Locality}: the successive cells along
the curve are mapped next to each other within the mesh;
\newline
${\it 3)}$~{\it Self-similarity}:~the curve is self-similar on
different scales. It can therefore be extended to arbitrarily
large size.

Figure \ref{fg_virgo} demonstrates the bijective mapping of
$16\times 16$ cells in a two-dimensional computational volume onto
the indexed Hilbert space-filling curve. The curve visits each
cell of the simulation volume exactly once.  By connecting the two
ends of the curve, the curve has the topology of a circle.  By
introducing $\donpc$ partitions along the circle (the {\it
partitioning state}) each being ascribed to one of the $\donpc$
processes in the parallel code, we specify the particle domain
decomposition of the whole simulation volume into $\donpc$ {\it
Local Region}s, each consisting of the cells along the curve
between two adjacent partitions and being assigned to one of the
$\donpc$ processes. Let us denote the local region of process $i$
defined by the partitioning state and the Hilbert curve by
$\dohclr^i$.

As we see, the space-filling curve provides an easy way of
bookkeeping for decomposition, since the local domains of each
process are completely specified by the Hilbert curve setup and
the $\donpc$ numbers that specify the partitioning state.

The surface to volume ratio of local domains defined by the
continuous segments of the Hilbert curve is small due to the
compactness property of the Hilbert curve.  This is the primary
reason for choosing a Hilbert curve as the space filling curve for
our domain decomposition. The small surface to volume ratio
significantly speeds up the reassignment of particles crossing the
$\dohclr$\ boundaries (\S~\ref{sec_adv}) and the PP-force
computation (\S~\ref{sec_hcpp}).  In the $\donp=800^3$ run
presented in \S \ref{sec_long}, the surface to volume ratio was on
average $0.1$ for the domains of voids.  In the Hydra code
\citep{mcpp98}, using a static $7\times 7$ two-dimensional cyclic
domain decomposition, the surface to volume ratio is $(4\times
7)/(7^2) \approx 57\%$, leading to more than five times as much
communication cost for the particle advancement and the
PP-calculation in comparison to our algorithm.

\subsection{Hilbert Curve Initialization}\label{sec_hci}
At the beginning of a simulation we set up the Hilbert curve
completely using the functions of \cite{doug} with an appropriate
choice of the HC mesh parameters. Only one parameter, the Hilbert
curve order $\dom$, is needed to completely specify the geometry
of a Hilbert curve filling an entire~$d$-dimensional cube of
volume $(2^\dom)^d$, which we will call~{\it the complete
HC-mesh}. Adding more parameters --- the~{\it HC mesh cell
spacings} $dx_h^i\,,\, i \in \{1,2,3\}$, the curve starting point
in the simulation volume, and the curve orientation --- completely
determines the Hilbert curve within the simulation volume.

While the real N-body simulation volume and the space filling curve are
three dimensional, two dimensional examples are used in figures throughout
this paper solely to simplify the presentation.

We use the Hilbert curve in our code only to specify the domain
decomposition for particle storage and computation.  The domain
decomposition does not affect any physical values computed. The
choice of the Hilbert curve order $\dom$ in our code is made based
solely on the parallel code performance considerations. From the
point of view of improving the resolution for particle domain
decomposition, higher $\dom$ is preferred. On the other hand each
local region cell costs additional memory, favoring lower $\dom$.
For a P$^3$M simulation, in order to simplify the force
calculation, we choose the HC mesh cells to coincide with the PP
chaining mesh cells:
\begin{equation}\label{eq_ppsimple}
  \dodxh^i \equiv \dodxc^i\ ,\quad \donh^i \equiv \donc^i\ .
\end{equation}


\begin{figure}[t]
  \begin{center}
    \includegraphics[scale=0.8]{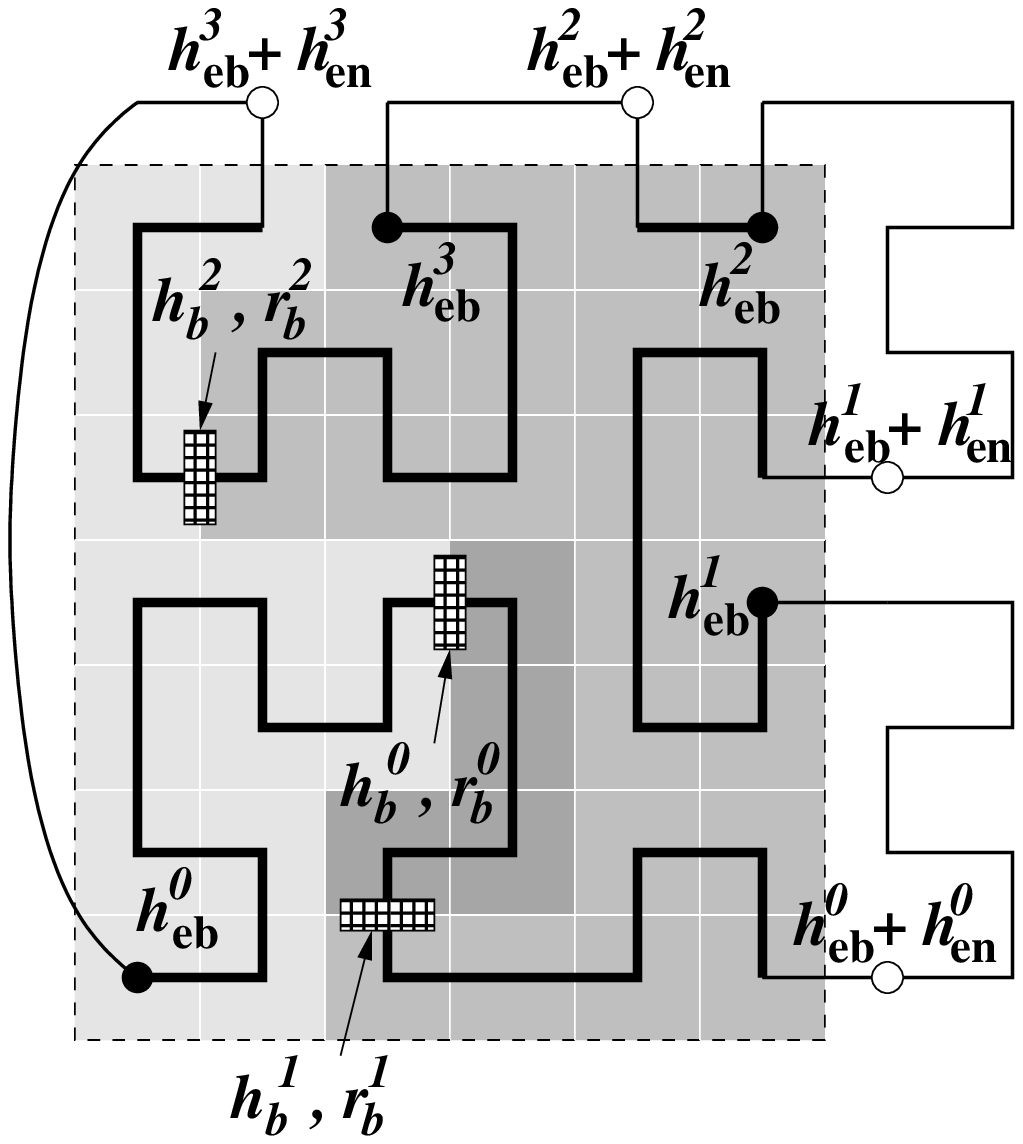}
  \end{center}
    \caption{Structure of the domain decomposition.
      A two-dimensional Hilbert curve ({\it solid line}) of order $m=3$
      fills a $6dx_h^0\times 7dx_h^1$ simulation box ({\it dashed
      line}).  By connecting the ends of the Hilbert curve, the resulting
      curve has a circular topology.  The number of processes is
      $\donpc=3$, so there are 3 partitions along the circle indicated by the
      cross-hatched bars.  We have $\dohb^i = \{11, 14, 58\}$, $\dohn^i =
      \{3, 44, 17\}$ and $\dorb^i = \{11, 14, 38\}$,  $\dorn^i=\{3, 24,
      15\}$.  }\label{fg_hc8}
\end{figure}

While the complete HC-mesh is a cube of length $2^\dom$ cells, the
chaining mesh length does not have to be a power of two. Therefore
we choose the HC order $m$ to be the smallest integer satisfying
\begin{equation}\label{eq_order}
  2^\dom \ge \donh^i\ , \quad i = \{x,y,z\}\ .
\end{equation}
From equations (\ref{eq_ppsimple}) and (\ref{eq_order}), the
complete chaining mesh is just a subset of the complete HC mesh.
If $\donh^i = 2^\dom$ for all~$i=0\ldots (d-1)$, as in Figure
\ref{fg_virgo}, the curve completely fits the simulation volume
and the two coincide.  If $2^m > n_h^i$ for some $i$, the complete
Hilbert curve mesh covers an extra space outside the chaining mesh
of the simulation volume as in Figure \ref{fg_hc8}, containing the
chaining mesh as a subset. We will refer to this submesh as the
{\it Simulation Volume HC mesh} or simply as the {\it Hilbert
curve mesh} where the context is clear.

Since the cells of the HC outside the simulation volume are
irrelevant, they do not take memory and their HC indices are
irrelevant too. Let us introduce a {\it raw HC index} along the
curve. For a HC mesh cell ${\bf \doc}$ which belongs to the
simulation volume, we define the {\it raw HC index \dor}, $\dor
\in [0, n_h^0n_h^1n_h^2)$, as the number of HC cells that the
curve spent within the simulation volume since its starting point
(HC index $\doh=0$).  In other words, while the HC index is
incremented each cell along the curve, the HC raw index is
incremented only at the cells along the curve that belong to the
simulation volume.

The mapping between the HC index $h$ and the HC raw index $\dor$
is specified completely by the {\it table of HC entries}.  Each
entry contains the HC index of an entry point $\doheb^k$ of the
curve into the simulation volume and the number of consecutive HC
cells that the HC spends within the simulation volume $\dohen^k$
before the next exit. Let $\doK$ be the number of entries in the
HC table, and let $\doK\equiv1$ if the HC mesh fits the simulation
box exactly [$\donh^i = 2^m $ for each $i=0\ldots (d-1)$]. Because
the Hilbert curve visits all the cells in the simulation box, we
have
\begin{equation}\label{domVh}
  \donh^0\donh^1\donh^2=\sum\limits_{k=0}^{\doK-1} \dohen^k\ .
\end{equation}
We denote the mapping of a cell ${\bf \doc}$ in the simulation box
into its HC raw index $\dor$ by $\domhctor({\bf \doc})$.

\begin{table}[t]
  $$
    {\scriptsize
  \begin{array}{|c|c|c|c||c|c|c|c||c|c|c|c||c|c|c|c||c|c|c|c||c|c|c|c|}\hline
    \doh&\doc^0&\doc^1&\dor&\doh&\doc^0&\doc^1&\dor&\doh&\doc^0&\doc^1&\dor&\doh&\doc^0&\doc^1&\dor&\doh&\doc^0&\doc^1&\dor&\doh&\doc^0&\doc^1&\dor\\\hline
    0  & 0  & 0  & 0  & 11 & 3  & 2  & 11 & 22 & 7  & 1  & -  & 33 & 4  & 5  & 25 & 44 & 5  & 7  & -  & 55 &  2 & 5 & 35 \\
    1  & 1  & 0  & 1  & 12 & 3  & 1  & 12 & 23 & 6  & 1  & -  & 34 & 5  & 5  & 26 & 45 & 5  & 6  & 28 & 56 &  1 & 5 & 36 \\
    2  & 1  & 1  & 2  & 13 & 2  & 1  & 13 & 24 & 6  & 2  & -  & 35 & 5  & 4  & 27 & 46 & 4  & 6  & 29 & 57 &  1 & 4 & 37 \\
    3  & 0  & 1  & 3  & 14 & 2  & 0  & 14 & 25 & 7  & 2  & -  & 36 & 6  & 4  & -  & 47 & 4  & 7  & -  & 58 &  0 & 4 & 38 \\
    4  & 0  & 2  & 4  & 15 & 3  & 0  & 15 & 26 & 7  & 3  & -  & 37 & 7  & 4  & -  & 48 & 3  & 7  & -  & 59 &  0 & 5 & 39 \\
    5  & 0  & 3  & 5  & 16 & 4  & 0  & 16 & 27 & 6  & 3  & -  & 38 & 7  & 5  & -  & 49 & 2  & 7  & -  & 60 &  0 & 6 & 40 \\
    6  & 1  & 3  & 6  & 17 & 4  & 1  & 17 & 28 & 5  & 3  & 20 & 39 & 6  & 5  & -  & 50 & 2  & 6  & 30 & 61 &  1 & 6 & 41 \\
    7  & 1  & 2  & 7  & 18 & 5  & 1  & 18 & 29 & 5  & 2  & 21 & 40 & 6  & 6  & -  & 51 & 3  & 6  & 31 & 62 &  1 & 7 & -  \\
    8  & 2  & 2  & 8  & 19 & 5  & 0  & 19 & 30 & 4  & 2  & 22 & 41 & 7  & 6  & -  & 52 & 3  & 5  & 32 & 63 &  0 & 7 & -  \\
    9  & 2  & 3  & 9  & 20 & 6  & 0  & -  & 31 & 4  & 3  & 23 & 42 & 7  & 7  & -  & 53 & 3  & 4  & 33 &    &    &   &    \\
    10 & 3  & 3  & 10 & 21 & 7  & 0  & -  & 32 & 4  & 4  & 24 & 43 & 6  & 7  & -  & 54 & 2  & 4  & 34 &    &    &   &    \\
    \hline
  \end{array}
}
  $$
  \caption{\label{tb_hcraw}
    Example of mapping of the 2-dimensional
    simulation volume shown in Fig.~\ref{fg_hc8} with a Hilbert curve. For
    each HC index $h$, the coordinates of the cell $(c^0,c^1)$ are shown
    as well as the HC raw index if the cells belongs to the simulation volume.
  }
\end{table}

Figure \ref{fg_hc8} gives an example of ${\donh}^0 \times
{\donh}^1 = 6\times 7$ simulation volume mapped by an $8\times 8$~
Hilbert curve ($\dom=3$) in two dimensions ($d=2$). Table
\ref{tb_hcraw} lists all the cells of the complete HC mesh
$\doh=[\,0,2^{d\dom}-1)$ along with the raw index of those of them
that belong to simulation volume HC mesh. The HC table of $\doK=4$
entries is
\begin{equation}\label{expldohib}
  \{ \{ \doheb^k,  \dohen^k\}: k=0\ldots (\doK-1)\} = \{\{0,20\}\,
    \{28,8\}, \{45,2\}, \{50,12\}\}\ .
\end{equation}
The simulation volume contains $42$ cells, in agreement with
equations (\ref{domVh}) and (\ref{expldohib}).

The space locality of the HC as a curve filling the simulation
volume is lost if $\doK\ne 1$.  Once the curve exits the
simulation volume, the next entry back into the simulation volume
may be far away (see Fig.~\ref{fg_hc8}).  The resulting $\dohclr$
may therefore consist of several disjoint parts, each having a
surface to volume ratio given by equation (\ref{eq_sv}). Since the
surface to volume ratio of a segment of HC decreases with
increasing number of cells in the segment, taken together those
subsegments have bigger surface to volume ratio than one big
segment of the HC of same volume.  A smaller value of
surface-to-volume ratio reduces the communication cost of PP-force
calculation by approximately the same factor (see
\S~\ref{sec_hcpp}).

\subsection{Local Regions and Partitioning State}\label{sec_partit}

To completely specify local regions $\dohclr^i$ of each process
$i\in[0,\donpc)$, we introduce $\donpc$ partitions along the
curve. A bottom partition of the process $i$ is set by the raw HC
index $\dorb(i)$ (also denoted $\dorb^i$) of the cell directly
above the partition along the HC. In Figure \ref{fg_hc8}, for
example, the entire domain is divided between three worker
processes by the three partitions with indices $\dorb^i = \{11,
14, 38\}$.

In general, a {\it partitioning state} and therefore all local
regions $\dohclr^i$, $i \in [0,\donpc)$ are completely specified
by a set of $\donpc$ numbers $\{\dorb(0)$, $\dorn(i)\hspace{-4pt}:
i=[0,\donpc-1)\}$, where $\dorn^i$ is the spacing between the
partitions $i$ and $i+1$.  This implies
$$
\dorb^i =  \left[\dorb^0+\sum_{j=0}^{i-1} \dorn^j\right]\mod
(\donhc)\ .
$$
We will denote a partitioning state symbolically by $\{\dorb,
\dorn\}$.  For the example in Figure (\ref{fg_hc8}), we have
$\dorb^i = \{11, 14, 38\}$ and $\dorn^i=\{3, 24, 15\}$.

One should always keep in mind the circular topology of the domain
decomposition data structures.  The set of Hilbert curve indices
is a circle with length $(2^\dom)^3$.  The set of the Hilbert
curve raw indices is a circle with length $\donh^0 \donh^1
\donh^2$. The set of partitions is again a circle of length
$\donpc$.

\section{Load Balancing}\label{sec_loadbal}

Having introduced the Hilbert curve, we next consider how to use
it, that is, how to choose the partitioning state each timestep.
We wish to do this so as to balance the workloads of all processes
so as to maximize the parallel efficiency.  This section discusses
details of our dynamic domain decomposition algorithm.

\subsection{Definitions of Workload, Load Imbalance, and Repartitioning}
\label{sec_repartit}

{\it Repartitioning} is the run-time (dynamic) change of particle
domain decomposition in order to solve the load balancing problem.
Repartitioning is performed by shifting the HC raw indices
$\dorb^i$ (i.e.\ the cross-hatched bars on Fig.~\ref{fg_hc8}) to
minimize the load imbalance by minimizing the resulting expected
maximum work load per process.

In a discrete time evolution problem like ours, the simulation is
synchronized among the processes each timestep, meaning that the
amount of time spent by a cluster of computers on a given timestep
is given by the maximum amount of wall clock time spent by any
process in the cluster doing its share of the problem.  We define
the {\it workload} of a process as the wall clock time that it
takes for the process to complete one timestep, including the
communication waiting time. The amount of wall clock time spent by
a process depends on the structure of the workload assignment.

Wall clock time is the number of elementary operations (clock
cycles) a processor performs for a given parallel process divided
by the CPU frequency.  During some of those cycles the processor
may be idle or working on other tasks; we call those
computationally useless periods waiting time and distinguish them
from CPU time. Because the different parallel processes must be
synchronized (at several points) each timestep, the workload of
each process is given by the wall clock time, and may be
decomposed as follows:
\begin{equation}\label{wctsum}
  \mbox{Wall Clock Time} = \frac{\mbox{CPU Time}}
    {\mbox{Average CPU Usage}} = \mbox{CPU Time} + \mbox{Waiting
    Time}\ .
\end{equation}
Wall clock time is measured using the system call {\tt
ntp\_\,gettime()}.

Ideally, we would like to eliminate the waiting time so that at
all times all CPUs are doing useful work. The waiting time has a
very complex and non-local structure as it depends on
communication and other factors unrelated to the computations done
by one process. (For example, on multiprocessor nodes, different
processes compete for memory access.) In our treatment, we balance
only the CPU time of different processes. Because the wall clock
times of all processes are forced to be the same by
synchronization, if the CPU time is balanced then there will be no
waiting time aside from the minimal amount required for
communication and memory access.

The CPU time of a process may be divided into two parts: one that
can be attributed entirely to the content of individual HC cells
(e.g. particle data) and all the rest (e.g. FFT). The dominant HC
cell-specific and CPU-intensive portions of the P$^3$M code are
the PM-density and force interpolation and the PP-force pair
summation. They execute at 100\% CPU usage (as they involve no
interprocess communication). All the contributions are summed to
define the P$^3$M {\it instantaneous CPU workload} at timestep
$\nstep$ for an HC cell at timestep $n$ as
\begin{equation}\label{workload_p3m}
  \begin{array}{lcl}
    \tilde{w}^{\rm PM}(\nstep)&=&\mbox{PM-density and force assignment wall clock time}\\[4pt]
    \tilde{w}^{\rm PP}(\nstep)&=&\mbox{PP pair summation wall clock time}\\[4pt]
    \tilde{w}^{\rm P3M}(\nstep)&=&\tilde{w}^{\rm PM}(\nstep)+\tilde{w}^{\rm PP}(\nstep)\ .\\
  \end{array}
\end{equation}
We use wall clock time to measure the CPU workload for these
portions of the computation because there is (ideally) no waiting
time.

Given a set of local {\it cell workloads} $w$ (which may differ
from $\tilde{w}^{\rm P3M}$) for all the cells $\dovc \in \dohclr$
local to each process, we define the {\it CPU workload of process
$i$} as
\begin{equation}\label{wk_pc}
  W_{\rm HC}(\dohclr^i,w)\equiv\sum\limits_{\dovc\in\dohclr^i}{w}\ .
\end{equation}
(Note that we use lower case $w$ for the workload of a single HC
cell and upper case $W$ for the total workload of all HC cells
assigned to one process.) We use a subscript HC because the total
CPU time of P$^3$M is dominated by the HC cell-specific PM and PP
computations and only these portions of the code need be included
in the workload. The other significant cost, the FFT, is
automatically load-balanced by FFTW. Note that $W_{\rm HC}$
depends on the local domains and other factors hence it may be
varied by repartitioning as discussed below.

The {\it load imbalance} is defined as a function of the set of
all CPU workload $W^i$ on each process as
\begin{equation}\label{eq_imbaltar}
  \mbox{Load Imbalance} \equiv \mathcal{L}(W) \equiv 1-\frac{\langle W
  \rangle}{\max\ W^i}\ ,
\end{equation}
giving the fraction of time that any processes are waiting instead
of computing.  The quantity $\langle W\rangle$ is the average of
$W^i$ over processes $i$.  In practice, we use $W_{\rm
HC}^i(\dohclr,w)$ for the workload $W^i$.



The cell workload defined by equation (\ref{workload_p3m}) ideally
should be proportional to the number of floating point operations
needed to compute the relevant parts of the force calculation.
However, the measured cell workload (wall clock time) is affected
by other factors.  For example, there are frequent, unpredictable
runtime changes in the efficiency of CPU cache memory management.
(Most CPUs have a speed much greater than the memory bandwidth.)
In addition, there may be multiple processes running on one
(single- or multi-processor) computing node and their competition
for system resources affects wall clock time. In addition, if some
CPUs in the cluster are slower than others, the workload
measurement for the same cell will be higher when measured by the
slower processes.

The result of these complications can be large fluctuations in the
cell workload measurements that are not repeatable from one
timestep to another and therefore interfere with our attempts to
load balance.  We represent these complications by noting that the
instantaneous cell workload defined by equation
(\ref{workload_p3m}) depends on several factors:
\begin{equation}\label{wk_cellarg}
  \tilde{w}(\mbox{particle positions},
  \mbox{CPU predictable factors},
  \mbox{CPU fluctuations})\ .
\end{equation}

To reduce our sensitivity to unpredictable CPU fluctuations, we
introduce {\it effective cell workloads} as
\begin{equation}\label{wk_cellrob}
  w(\nstep) =
  \left\{
  \begin{array}{ll}
    \frob w(\nstep-1),  & \tilde{w}(\nstep) > \frob w(\nstep-1)\\
    (1/\frob)w(\nstep-1), & \tilde{w}(\nstep) < (1/\frob)w(\nstep-1)\\
    \tilde{w}(\nstep),      & \mbox{otherwise}\ ,\\
  \end{array}
  \right.
\end{equation}
where $\frob$ is a constant parameter and $\nstep$ is the
timestep. The effective cell workload is a time average with
clipping to eliminate large fluctuations.  It is slightly more
accurate than the instantaneous workload for predicting the
workload of the next timestep. A series of tests with large
simulations showed that the optimal value parameter is $\frob
\approx 2.0$.

The {\it instantaneous} and {\it effective load imbalance} are
defined by equation (\ref{eq_imbaltar}) using equations
(\ref{workload_p3m}) and (\ref{wk_cellrob}) respectively for the
cell workloads. The instantaneous load imbalance represents the
fraction of time that the parallel processes spend idle, while the
effective load imbalance is an estimate of the same fraction in
the absence of CPU fluctuations.

Each timestep $n$, we compute the values of instantaneous
$\mathcal{L}_{\rm ins}^n$ and effective $\mathcal{L}_{\rm eff}^n$
load imbalance. We perform repartitioning each time when the value
of the effective load imbalance exceeds the maximum tolerance
value.  The {\it target partitioning state} $\{\dorb', \dorn'\}$
(see \S~\ref{sec_partit}) should be chosen so as to minimize the
expected value of the instantaneous load imbalance during the
force evaluation next timestep.  Aside from the target
partitioning state, that value also depends on the unknown cell
workloads at the next timestep. To find the optimal partitioning
state, one may estimate the cell workload in the next timestep
very well using its latest measured value
\begin{equation}\label{eq_wknext}
  w(\nstep+1) \approx w(\nstep)\ .
\end{equation}

As illustrated by equation (\ref{wk_cellarg}), the cell workload
during the next timestep is a function of the unknown particle
positions at the next timestep. However, since particles do not
move far in one timestep compared to the size of a HC cell, we can
ignore this dependence for now. The other two arguments factors
determining the cell workload are due mainly to the effectiveness
of CPU cache memory management, which depends on the memory layout
and is hard to predict.  The main change in the memory layout
during the next timestep is a different partitioning state which
means different local regions. By introducing the technique
described in \S \ref{sec_adv}, we eliminate the dependence of the
second argument in equation (\ref{wk_cellarg}) on local region
assignment. The third argument of equation (\ref{wk_cellarg}) can
not be eliminated and is the main cause of inaccuracy of equation
(\ref{eq_wknext}), as demonstrated in \S\S \ref{sec_long} and
\ref{sec_sclb} using test simulations.

The {\it residual load imbalance} is defined as the minimum
possible load imbalance, computed with equations (\ref{wk_pc}) and
(\ref{eq_imbaltar}) allowing for arbitrary repartitioning, based
on the effective cell workloads of the current timestep:
\begin{equation}\label{eq_resimbal}
  \mathcal{L}_{\rm res}(W') = \min\limits_{\{\dorb', \dorn'\}}
  \mathcal{L}(W) \ .
\end{equation}
We seek to find the partitioning state that minimizes
$\mathcal{L}_{\rm res}(W')$, called the {\it target partitioning
state}.  With this choice of partitioning, $\mathcal{L}_{\rm
res}(W')$ will become an estimate for the effective load imbalance
of the next timestep.

Even in the absence of CPU fluctuations, the residual load
imbalance cannot be reduced to zero because of the granularity of
the workload distribution across HC cells.  For an extremely
clustered matter distribution, the workload $w_{\max}$ of the
densest HC cell within the simulation volume may be greater than
the average workload of all processes, $w_{\max} \ge \langle W
\rangle$. (This requires extreme inhomogeneity because most
processes have thousands or even millions of HC cells associated
with them, while the slowest to finish may have only one HC cell.)
The granularity of the HC method requires that each process have
at least one HC cell. In this case, the residual load imbalance is
bounded by
\begin{equation}\label{eq_resimbalcon}
  \mathcal{L}_{\rm res} \ge 1-\frac{\langle W \rangle}{w_{\max}}\ .
\end{equation}
In this regime there is no point in extending the problem to a
larger number of processes, since the wall clock time will be
given by that of the process holding the cell $w_{\max}$
(\S~\ref{sec_sclb}).
In general, the N-body problem is scalable only up to a number of
processes given by
\begin{equation}\label{eq_npclim}
  \donpc \le \frac{W_{\rm tot}}{w_{\max}}\ .
\end{equation}
Improved load balance can be achieved by further subdividing the
computation of short-range forces using an adaptive mesh
refinement technique, as we will demonstrate in a later paper.

\subsection{Repartitioning and Memory Balancing}\label{sec_rep}
As discussed in \S \ref{sec_partit} the local regions at any given
time are completely specified by the current partitioning state
$\{\dorb, \dorn\}$.
The target partitioning state is given by a primed set $\{\dorb',
\dorn'\}$. The target partitioning state can be reached from the
initial one by a sequence of sets of $\donpc$ non-overlapping {\it
elementary partition shifts} $\Delta \dorb^i$ along the circle
indexed with the HC raw indices, so that
$$
\dorb^{\prime\,i} = \dorb^i+\sum_{i=0}^{\donpc-1}\Delta \dorb^i\ .
$$
It is efficient to perform each set of the elementary partition
shifts in two stages: first by moving simultaneously all the even
partitions followed by the movement of all the odd ones.  This
way, during each of the two stages, the entire process group will
decouple into pairs of adjacent processes each involved with an
elementary partition shift exchanging particles with the other
process in the pair.

Given the initial and target partitioning states, each partition can be
moved from its starting to its target state in one of two possible
directions along the circle.
We define a parametric isomorphic linear mapping $R_b$ that takes
the initial partitioning state $\{\dorb, \dorn\}$ into the target
one $\{\dorb', \dorn'\}$ as the parameter $\alpha$ goes from zero
to one:
\begin{equation}
  \begin{array}{l}
    R_b^0(\alpha)\equiv\dorb^0 + \alpha \,\left[
    (\dorb^{\prime,0}-\dorb^0) \mod{(\donhc)}\right]\ ,\\[5pt]
    R_n^i(\alpha)\equiv\dorn^i + \alpha \left[\; \dorn^{\prime\,i}-\dorn^i
      \;\right]\ ,\\[5pt]
  \end{array}
\end{equation}
where $\donhc$ is the total number of HC cells and the partition
$i=0$ is treated so as to ensure a circular topology.  It follows
that
\begin{equation}\label{eq_rbalpha}
    R_b^i(\alpha) = R_b^0(\alpha)+\sum\limits_{i=0}^{j-1}R_n^j(\alpha)\ .
\end{equation}
The initial and target partition state starting indices are given
by $r_b^i = R_b^i(0)\,\mod{(\donhc)}$ and $r^{\prime\,i}_b =
R_b^i(1)\,\mod{(\donhc)}$, respectively.  The direction of
movement of the individual partitions along the circle in our code
is given by differentiating equation (\ref{eq_rbalpha}) with
respect to $\alpha$.

The target partitioning state is reached from the initial one by
the sequence of maximal non-overlapping elementary partition
shifts in the directions specified by the above procedure until
the target partitioning state is achieved.
All of the partition-dependent data are adjusted to reflect the
change of partitioning state. The corresponding particle sends and
receives are performed and the relevant cell data are exchanged.
In addition, the irregular particle domains are reallocated for
each process participating in any of the resulting elementary
partition shifts.

In order to avoid paging one needs to impose a total memory
constraint for repartitioning. Since the memory associated with
particles dominates the problem, while doing repartitioning we
check whether the reallocation of the particle array on the
receiving processes succeeds. If it does not, we divide the
requested number of cells $|\Delta \dorb^i|$ by two and try the
repartitioning again.  This procedure guarantees that we satisfy
the memory limit on each process.

Another practical consideration arises when using a cluster with
multi-processor or multi-process nodes.   As a result of Hilbert
curve domain decomposition the memory loads and cache usage of
sequential processes are correlated.  These correlations can make
it more difficult to achieve load balance.  One should therefore
avoid assigning sequential processes to the same computational
node.

\subsection{Finding the Optimal Target Partitioning State}\label{sec_reptar}

In this section, we show how to find the target partitioning state
$\{\dorb', \dorn'\}$ that minimizes load imbalance (eq.\
\ref{eq_imbaltar}), given the current HC cell workloads and the
current partitioning state $\{\dorb, \dorn\}$.  As discussed in \S
\ref{sec_repartit}, we assume that the current cell workloads are
an adequate predictor of those at the next timestep, equation
(\ref{eq_wknext}).

\subsubsection{Cell Workload Data Compression}\label{sec_wdis}

\begin{figure}[t]
  \begin{center}
    \includegraphics[scale=0.092]{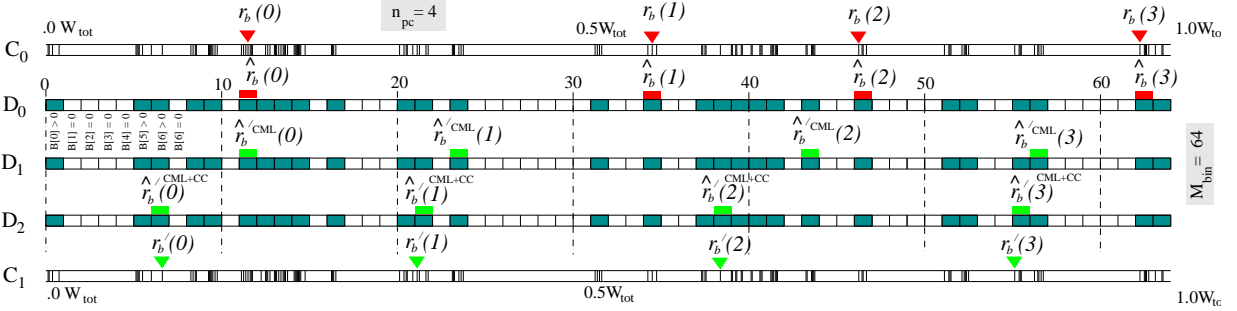}
  \end{center}
    \caption{Representation of the HC cell workloads using
      continuous ($C_0$ and $C_1$) and discrete ($D_0$, $D_1$, and
      $D_2$) workload bars, as described in the text.  This example
      is for a simulation on $\donpc=4$ processes, with HC-mesh of
      size $\donh^0=\donh^1=\donh^2=100$ and $N_{\rm bin} = 16$.
      $C_0$ and $C_1$ are the continuous total workload bars before
      and after repartitioning, respectively.  The filled triangles
      give the locations of the initial ($C_0$) and target ($C_1$)
      partitions.  A bin $B(k)$ along the bar $D_0$ is filled if and
      only if the number of dashes in the same interval of bar $C_0$
      is non-zero. The discrete partitioning states are marked by the
      filled rectangles above the filled bins of bars $D_0$--$D_2$.
      The solution in the discrete space marked on bar $D_2$ is
      obtained by first repartitioning $D_0\rightarrow D_1$
      [holding $r_b(0)$ fixed] and then shifting $D_1\rightarrow D_2$.
      Finally, the continuous target partitioning state $\{\hat{r}'_b,
      \hat{r}'_n\}$ marked on bar $C_1$ follows from $D_2$.  Note that
      the topology of each bar is a circle formed by connecting
      its ends.}\label{fg_partit}
\end{figure}

The optimal target partitioning state depends on the workloads of
every HC cell on every process, $w(j)$ for $j\in[0,\donhc)$.  This
information can be represented as a one-dimensional {\it
continuous total workload bar} of length $W_{\rm tot}$ equalling
the total work summed over all cells.  For each HC cell we mark
the bar with vertical dashes at positions
\begin{equation}\label{eq_u}
  u(r)=\sum_{j=0}^r w(j), \quad r =0,\ldots,\donhc-1\ ,
\end{equation}
which gives the cumulative workload of cells up to the one with
raw index $r$.  Figure \ref{fg_partit} illustrates this with
continuous total workload bars $C_0$ and $C_1$. The horizontal
spacings between the adjacent dashes (the white stripes) represent
the cell workloads of each cell: $w(r)=u(r+1)-u(r)$. Each white
stripe is due to the cell workload associated with one cell. A
single dash however may be an overlap of thousands of very close
dashes showing up as one due to the limited resolution of the
figure.

In a large N-body simulation, the total number of HC cells is
huge. For example, in the simulation described in \S
\ref{sec_lcdm}, $\donhc = 2.36\times 10^7$, which
requires~$\donhc\times{\tt sizeof(float)} = 94.5\mbox{MB}$ to hold
the values of the workloads.  This memory requirement grows with
the volume of the simulation box and if the mesh is large enough
the problem of finding the optimal partitioning state is
impossible to process serially (i.e. on one of the cluster nodes).

To solve this problem we compress the cell workload data by
discretizing it.  The total workload bar is divided into $N_{\rm
bin}$ segments per process, or $M_{\rm bin}=\donpc N_{\rm bin}$
segments in total.  The continuous total workload array $u(r)$ is
replaced the much smaller array $B(k)$ with $k\in[0, M_{\rm
bin})$. Figure \ref{fg_partit} illustrates this with the bars
$D_0$, $D_1$, and $D_2$. Each array member $B(k)$ is assigned to
the subinterval $[k\,\Delta W , (k+1)\Delta W )$ of the total
workload bar, where~$\Delta W \equiv W_{\rm tot}/M_{\rm bin}$. The
value $B(k)$ is defined as the number of cell boundaries (the
dashes) within the corresponding subinterval of the total workload
bar. The non-zero members $B(k)> 0$ correspond to the filled
rectangles of bars $D_0$--$D_2$ in Figure~\ref{fg_partit}.

Suppose we start from the initial partitioning state $\{\dorb,
\dorn\}$ marked by triangles above $C_0$ in Figure
\ref{fg_partit}. We define a {\it discrete partitioning state}
$\{\hat{r}_b, \hat{r}_n\}$ in the discrete workload space by
$\hat{r}_b^i \equiv [u(\dorb^i)/\Delta W]$, $0 \le i < \donpc$,
where the square brackets signify taking the integer part;
$\hat{r}_n^i$ is the spacing between the consecutive $\hat{r}_b^i$
along the binned bar of length $M_{\rm bin}$.  We define the
workloads in the discretized problem as
$\hat{W}^\equiv\hat{r}_n^i$.  Following equation
(\ref{eq_imbaltar}), the load imbalance of a discrete partitioning
state is defined by
\begin{equation}\label{eq_disimbal}
  \mathcal{\hat{L}}[\hat{r}_n] = 1-\frac{\langle \hat{r}_n \rangle}
    {\max \hat{r}_n}\ .
\end{equation}
The residual load imbalance is redefined in the discrete space as
[cf. eq.~(\ref{eq_resimbal})]
\begin{equation}\label{eq_probdisimbal}
  \mathcal{\hat{L}_{\rm res}}[\hat{r}'_n] = \min\limits_{\{\hat{r}'_b,
  \hat{r}'_n\}:\;B(\hat{r}'_b) > 0} \mathcal{\hat{L}}[\hat{r}'_n] \
  .
\end{equation}
The problem of load balancing is posed in the discrete space as
finding the discrete target partitioning state $\{\hat{r}'_b,
\hat{r}'_n\}$ that will minimize the load imbalance.  We discuss
how this is done in the next subsection.

Once the discrete target partitioning state $\{\hat{r}'_b,
\hat{r}'_n\}$ is found, the continuous target partitioning state
$\{\dorb', \dorn'\}$ is also found by setting $\dorb^{\prime\,i} =
r^i$, where $r^i$ is the raw HC index of any cell such that $
u(r^i) \in [\,\hat{r}^{\prime\,i}_b\Delta W,
(\hat{r}^{\prime\,i}_b+1)\Delta W]$.
There are, in general, many HC indices that will accomplish this.
For example, in Figure \ref{fg_partit}, the final triangles for
bar $C_1$ may be placed at any dash lying beneath the rectangles
above bar $D_2$.  The choice is arbitrary and this freedom in
setting the target partitioning state will result in negligible
differences in the residual load imbalance $\le 2/M_{\rm bin}$.
In practice, we set the partition at the first HC cell that lies
in the desired interval.

\subsubsection{Finding the Target Partitioning State in the Discrete Case}
\label{sec_disreptar}

There are two practical approaches to solving the discrete target
partitioning state problem of equation (\ref{eq_probdisimbal}).


In the {\it cumulative repartitioning} approach we keep the zeroth
partition fixed  while setting the other ones as close as possible
to being equally spaced along the discrete workload bar, subject
to the constraints $B(\hat{r}^{\prime\,i}_b)\ne 0$. It is evident
that the resulting target partitioning state is a function of only
the initial position of the zeroth partition $\hat{r}^0_b$ and the
discrete workload array $B^k,\ k\in[0,M_{\rm bin})$.

The cumulative approach alone is not satisfactory for optimizing
the discrete load imbalance equation (\ref{eq_disimbal}) when the
cell workloads of some of the HC cells far exceed the
discretization load $w(j) \gg \Delta W,\; j \in [0,\donhc)$.
Indeed this problem is illustrated in Figure \ref{fg_partit}. The
initial discrete partitioning state is given by $\hat{r}_b = \{11,
34, 46, 62\}$ as shown by the rectangles above the workload bar
$D_0$. Applying the cumulative approach using the above rule, we
have $\hat{r'}_b^{\rm CML} = \{11,23,43,56\}$, and
$\hat{r'}_n^{\rm CML} =\{12,20,13,19\}$, yielding load imbalance
$\mathcal{L}^{\rm CML} = 1-(16/20)=0.2$, which is relatively poor.
(The superscript CML is used for partitions found with cumulative
repartitioning.) This approach uses only the position of the
zeroth partition and the discrete cumulative workload array. It is
insensitive to differences in the adjacent workloads, e.g.
$\hat{r}_n^i$ and $\hat{r}_n^{i+1}$.

In the {\it circular cyclic correction repartitioning} approach
(denoted by superscript CC), we start from a partition $i$ and
shift it to the bin $\hat{r}_b^i=k$ such that it is the closest
possible distance to the bin in the middle of the two adjacent
partitions, $k = ([\hat{r}_b^{i-1}+\hat{r}_b^{i+1}]/2$. After the
correction of the partition $i$ is done, we move on to the next
partition $i+1$, applying the same technique but using the already
corrected value for the position of partition $i$. We then
continue applying the same scheme for all the other partitions in
cycles along the circle $i \in [0, \donpc)$ until the resulting
shifts for all partitions $i \in [0, \donpc)$ become zero.  The
resulting positions of the partitions will define the target state
in the circular cyclic correction repartitioning approach. This
approach if used alone is not satisfactory just as for the
cumulative partitioning approach above, however the nature of the
problem is completely different. If a large variation in workload
$\hat{r}_n^i$ develops across a large range of indices $i$ (e.g.
between $i$ and $i+\donpc/d$), this variation will not be
suppressed by the circular cyclic correction scheme since only the
adjacent partitions $\hat{r}_b^{i-1}$ and $\hat{r}_b^{i+1}$ are
used for correction of any given partition $\hat{r}_b^i$.  On the
other hand, all the local fluctuations in workload will be
suppressed very effectively.

As we see, the cumulative repartitioning approach and the cyclic
circular partitioning approaches smooth the large scale and small
scale (in terms of the range of indices) workload fluctuations
respectively. Applying the two approaches in sequence works well
to provide a nearly optimal solution for the discrete workload. In
the example of Figure \ref{fg_partit}, the bar $D_2$ shows the
result of applying the circular partition correction approach to
the output of the cumulative approach (bar $D_1$) obtained from
the initial discrete partitioning state (bar $D_0$). As follows
from the bar $D_2$ of Figure \ref{fg_partit}, the resulting target
partitioning state is $\hat{r}_b^i=\{6,21,38,55\}$ and
$\hat{r}_n^i = \{15,17,17,15\}$.  The resulting discrete load
imbalance is $\mathcal{\hat{L}} = 1-(16/17)=0.06$ is 3.4 times
smaller than the load imbalance obtained using only the cumulative
method. Our experiments show that the combination of the two
approaches results in a good approximation to the load-balanced
target partitioning state.  The residual load imbalance is
generally limited not by our ability to find the optimal solution
but instead by the CPU time fluctuations due to variations in
cache usage.

\section{Particle Data Layout and Communication}\label{sec_layout}

In a serial code, the array of particle structures (\ref{eq_part})
is static, that is, it remains fixed length with unchanging
particle labels.  In a parallel code with domain decomposition,
particles may move from one process to another. This not only
requires interprocessor communication, it also complicates the
storage of particle data.  This section discusses our solutions to
these problems.

\subsection{Linked List Structure, Particle Movement, and Sorting}
\label{sec_adv}

\begin{figure}[t]
  \begin{center}
    \includegraphics[scale=0.39]{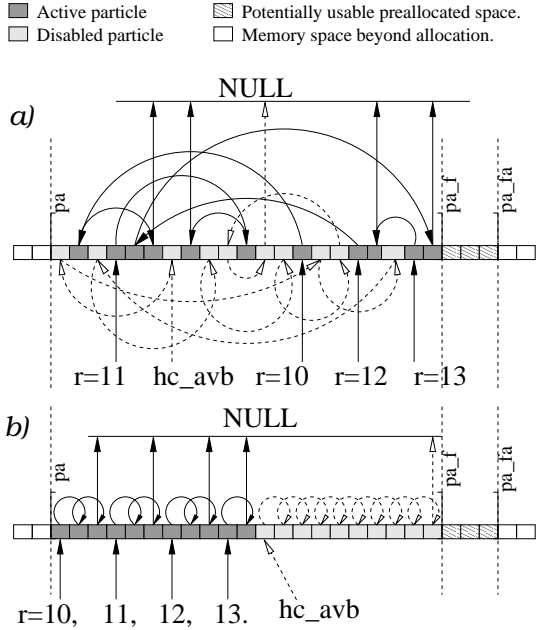}
  \end{center}
    \caption{Particle array structure and access in the parallel
    code.  This example corresponds to process $i=0$ of
    Fig.~\ref{fg_hc8}.  The HC cells associated with this process
    are $r=(10,11,12,13)$.  The particle arrays are the horizontal
    bars (with disabled particles corresponding to gaps
    in the array opened up when particles moved to other
    processes).  The linked list is given by the arrows going from
    one particle to another; the solid (dashed) arrows give the
    linked list for the active (disabled) particles.  The linked
    list roots are the pointers \dohcavb\ (for disabled particles)
    and $r=(10,11,12,13)$ (for active particles) beneath the
    particle array bars.  Each linked list begins at a root and
    ends with the {\tt NULL} pointer.  The particle array is
    allocated slightly more storage (\dopafa) than needed
    (\dopafa).
    a) The particle array and linked list before sorting.
    b) The same particles and the linked list after sorting.
      }\label{fg_linked}
\end{figure}

The particle data are stored as a single local particle array of
pointer ${\tt [pa, pa\_f)}$ on each process.  A slightly larger
range ${\tt [pa, pa\_fa)}$ is allocated to avoid reallocation
every timestep. In addition to the particle array, we have a
linked list that tells which particles lie in each HC cell.  For
each HC cell there is a pointer (the root) that (if it is
non-null) points into the particle array to the first particle in
that HC cell. A complete list of particles within a given local HC
region $\dohclr^i$ is obtained by dereferencing the appropriate
linked list root and then following the linked list from one
particle to the next, as illustrated in Figure \ref{fg_linked}.
The linked list also has a root \dohcavb\ that points to disabled
particles.

There are several challenges associated with this simple linked
list method of particle access.  First, one must transfer
particles between processes. Second, HC cells are themselves
exchanged between processes as a result of repartitioning.  Third,
one must optimize the traversal of the linked lists to optimize
code performance.  Finally, one must specify which HC cells are
associated with a given process.  We discuss these issues in the
remainder of this section.

During each position advancement equation (\ref{eq_leapfrog}),
twice every timestep some particles move across the boundary of
their local particle domain.  As a result, such a particle is sent
from a process $i$ to another process~$j$ whose local region
$\dohclr^j$ it entered. Particles may cross the boundary of any
pair of domains.
The associated communication cost scales linearly with the
$\dohclr$\ surface area.  The Hilbert curve domain decomposition
minimizes this cost because of the low surface to volume ratio
(\S~\ref{sec_hc}).

When a particle ${\tt p}$ moves outside the local region
$\dohclr^i$, it leaves a gap in the local particle array.  We set
the particle mass to $-1$ and call this particle array member a
disabled particle.  All the disabled particles on each process
form a separate linked list with root \dohcavb.  The particles
entering $\dohclr^i$ from other processes replace the disabled
particles or are added to the end of the particle array.

As a particle initially in process $i$ crosses a boundary to
another process, the id of the target process $j$ should be
immediately found in order to send this particle to the new
process. Dividing the new particle coordinates by the HC mesh
spacing gives the new Hilbert curve mesh cell coordinates ${\bf
\doc}$.  The target process id can then be found calling Moore's
function for the new HC index $h=\domhctoi({\bf \doc})$. By using
the current Hilbert curve partitioning, one finds the id of the
target process $j$ from $h$. Once all particles to move have been
identified, the particles are transferred between processes.

\begin{figure}[t]
  \begin{center}
    \includegraphics[scale=0.7]{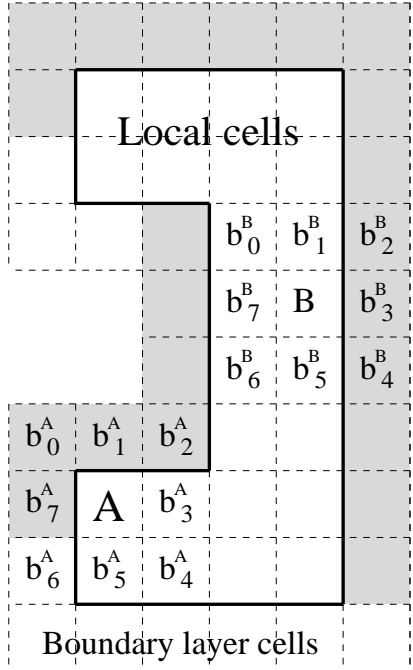}
  \end{center}
  \caption{Hilbert curve mesh cells. The cells within the
    solid line are the $\dohclr^i$\ cells containing all the
    particles assigned to process $i$.  Information about the
    layer of boundary cells (all gray and white cells outside the
    local region) is also stored by process $i$.  This information
    is used both when particles are transferred between processes
    and during the short-range (particle-particle) force
    computation.  In the latter case, the particle data for the
    shaded cells is used to compute forces on particles in cells A
    and B as discussed in \S~\ref{sec_hcpp}.
    }\label{fg_pp}
\end{figure}

As we show in Appendix \ref{sec_moore}, Moore's function calls are
relatively expensive. To avoid having this cost each time a
particle crosses the boundary, we allocate an extra one layer of
HC cells surrounding the boundary of $\dohclr^i$, as shown in
Figure \ref{fg_pp}, and we mark the surrounding cells with the ids
of the appropriate processes $j$ by calling Moore's function for
each of them exactly once. By doing this once, we avoid calling
Moore's functions in the future. However we still have to call the
function for the very small fraction of the boundary-crossing
particles that went further than one boundary layer cell in one
timestep.  The extra layer of HC cells surrounding the local
region is also used with the particle-particle force computation
as described in \S \ref{sec_hcpp}.

We maintain the particle linked list throughout the simulation
instead of reforming it each timestep. As particles cross from one
HC cell to another --- even if they are in the same local region
$\dohclr^i$ --- the linked list is updated to reflect these
changes.
The particle array is reallocated whenever the fraction of
disabled particles exceeds a few percent (the exact value is a
parameter set by the user), or the amount of particles exceeds the
boundary of the pre-allocated particle array ${\tt pa\_fa}$.

In addition to the pointer to the root of the linked list that
contains all the particles within each HC cell, each cell of the
local region contains other structure members: the process number
the cell belongs to, the current and previous timestep cell
workloads required by equation (\ref{wk_cellrob}), the number of
particles in this cell, etc.  We will refer to this structure as
the {\it HC cell structure} and the array of structures for all HC
cells the {\it HC cell array}.  One member of this array has size
16 bytes.  When repartitioning occurs, we send and receive the
relevant HC cell array members and the particles they contain to
the appropriate processes.

Some program components, such as particle position advancement,
require access to the complete particle list on each process. All
local particles can be accessed using the particle array and
filtering out the passive particle array members as follows:
\begin{equation}\label{forloop}
  \begin{array}{l}
    \tt
    for( p = \dopa;\; p < \dopaf;\; p++ )\{\\\tt
    \quad if(p\ptr mass =\hspace{-.2mm}= -1.)\; continue;\\\tt
    \quad \ldots\\\tt
    \}\ .
  \end{array}
\end{equation}
We found that because of cache memory efficiencies, it is up to
ten times faster to use a simple array to access every local
particle than it is to dereference the three-dimensional linked
list roots for each of the local cells of $\dohclr^i$. The reason
for such difference is that simple array members are sequential in
the machine memory, while the successive linked list members are
not, and the CPU cache memory is more effectively used when data
are accessed sequentially in an array.  The improvement in
efficiency is especially important in the particle-particle
calculation because each particle is accessed many times during
one force computation.

Here we introduce a fast sorting technique that places the
particle data belonging to the same HC cell sequentially within
the segments of the particle array, ordered by increasing HC-cell
raw index.  This sorting procedure is performed each timestep
before the force computation.

Every timestep, before a force calculation, we follow all
the~$\dohclr$ cells in the order of their raw HC index, and
concatenate their linked lists, resulting in just one linked list
of all the particles in the local particle array. Then, using the
unnecessary acceleration {\tt g0} and {\tt g1} members of the
particle structure as pointers, we form an extended linked list
replacing the old one. The result is a new linked list which can
be traversed both forward (using {\tt g1}) and backward (using
{\tt g0}). Then, starting from the first particle of a simple
array of particles, we swap it with the first particle in the
extended linked list while the forward and backward pointers of
the immediately adjacent within the extended linked list particles
being updated. We then proceed to the next particle in the simple
array and in the linked list doing the same, until we have sorted
the entire particle list. The result of this sorting is
illustrated by Figure \ref{fg_linked}b.

In addition to optimizing the CPU cache memory usage, the above
sorting technique eliminates the need to allocate an additional
buffer for sending and receiving particles while repartitioning,
because all the particles to be moved as the result of
repartitioning will occupy contiguous segments in the simple
particle array.  When the sorting is completed the original linked
list is unnecessary and is deallocated in order to be formed again
directly using the sorted particle array, before the particle
advancement and repartitioning take place.

To transfer particles between processes we use a modification of
{\tt MPI\_Alltoallv} that assures no failure will occur if
insufficient memory has been pre-allocated for the send and
receive buffers. This achieved by using {\tt MPI\_Alltoall} to
exchange the numbers of particles to be sent and received and then
using as many {\tt MPI\_Alltoall} and {\tt MPI\_Alltoallv} calls
as necessary to avoid overflowing the available memory of each
processor.

\subsection{Scalable Allocation Local Region Access}
\label{sec_voids}

As mentioned above, during particle exchange and force computation
one needs frequent access to a cell's particle list and other cell
data, given the indices ${\bf \doc}$ of the cell in the HC mesh.
The most obvious method is to call Moore's function $h =
\domhctoi({\bf \doc})$ to get the global HC index and then use our
table of HC entries (\S~\ref{sec_hci}) to convert $h$ into the raw
index $r$. The raw index then gives the root to the particle
linked list as shown in Figure \ref{fg_linked}.  This method is
unsatisfactory because of the expense of calling Moore's function
many times during the force evaluation.

\begin{figure}[t]
  \begin{center}
    \includegraphics[scale=0.44]{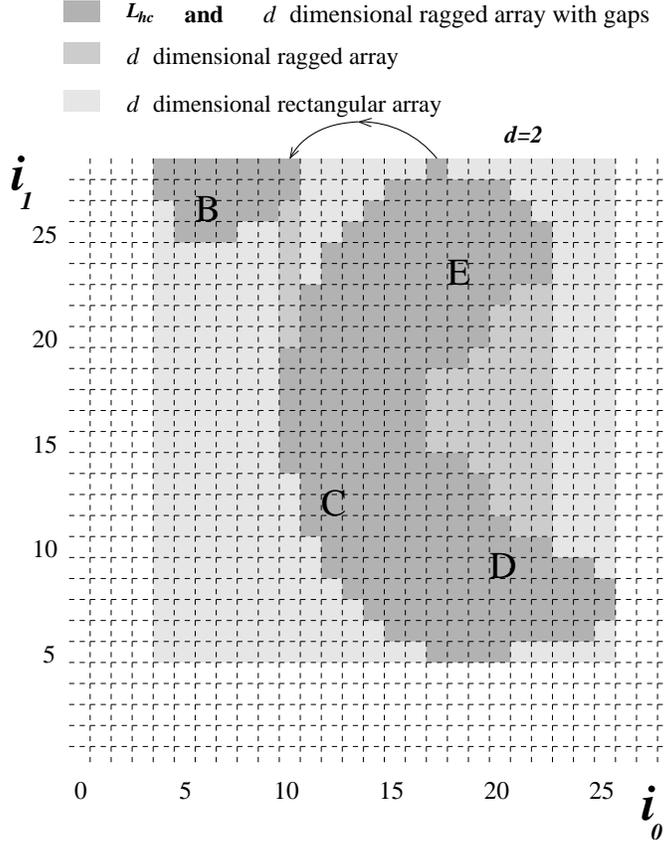}
  \end{center}
    \caption{Schematic illustration of the HC local region $\dohclr$
    (dark gray) assigned to one process and the rectangular array
    that includes it (light gray combined with dark gray).  The
    ragged array (middle gray combined with dark gray) requires
    much less storage but only the ragged array with gaps (dark gray)
    corresponds exactly to $\dohclr$.   Four cells belonging to
    $\dohclr$ are randomly selected and labelled B, C, D and E.
    }\label{fg_lrbad}
\end{figure}

Another simple method of allocation for the $\dohclr$\ cells would
be a {\it $d$-dimensional rectangular array} of cells holding the
frequently used roots of the linked lists to the particles
contained in this cell and the total number of particles within
it. The access to a HC cell given its coordinates ${\bf \doc}$ in
this case is given by dereferencing the array $r=A[c_0][c_1][c_2]$
in the case of $d=3$, where $A$ is an array of HC cell raw indices
(or pointers to HC cells) updated after each repartitioning. The
problem here, illustrated in Figure \ref{fg_lrbad}, is that many
of the entries of $A$ are wasted because the HC local regions are
not rectangular parallelpipeds.  This can be improved by adjusting
the bounds of the array indices $(c_0,c_1,c_2)$ to the extremal
values for cells in the local region.  The result is a simple {\it
$d$-dimensional ragged array}, also illustrated in Figure
\ref{fg_lrbad}.

The optimal method of local region HC cell allocation and access
is to add one more dimension to the array of HC cell pointers $A$
used in a simple ragged array.  The extra dimension accounts for
variable number of disjoint parts in the last dimension.  This
method allocates the minimal storage needed beyond the number of
HC cells in $\dohclr^i$.  We call this a {\it d-dimensional ragged
array with gaps}.  The HC cell is then obtained by dereferencing
the $(d+1)$-dimensional array $A$.

To access a cell with coordinates $c_0 \ldots c_{d-1}$ using a
$d$-dimensional ragged array with gaps, we use
$r=A[\,c_0][\,c_1]\ldots[c_{d-2}][\mathcal{M}][\,c_{d-1}]$, where
$\mathcal{M}=\mathcal{M}(c_0 \ldots c_{d-1})$ is the integer
function equal to the number of the completed contiguous intervals
in the $c_{d-1}$-\,ordered set of all the HC cells in the local
region having coordinates $c_0 \ldots c_{d-2}$ and having
$(d-1)$-th coordinate less than $c_{d-1}$. For example, in the
case $d=2$ of Figure \ref{fg_lrbad}, access to the cells B, C, D,
and E is given by $r_B=A[6][0][26]$, $r_C=A[12][0][12]$,
$r_D=A[20][0][9]$ and $r_E=A[18][1][23]$. The disadvantages of the
other methods considered above do not apply now: the $(d+1)$-array
dereference call is exponentially faster than the function call,
and the space allocated exactly equals the required number of
$\dohclr$ cells. For $d=3$, the function evaluation
$\mathcal{M}(c_0,c_1,c_2)$ takes a time that grows only
logarithmically with the number of disjoint parts along the last
dimension for a give $c_0$ and $c_1$.

\section{Force Calculation}\label{sec_hcforce}

In this section, we present an efficient method for parallel PM
and PP computation of forces for particles within the HC local
regions.  By using the techniques developed in \S\S
\ref{sec_loadbal} and \ref{sec_layout}, we have made our
algorithms load balanced and efficient.

\subsection{PM Force Calculation}\label{sec_hcpm}

The PM force calculation requires communication between two
different data structures with completely different distributions
across the processes.  The particles on one process are organized
into irregularly-shaped HC local regions.  The density and force
meshes, on the other hand, have a one-dimensional slab
decomposition based on FFTW.  The parallel computation is an SPMD
implementation of the five PM steps presented in \S \ref{sec_pm}.

\begin{figure}[t]
  \begin{center}
    \includegraphics[scale=0.45]{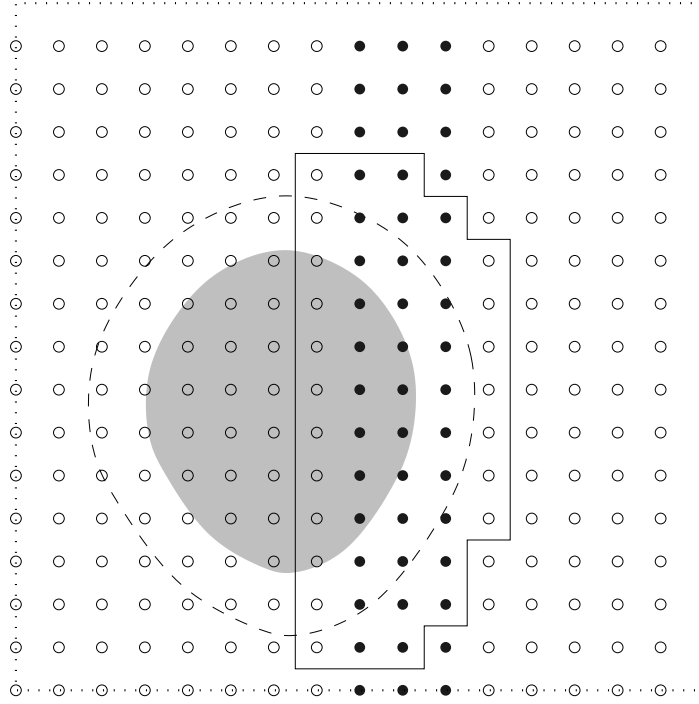}
  \end{center}
    \caption{Schematic representation of the sets used in the PM
      force calculation. The volume within the dotted line is the
      total simulation volume $V_0$, the small circles are the
      discrete set of PM density gridpoints $G_0$.  The filled
      circles are the PM gridpoints within a FFTW slab $j$,
      $G_{\rm  sl}^j$.  The gray filled region is the HC local region
      $\dohclr^i$.  The set of all  circles within the dashed
      line is $\gfun(\dohclr^i)$; the set of filled circles within
      the dashed line is $G_{\rm sl}^j \cap\gfun(\dohclr^i)$.
      Extending this last set slightly gives the continuous set
      within the solid line, $\lfun(G_{\rm sl}^j \cap \gfun(\dohclr^i))$.
      Eq.~(\ref{eq_lmes}) gives the intersection of this last set
      with the gray region$\dohclr^i$.
      }\label{fg_lg}
\end{figure}

\subsubsection{Definitions}
We define a few concepts that will be needed in order to describe
and implement the data exchange between the two different data
structures during the parallel PM force calculation.  The various
sets used in the calculation are illustrated in Figure
\ref{fg_lg}.

The FFTW parallel Fast Fourier Transform implementation
\citep{fftw} allows one to compute forward and inverse Fourier
transforms of the complete three dimensional array of $n_0 n_1
n_2$ mesh points distributed among the processes~$j$ in the form
of slabs of $\donloc(j)\, n_1 n_2$ grid points, where $\sum_j
\donloc(j) = n_0$, each slab starting at the position~$\dosloc(j)
= \sum_{i=0}^{j-1}\donloc(i)$ along the 0-th dimension.  We will
call these slabs the {\it density} or {\it force mesh slabs}
(depending on the context) and denote them by $\gsl^j$. The
geometry of the slab~$\gsl^j$ is calculated once and for all at
the start of the run by calling the FFTW Fourier transform plan
initialization routine.

Let us denote the complete discrete set of all density mesh
gridpoints needed for a complete Fourier transform by $\grid$, and
the complete continuous set of all positions within the whole
simulation volume by $\vol$. We have
\begin{equation}
  \begin{array}{ll}
    \gsl^j \in \grid, & \bigcup\limits_j \gsl^j = \grid\\
    \dohclr^i \in \vol, & \bigcup\limits_i \dohclr^i = \vol\ .\\
  \end{array}
\end{equation}
Here, $i$ labels the process holding the HC local region while $j$
labels the process holding a given density/force mesh slab.

For a continuous set of positions $L\in\vol$, let us define
$\gfun(L)$ to be the minimal complete subset of the density grid
points ${\bf X}_{\rm gr}\in\grid$ such that equation
(\ref{eq_int}) is satisfied for any position vector ${\bf X} \in
L$. By this definition, if all the local particles are contained
within $L$, after the density assignment of Step 1 of the PM force
calculation, the only non-zero PM-density grid points of $\grid$
are in fact within a subset $\gfun(L) \in \grid$.

For a discrete subset $G\in\grid$ of the density gridpoints, let
us define $\lfun(G)$ to be the minimal complete continuous set of
points ${\bf X}\in\vol$ such that equation (\ref{eq_int}) is
satisfied for any ${\bf X}_{\rm gr} \in G$. Now, if all the grid
points local to a process are within a subset $G \in \grid$ of all
the particles in the simulation volume $\vol$, only the particles
of the subset $\lfun(G)$ may acquire any non-zero force
contribution from those gridpoints during {\it Step~5} of the
PM-force calculation.

\subsubsection{Optimal PM Communication Strategy}\label{sec_pm_denfor}

As we discussed in \S\ref{sec_pm}, Step 1 of the PM force
calculation involves filling the density grid points in
$\gsl^j\in\grid$ using the particles distributed in the volumes
$\dohclr^i\in\vol$. Steps 2--4 involve working only with $\gsl^j$
and are straightforward since they do not require any
interprocessor communication aside from the parallel FFT. During
Step 5 the information flows in the exactly opposite direction,
therefore an algorithm for Step 1 applies to Step 5 as well with
the direction of the information flow reversed. The problem
remaining now is for Step 1 of the PM force calculation to decide
how to fill the local density grids $\gsl^j$ from the particles
distributed within the local regions $\dohclr^i$. To solve this
problem we considered a number of approaches described briefly
below, but only the last one is implemented in our code and is
effective over the entire range of clustering.

\paragraph{\it a) Sending Particles.}
Under this method, each pair $ij$ of processes sends the
appropriate portion of the particle data from process $i$ to
process $j$ to fill the density mesh $\gsl^j$ of slab $j$. For
each pair of processes the set of the density gridpoints
\begin{equation}\label{eq_gmes}
  \gsl^j \cap \gfun\left(\dohclr^i\right)
\end{equation}
on process $j$ will be updated with the particles brought from the
volume
\begin{equation}\label{eq_lmes}
  \dohclr^i \cap \lfun\left( \gsl^j \cap \gfun\left(\dohclr^i\right) \right)
\end{equation}
within the HC local region of process $i$.

This method is very efficient for the pairs where the particle
sender processes $i$ have low particle number density, thus
reducing the number of particles to be sent and the communication
cost.

\paragraph{\it b) Sending Grid Points.}

Under this method, each pair $ij$ of the processes fills the
portion (\ref{eq_gmes}) of the grid points using the local
particles within (\ref{eq_lmes}), then sends the filled gridpoints
to process $j$.

This method performs poorly when the particle number density is
low on the sender process, because most of the density values in
the message are zero.  This method is very efficient for the pairs
where the particle sender processes $i$ have a high particle
number density: each gridpoint of the sender process contains the
contributions from many particles.

\paragraph{\it c) Combined Particle and Grid Point Send.}
Method {\it a)} is effective with low particle number density
while method {\it b)} is effective with high particle number
density on the particle sender process. The idea of the combined
particle and grid point send method is to choose for each pair of
processes the approach that requires sending the least data.

\paragraph{\it d) Sending Compressed Grid Points.}
This approach optimizes the communication cost in both the extreme
cases of low and high number density of the particles on the
sender process $i$. The idea behind this method is to use the
approach {\it b)} above and apply {\it sparse compression} to the
gridpoint messages (\ref{eq_gmes}).  As we know, the grid point
approach performs poorly when the particle number density is low
on the sender process. Using sparse compression as we explain in
the following subsection significantly alleviates this problem by
reducing the message size for the underdense regions $\dohclr^i$.

\begin{figure}[t]
  \begin{center}
    \includegraphics[scale=0.45]{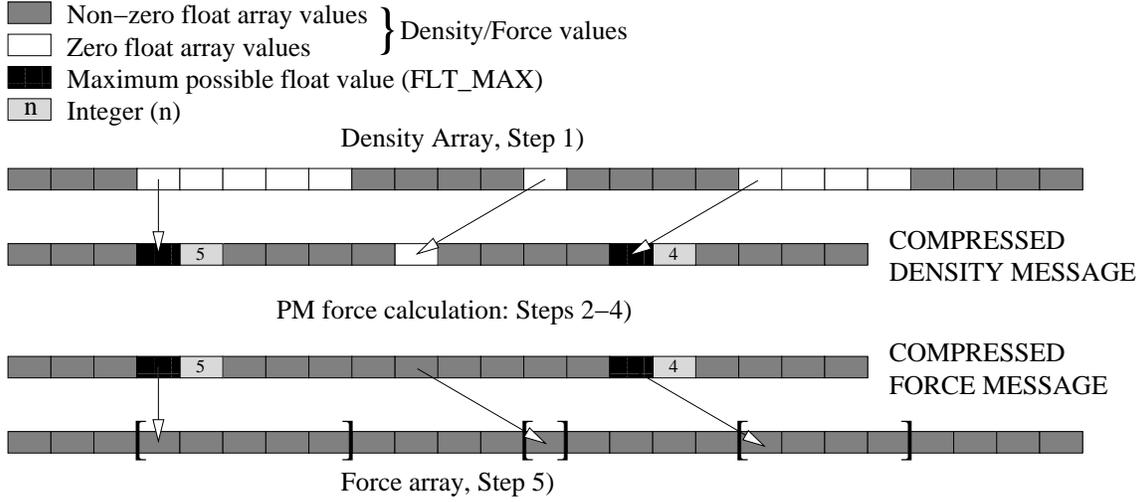}
  \end{center}
  \caption{Sparse array compression of density and force messages during
    PM force computation. The two density arrays (top two bars)
    are equivalent but the lower one is compressed by run-length
    encoding.  Compression is signalled by a special data value
    ({\tt FLT\_MAX}) followed by the number of zeros.  The
    compressed array on process $i$ is sent to process $j$ using an
    MPI function call.  A compressed force message is constructed on
    process $j$ using the template given by the density message.
    The forces are sent back to process $i$ and expanded.  The
    bracketed values in the bottom array can be ignored because
    there are no particles nearby the relevant grid points. }
  \label{fg_sparse}
\end{figure}

\subsubsection{Sparse Compression of  Grid Point Messages}\label{sec_compress}

In a cosmological simulation, the overdense regions have small HC
local regions with every grid point having many nearby particles
so that the force and density messages are small.  On the other
hand, low-density regions have large HC local regions with many PM
grid points but the density and force messages are made small by
the compression method illustrated in Figure \ref{fg_sparse}.

During Step 1 of the PM computation, if a number of binary zeros
are encountered in the grid message, they all are substituted by a
pair of numbers before sending packets: the first number is a
delimiter (an illegal density or force value such as {\tt
FLT\_MAX}) and the second number is an integer giving the number
of zeros to follow in the original uncompressed message. This
technique is called {\it run-length encoding}.  The resulting
compression factor is unlimited and depends on how frequent and
contiguous the zero values are positioned in the grid message. The
receiver process $j$ simply uncompresses the message by filling
the gridpoints within $\gsl^j \cap \gfun(\dohclr^i)$.

During Step 5, the force values are sent from process $j$ to $i$
three times (once for each of the three dimensions). The force
array message is identical in the size to the density message that
was sent during Step 1 for each pair $ij$ of processes. We
compressed the density values in Step 1 using run-length encoding
of zero value densities. In the force message the technique runs
into a difficulty because the gravitational forces are long range
forces by nature and their values are nowhere equal to zero. If we
do not compress the force values, there is no advantage in
choosing the compressed gridpoint approach, since the force
messages would have the same length as the uncompressed density
messages.

By using packet information obtained while receiving the density
array, we can compress the forces using exactly the same pattern
formed by the packets of the density message, as shown in Figure
\ref{fg_sparse}.  The receiving process will decompress the force
and obtain exactly the initial force array excluding the values of
force at the array members which were skipped in the density
assignment (the square bracketed force values in
Fig.~\ref{fg_sparse}). This loss of information is however
completely irrelevant for interpolation of the force values to the
particles in {\it Step 5} because the square bracketed force
values in the force array belong to grid points which earlier
acquired absolutely no density values from the surrounding
particles, which means that for that grid point and for any
particle within $\dohclr^i$, the gridpoint has no nearby particles
[the condition (\ref{eq_int}) is not satisfied].  Thus the force
values at that grid point will not be interpolated to any
particles during Step 5.

The idea of sparse array compression is not implemented in the
Hydra code \citep{mcpp98}. Once implemented it will significantly
reduce their communication and memory costs.

\subsection{Practical PM Implementation}\label{sec_pracpm}

Equation (\ref{eq_gmes}) gives the minimal set of density grid
points on process $j$ needing to be filled with values from
particles on process $i$.  This set is impractical to work with
because of its irregular shape.  For a practical implementation we
embed this region within a rectangular submesh of $\grid$ during
Steps 1 and 5 of the PM force computation, as follows.

For a continuous set of positions inside the simulation volume
$L\in\vol$, let us define $\rfun(L)$ to be the minimal rectangular
subset of density grid points such that $\gfun(L) \in \rfun(L)$.
For grid points with $\rfun(L)$ but outside $\gfun(L)$ we set the
density values to zero.  It follows at once that if we use
$$
\gsl^j \cap \rfun\left(\dohclr^i\right)
$$
instead of equation (\ref{eq_gmes}) for the definition of PM grid
point messages, we will have the rectangular mesh
$\rfun\left(\dohclr^i\right)$ for interpolation of density for
particles within $\dohclr^i$, and this still give the correct
result.
However, since the extent of the local region $\dohclr^i$ inside
the simulation box is not limited, neither is the extent of
$\rfun\left(\dohclr^i\right)$. For example, when $\dohclr^i$
consists of just two cells with the coordinates $(0,0,0)$ and
$(\donh^0-1,\donh^1-1,\donh^2-1)$, it is easy to see from the
definition that $\rfun\left(\dohclr^i\right)$ encloses the whole
simulation density mesh $\grid$ as a subset and this is too much
memory space for allocation on a process.

To avoid this problem, we dissect the local region $\dohclr^i$
uniformly into $n_k^i$ slices $M^{ik}$ along the 0-th dimension so
that the extent of each slice along the 0-th dimension will not
exceed $n^0/\donpc$.  Using the previous equation we have, now
summed for all the receiving processes~$j=0\ldots\donpc-1$
\begin{equation}\label{eq_pmmsg}
  \sum_{j=0}^{\donpc-1} \gsl^j \cap \rfun\left(\dohclr^i\right)
  = \sum_{j=0}^{\donpc-1} \gsl^j \cap \rfun(\,\sum_{k=0}^{n_k^i-1} M_h^{ik}\,)
  = \sum_{k=0}^{n_k^i-1} \left( \sum_{j=0}^{\donpc-1} \gsl^j \cap \rfun(M_h^{ik}\,)
    \right)\ .
\end{equation}
For each slice $M_h^{ik}$ of the HC local region $\dohclr^i$, the
density is interpolated onto the rectangular mesh
$\rfun(M_h^{ik}\,)$ which is small enough to be allocated since
its extent in the 0-th dimension is limited by roughly
$n^0/\donpc$ grid points. Then, the messages under the inner sum
of equation (\ref{eq_pmmsg}) are sent to processes
$j=0..\donpc-1$. The procedure is repeated for each slice
$M_h^{ik}$.

In the code presented in this paper we use the blocking MPI
routines for PM message communication, which requires
synchronization between each pair of processes exchanging the
message.  In order to reduce waiting time, MPI allows
bi-directional blocking communication using {\tt MPI\_SendRecv}.
In the above equation the process $i$ is described as the {\it
sender} of the PM-grid messages obtained by interpolation from the
particles within $\dohclr^i$ to the processes $j$ in order to
update their FFTW-slabs $\gsl^j$. Note however, that the same
process $i$ also behaves as a {\it receiver} of the PM grid
messages from the other processes $j$ in order to update the
FFTW-slab $\gsl^i$. The set of the received messages is obtained
by simply swapping the indices $i$ and $j$ in the above equation.
Adding the two together we have for the set of gridpoints
participating in the communication on process $i$ in both
directions
\begin{equation}\label{eq_pmmsg2}
  \sum_{j=0}^{\donpc-1} \left[\gsl^j \cap \rfun\left(\dohclr^i\right)+\gsl^j
  \cap \rfun\left(\dohclr^i\right) \right] =
  \sum_{k=0}^{n_k^{ij}-1}
  \sum_{j=0}^{\donpc-1}
  \left[\; \gsl^j \cap \rfun(M_h^{ik}\,) +
    \gsl^i \cap \rfun(M_h^{jk}\,)
  \right]\ ,
\end{equation}
where ${n_k^{ij}}\equiv \max(n_k^i, n_k^j)$ and the $M_h^{ik}$ is
defined to be an empty set for $k\ge n_k^i$ .

In order to access particles in a given slice $M_h^{ik}$ of the
local region we use the particle access technique described \S
\ref{sec_voids}.  The sorting technique described in \S
\ref{sec_adv} speeds up the density and force interpolation. The
timing of the interpolation for each HC cell gives the PM part of
the HC-cell workloads in equation (\ref{workload_p3m}).

The above procedure is used for both density and force
interpolation in the PM force calculation. In the current
implementation, the MPI messages are blocking, which means
additional waiting time. In a subsequent paper we describe the
implementation of non-blocking communication resulting in a
significant speedup of the PM calculation.

\subsection{PP Force Calculation}\label{sec_hcpp}

The particle-particle (PP) force calculation increments forces
acting on each of the particles in a pair if the particles are
closer than $\dormax$.

The method of particle access developed in \S \ref{sec_voids}
allows one to access all the particles within a given HC cell.
From equation (\ref{eq_ppsimple}), HC cells are coincident with
the chaining mesh cells needed for the PP force calculation.  To
see how the communication and computation work, consider the
example of Figure \ref{fg_pp}.  To compute the PP force for a
particle ${\tt p}$ within chaining mesh cell $A$, the particle
data in the surrounding cells $b^A_0\ldots b^A_3$ are required.
The particle data within the cell $b^A_3$ are locally available.
However one needs to bring the positions and masses from the other
processes to get the particle data for the boundary layer cells
$b^A_0\ldots b^A_2$.  Once the particle data from the boundary
layer cells are gathered, the PP force calculation may be
performed by pair summation, after which the resulting forces for
the particles within $b^A_0\ldots b^A_2$ are sent back to their
processes where the PP forces of the original particles are
incremented.

The same algorithm applies to any other cell within $\dohclr$, for
example the cell $B$ of Figure \ref{fg_pp}, for which the particle
data for $b^B_0$ and $b^B_1$ are available locally while the
particle data for cells $b^B_2$ and $b^B_3$ must be brought to the
local process from the others. Because of its pairwise nature only
half the surrounding cells are needed for the PP force calculation
for each HC cell. In total, the particle data for the non-local
cells shaded in Figure \ref{fg_pp} are required for the PP force
calculation for each particle within $\dohclr$. The amount of
communication needed for a complete PP force calculation is
proportional to the number of particles in the cells required to
be brought from the other processes through the boundary layer
cells.

If the PP pair summation step is started synchronously on all
processes, it will finish at approximately the same time on all
processes if the load imbalance is low.  Otherwise, the processes
that complete the PP force computation first will have to wait for
the remaining processes to finish their pair summation. Since the
pair summation is the most time-consuming step of P$^3$M, it is
crucial that the procedure be load-balanced.  This is accomplished
using the methods of \S \ref{sec_loadbal}.  The CPU time of the
pair summation step is used in the cell workload calculation of
equation (\ref{workload_p3m}).  The particle access time in the
pair summation loop is minimized by pre-sorting the particles as
described in \S \ref{sec_adv}.

\subsection{Memory Management}\label{sec_llmem}

In early versions of our code, the memory often exceeded the
available resources causing the code to crash.  By implementing
runtime tracking of memory usage we were able to identify the
problems and optimize the memory requirements. Memory usage was
reduced largely in three ways: the irregular shaped local domain
memory technique of \S \ref{sec_voids}, the elimination of
particle buffer allocation while repartitioning, and memory
balancing when necessary during repartitioning as described in \S
\ref{sec_rep}.

\noindent
\begin{table}[t]
    \begin{tabular}{|l|l|l|l|}\hline
                                       & Notation    & Memory Size, per process~$i$                    & Total Memory size                       \\[8pt]\hline
      Particle array                   & $M_{\rm P}$   & $4\,\hbox{bytes} \times 11 \donp(i) $                       & $4\,\hbox{bytes} \times 11\donp $       \\
      Particle linked list             & $M_{\rm L}$   & $4\,\hbox{bytes} \times \donp(i)          $                 & $4\,\hbox{bytes} \times \donp$        \\
      HC mesh                          & $M_{\rm HC}$  & $4\,\hbox{bytes} \times 5 \dorn^i$                          & $4\,\hbox{bytes} \times 5n_h^0n_h^1n_h^2 $       \\
      Green's function                 & $M_{\rm G}$   & $4\,\hbox{bytes} \times  \donloc(i)\, n^1(n^2/2+1)$         & $4\,\hbox{bytes} \times  n^0n^1(n^2/2+1)$      \\
      Density and Force meshes         & $M_{\rm PM}$  & $4\,\hbox{bytes} \times  2\,\donloc(i)\, n^1(n^2+2)$       & $4\,\hbox{bytes}\times 2 n^0n^1(n^2+2)$  \\
      FFTW scratch space               & $M_{\rm FFT}$ & $4\,\hbox{bytes} \times  \donloc(i)\, n^1(n^2+2)$         & $4\,\hbox{bytes} \times  n^0n^1(n^2+2)$      \\
      PP boundary layer particles      & $M_{\rm PP}$  & $4\,\hbox{bytes} \times  8\Delta n_{\rm PP}$                & $4\,\hbox{bytes} \times  8\sum\Delta n_{\rm PP}$ \\\hline
    \end{tabular}
    \caption{\label{tb_parmem}
      Dominant memory requirements of the parallel~\parhc\ code.
      Here $\don(i)$ is the number of particles and $\donloc(i)$
      is the thickness of the PM slab , both on process $i$.
    }
\end{table}

In Table \ref{tb_parmem} we list the main memory requirements for
our \parhc\ code.  Compared with the memory requirements for the
serial code in Table \ref{tb_sermem}, we were able to reduce the
size of the particle linked list by 50\%. Note that the HC mesh is
the parallel equivalent of the serial chaining mesh but requires 5
times more storage.  (This is still less than the storage required
for the linked list, so we have accomplished a net savings.) The
FFTW scratch space is optional but significantly improves FFTW
performance, so we allocate it. The maximum memory allocated per
process each timestep can be obtained by combining the tabulated
values in the sequence that follows their actual allocation and
release in the code. For example, the memory spaces $M_{\rm PM}$,
$M_{\rm FFT}$ and $M_{\rm PP}$ are allocated when the memory space
$M_{\rm L}$ is released. The actual memory usage along with the
detailed measurements from a $800^3$ run are described in \S
\ref{sec_long}.

\section{Tests}\label{sec_test}

The first and most important test was to verify that our parallel
PM and P$^3$M codes give identical results to the serial codes
(both the original Fortran codes and their C translations) when
given identical inputs, to within the precision of machine
roundoff error.  The serial codes have been thoroughly tested by
\cite{gb94}.

The remaining tests presented in this section test the performance
of the parallel codes in order to optimize the performance.
Several of the innovations described in the preceding sections
were devised in response to performance tests.

The PM test presented in \S \ref{sec_testpm} was performed on a
beowulf cluster consisting of 7 nodes each with a 1.7 GHz
Pentiun-4 processor with 256 KB L2 cache memory, 1 GB RAM memory,
and 34 GB of hard drive, connected by 100 Mb/s ethernet.  The
Linux {\tt gcc} compiler was used.  This cluster has a Linpack
performance of 15 GFlops.

The rest of the test runs were performed on a beowulf cluster
consisting of 20 dual Xeon 2.4 GHz Pentium-4 and 512 KB L2 cache
memory processor computing nodes each containing 4 GB RAM memory
and 360GB of disk, connected by gigabit Ethernet.  The Intel {\tt
icc} compiler was used.  This cluster has a Linpack perfomance of
70 GFLops.

\subsection{PM Simulation of Extremely Clustered Matter}
\label{sec_testpm}

Cosmological initial conditions for cold dark matter were
generated for a simulation with $512^3$ particles and grid points,
with a power spectrum having a Gaussian cutoff at a wavelength
equal to one-fourth of the box size and white noise on larger
scales. The resulting nonlinear evolution, shown in Figure
\ref{fg_vol}, leads to the formation of two massive particle
clusters displaying many phase space caustics. The initial
conditions were evolved using both the
\parsl\ and \parhc\ codes to compare their timing performance.

\def\tstep{5693}
\begin{figure}[th]
  \begin{center}
    \includegraphics[scale=0.6]{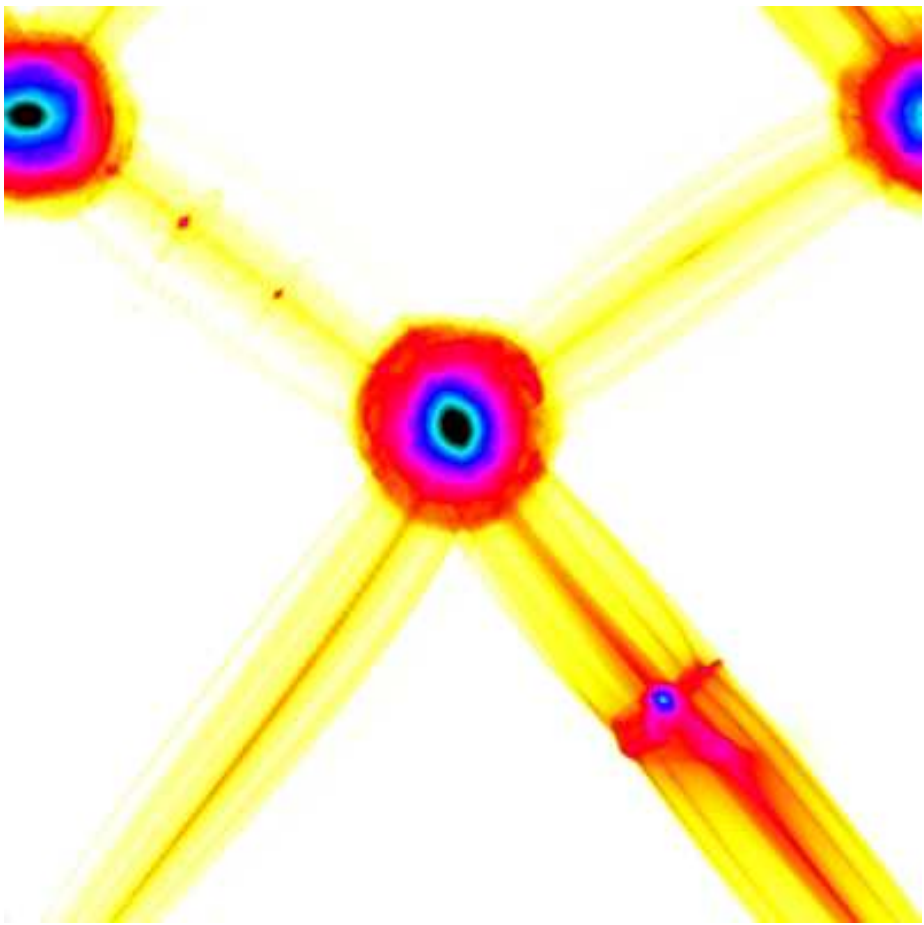}
  \end{center}
    \caption{Projected particle distribution for the entire
      PM simulation volume of $512^3$ particles at timestep \tstep.
      False colors scale with the logarithm of projected mass
      density.  Strong clustering like this favors dynamic rather
      than static domain decomposition methods.
    }\label{fg_vol}
\end{figure}

This test used an early version of \parhc\ with only PM forces.
Since we did not compute PP forces the constraint given by
equation (\ref{eq_ppsimple}) was not in effect. Instead we set our
HC mesh spacing to $\dodxh^i \ge 1.5$. The sorting technique
described in \S \ref{sec_adv} was not implemented.  Instead of
equation (\ref{workload_p3m}), we used the number of particles in
a cell to define the HC cell workload.  Repartitioning therefore
resulted in an approximately equal number of particles on all
processes at each timestep.

\begin{figure}[th]
  \begin{center}
    \includegraphics[scale=0.6]{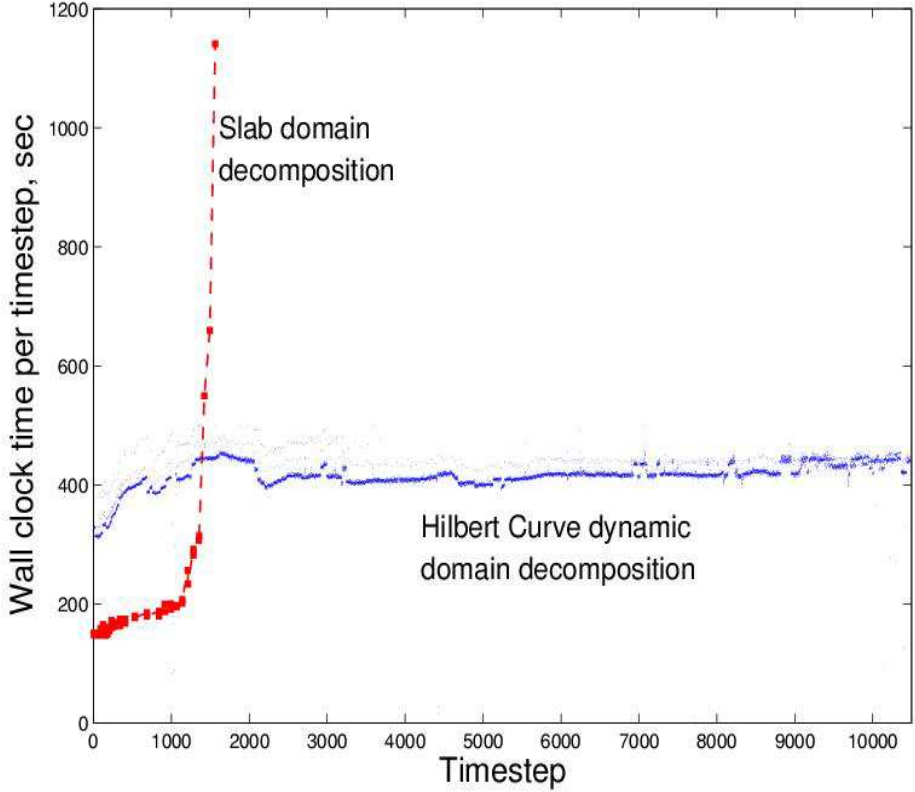}
  \end{center}
    \caption{Timing performance comparison of a Hilbert curve dynamic
      domain decomposition code \parhc\  and a fixed (slab) domain
      decomposition parallel code \parsl\ for the identical run
      showed in Fig.~\ref{fg_vol}.  The runs start from the linear
      regime and are evolved using only PM forces.
    }\label{fg_wall}
\end{figure}

Figure \ref{fg_wall} shows the wall clock time per timestep for
the Hilbert curve code \parhc\ and the fixed slab domain
decomposition code \parsl.  As we see from the Figure, the
HC-based PM code evolves very far into the regime of strong matter
clustering without any significant slowdown.  On the other hand,
the slab decomposition code grinds to a halt because of the
growing memory imbalance arising in any fixed domain decomposition
method.  As more and more particles end up on one process, not
only does its CPU workload grow, but the process eventually runs
out of memory and starts paging to disk, slowing down the
evolution by orders of magnitude.  Only a dynamic domain
decomposition can handle clustering as extreme as that shown in
Figure \ref{fg_vol}.

Even though the HC code is vastly superior to slab decomposition
under strong clustering, it is slower at early times.  This is
mainly because the local regions are displaced from the FFTW slabs
in the \parhc\ code, therefore more communication is required. In
addition, since the non-blocking communication was not implemented
(see \S~\ref{sec_pracpm}), there is some unnecessary waiting time
in the HC code.

\subsection{P$^3$M Simulation of $\Lambda$CDM without Repartitioning}
\label{sec_lcdm}

An extensive series of tests were performed using the \parhc\ code
to assess its behavior under a wide range of clustering
conditions. All of the runs use one particle per PM mesh cell in a
cube of size $L^0=L^1=L^2=200$ Mpc.  The Plummer softening length
was set to $\tilde{\epsilon} = 0.4$ (i.e. 40\% of the PM mesh
spacing).  We generated the initial conditions for the
$\Lambda$CDM model (with $\Omega_m=0.27$, $\Omega_{\Lambda}=0.73$,
$H_0=71$ km s$^{-1}$ Mpc$^{-1}$, $\sigma_8=0.84$, $n=0.93$ from
Bennett et al.\ 2003) using the BBKS transfer function in a C
parallel version of {\tt grafic1} \citep{b95}.  The timestep
parameter of equation (\ref{eq_timestep}) was set to
$\eta_t=0.05$.

\begin{figure}[t]
  \begin{center}
    \includegraphics[scale=0.35]{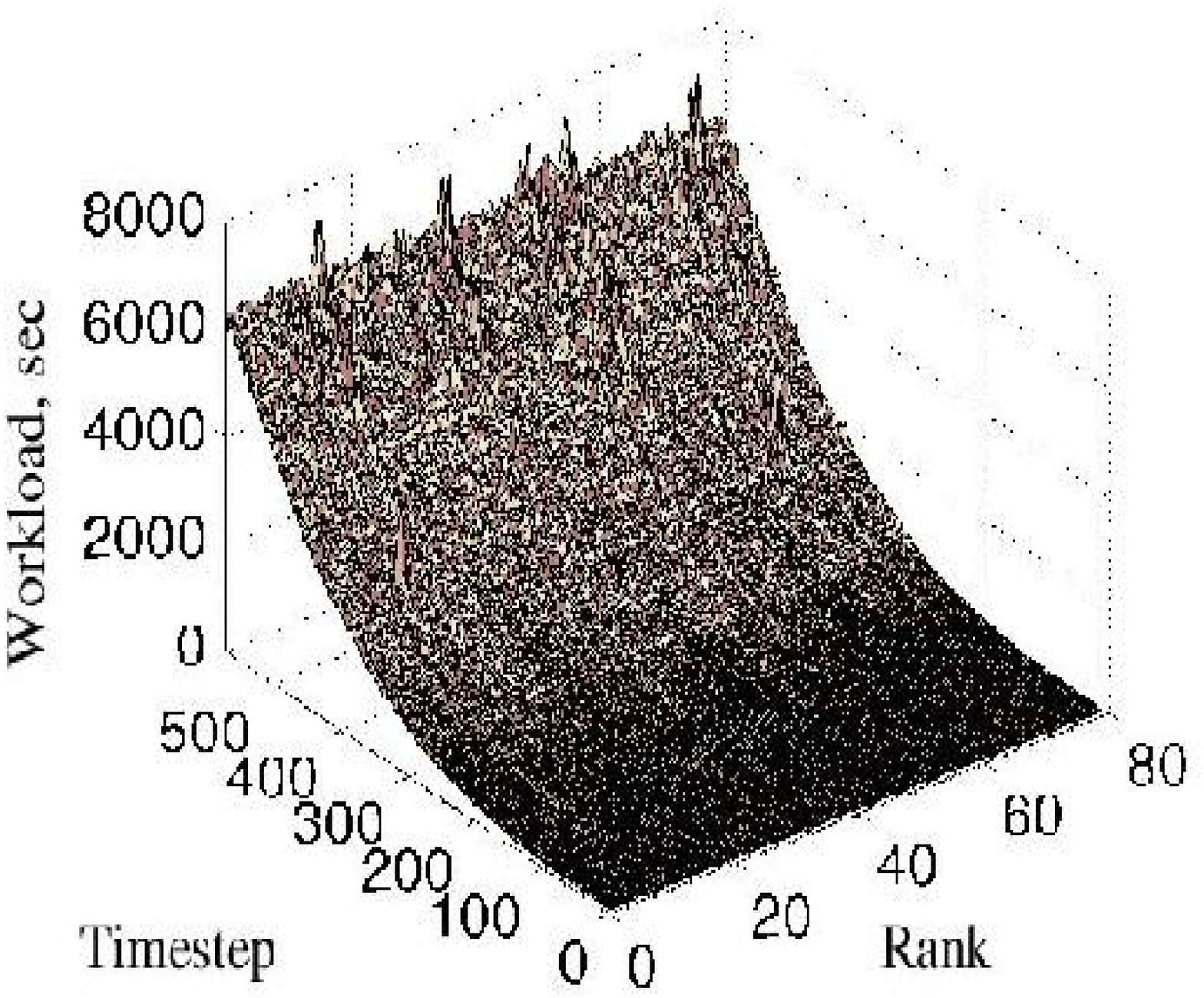}
    \hspace{0.5cm}
    \includegraphics[scale=0.35]{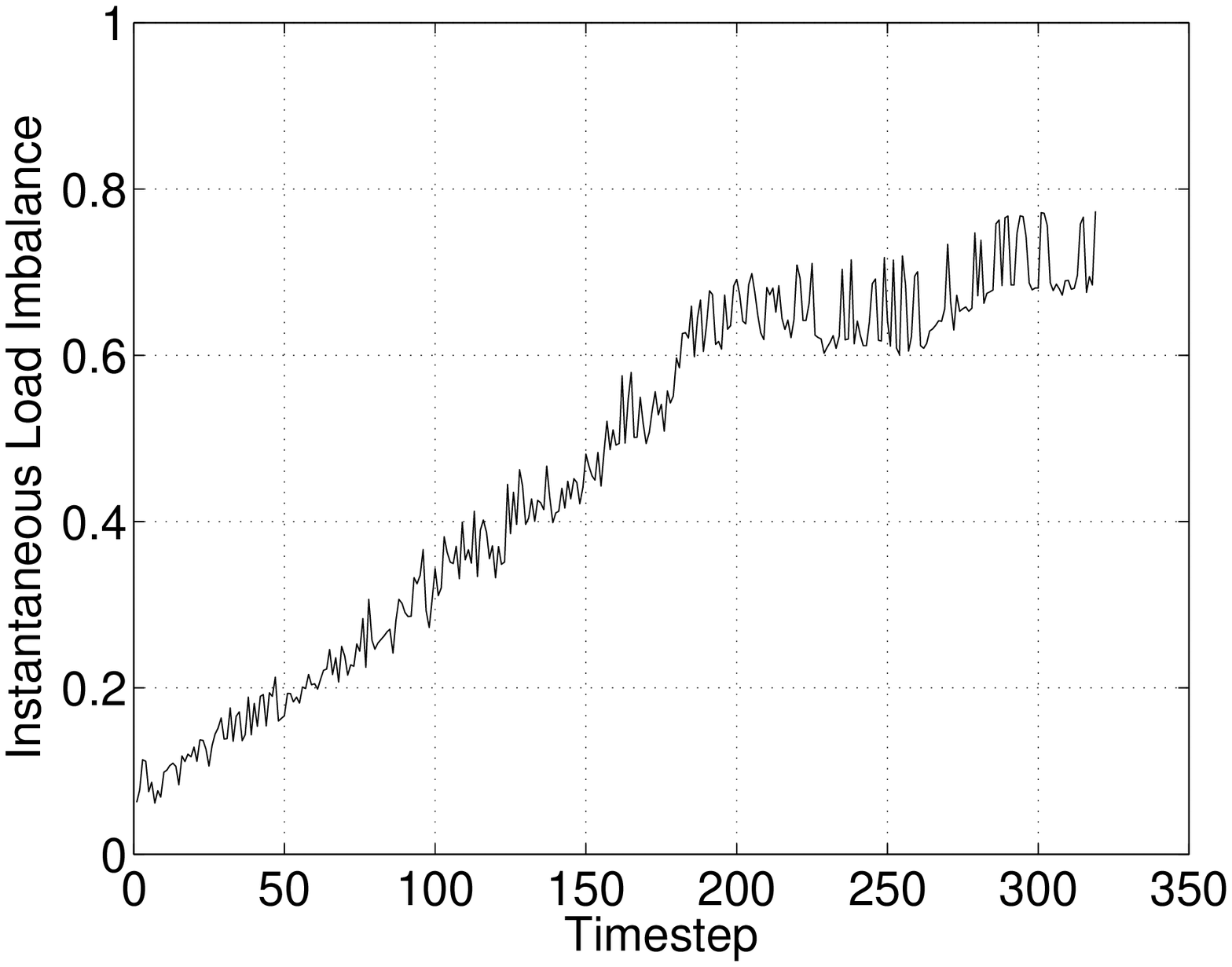}
  \end{center}
  \caption{Left: Time required to complete the pair summation step on
    each process (labelled by MPI rank) as a function of timestep during
    a $800^3$ P$^3$M run without  repartitioning.  Right: Instantaneous
    load imbalance as a function of timestep for the same run.
  }\label{fg_norep}
\end{figure}

As a first test of the full \parhc\ code we ran a simulation with
$800^3$ particles and grid points with no repartitioning. This run
was performed with 80 processes on 20 nodes (40 CPUs using
hyperthreading, which treats a physical CPU as two virtual CPUs
with improved performance).  Without repartitioning the HC local
regions on each process remain the same throughout the run. The
results appearing in Figure \ref{fg_norep} are predictable.  A few
processes require much longer time to finish the pair summation,
leading to a large load imbalance.  Late in the simulation, only
about $1-\mathcal{L}=25\%$ of the net wall clock time is spent
doing computation; most of processes sit idle most of the time
waiting for the heavily loaded processes to finish the PP pair
summation.

\subsection{P$^3$M Simulation of $\Lambda$CDM with Repartitioning:
Load Balancing} \label{sec_long}

We reran the $800^3$ simulation of \S \ref{sec_lcdm} on 80
processes with repartitioning enabled in order to load balance the
computation. Because of the strong increase in clustering and the
resulting growth of the PP pair summation time, the wall clock
time to complete one timestep increased from just over 4 minutes
at the beginning of the simulation to 2 hours at the end (timestep
569, when the expansion factor was $a=0.7$, or a redshift of
$z=0.43$). The simulation,
finished on August 29~2004, 
took two weeks to get to this point and would have required
another month to evolve to $a=1$ provided it remained well load
balanced.  In a subsequent paper we introduce an adaptive
technique that substantially decreases the PP workload enabling
longer and more highly clustered simulations to be performed in
much less time.  A projection of the particle distribution at
timestep 566 for this simulation was used in Figures
\ref{fg_virgo_sl} and \ref{fg_virgo}.  At the end of the
simulation the Layzer-Irvine energy conservation check
(eq.~\ref{eq_econ}) was satisfied to a precision $\tilde E_{\rm
con}/\tilde E_g=5\times10^{-5}$.  (Energy conservation can be as
much as 100 times worse when clustering is very weak and the PM
force contributions dominate over PP.)

\begin{figure}[t]
  \begin{center}
    \includegraphics[scale=0.6]{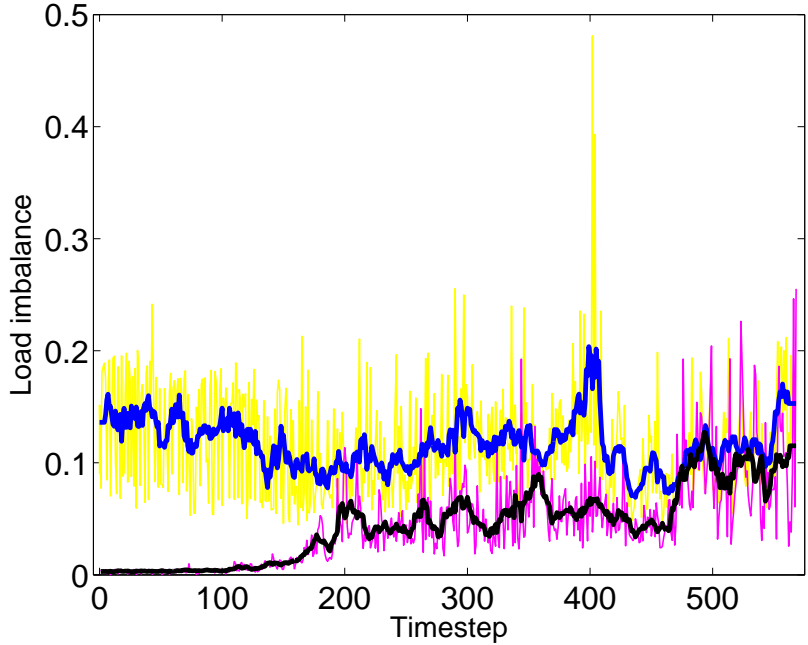}
  \end{center}
  \caption{
    Instantaneous (blue and yellow curves) and residual
    (black and pink curves) load imbalance as a function of
    timestep for the $800^3$ P$^3$M run with Hilbert curve
    repartitioning on 80 processes. The yellow and pink curves
    give the load imbalance of every timestep; the blue and
    black curves apply boxcar averaging over 10 timesteps to
    reduce the fluctuations. The residual load imbalance is
    the minimum possible load imbalance that could be achieved
    by repartitioning in the absence of fluctuations.
  }\label{fg_imballong}
\end{figure}

In Figure \ref{fg_imballong}, we present the instantaneous and
residual load imbalance as functions of timestep.  The effective
load imbalance (not shown) is almost identical to the
instantaneous load imbalance but is lower by 10\% for the highest
spike at timestep 403.  The differences between these various
measures of load imbalance are explained in \S \ref{sec_repartit}.

One of the main results of this paper is that the time-averaged
instantaneous load imbalance generally remains between 10 and 15\%
(averaging 12\%) and does not grow steadily worse with time.  By
contrast, without Hilbert curve dynamic domain decomposition, by
timestep 300 the load imbalance exceeded 70\%
(Fig.~\ref{fg_norep}).

At early times the residual load imbalance is much less than the
instantaneous load imbalance because fluctuations in cache usage
limit our ability to predict the optimal repartitioning for the
next timestep.  Later in the run, the residual load imbalance
grows when a small number of highly-occupied HC cells begin to
dominate the CPU time in the PP force calculation, as we discuss
further below.

\begin{figure}[t]
  \begin{center}
    \includegraphics[scale=0.33]{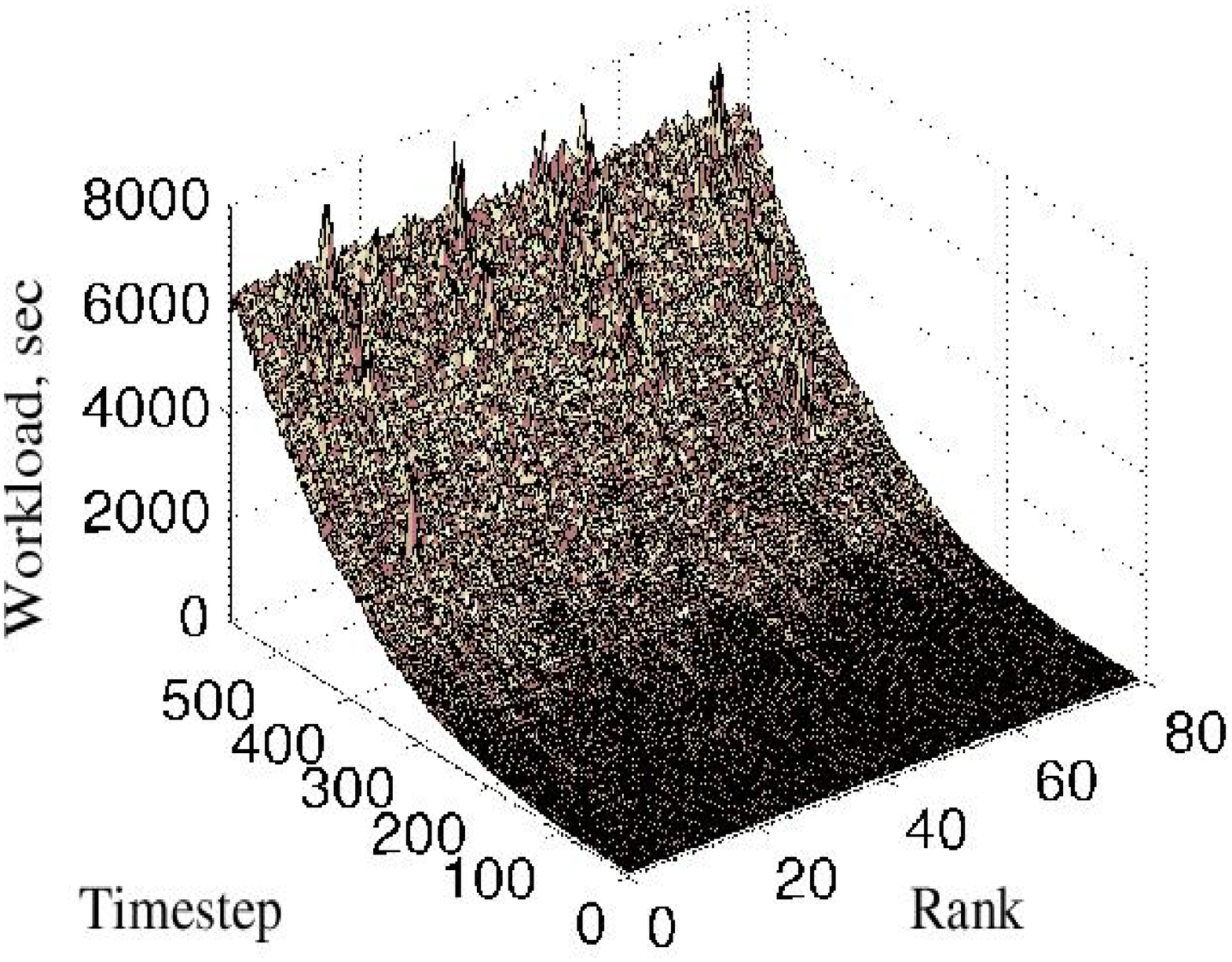}
    \hspace{0.5cm}
    \includegraphics[scale=0.3]{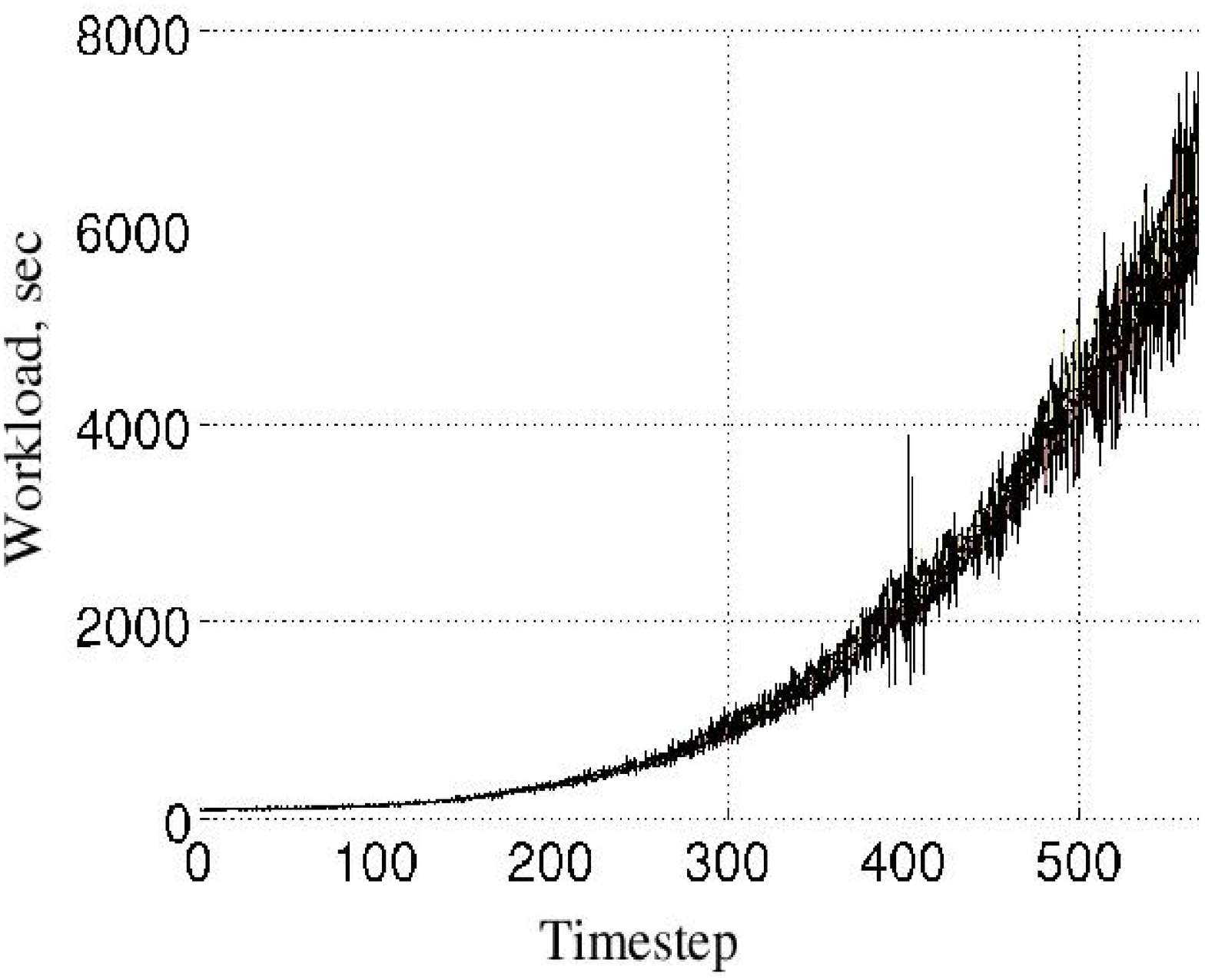}
  \end{center}
  \caption{
    Instantaneous workload of each process as a function of timestep.
    Left: view from an oblique angle.  Right: projected view down
    the rank (process number) axis.
  }\label{fg_long_lbal}
\end{figure}

To analyze the limitations of our load balancing algorithm, in
Figure \ref{fg_long_lbal} we plot the instantaneous workload
measured every timestep for every process using equations
(\ref{workload_p3m}), (\ref{wk_pc}), and (\ref{wk_cellrob}). The
workload is approximately the CPU time required for PP pair
summation summed over all the local HC cells on each process. The
main pattern seen is the steady rise of the average PP workload
with timestep due to the increase in clustering caused by gravity.
After that, one sees in the left plot four spikes in processes 14,
34, 54, and 74. Given our assignment of processes to nodes, these
processes reside on the same physical node 14 and are probably
caused by competition of these processes for memory access.
Fluctuations in cache memory usage are probably also responsible
for the smaller fluctuations in workload superposed on the steady
rise with timestep in the right plot.

\begin{figure}[t]
  \begin{center}
    \includegraphics[scale=0.8]{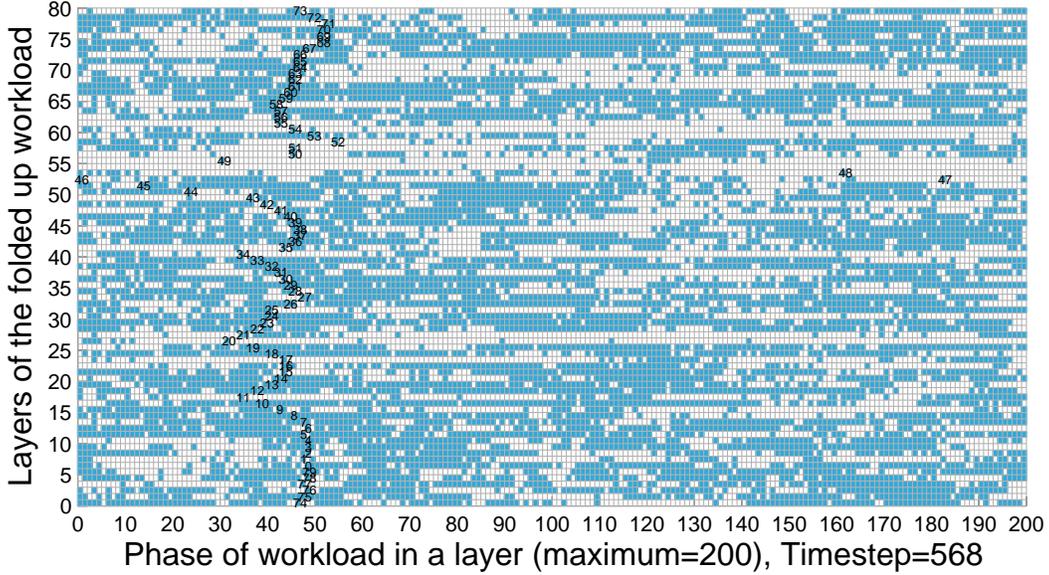}
  \end{center}
  \caption{The discrete workload array for timestep 568 of the same
    simulation analyzed in Figs.~\ref{fg_imballong} and
    \ref{fg_long_lbal}.  The discrete workload array is a
    one-dimensional array of $M_{\rm bin}=1600$ cells folded into
    a set of 80 layers (along the vertical axis) of length 200
    (along the horizontal axis).  Blue cells contain at least one
    boundary between HC cells; a continuous white segment
    represents a single HC cell. The target discrete partitions
    $\hat{r'}_b^i$ are marked with process rank $i$  (cf.~ bar
    $D_2$ in Fig.~\ref{fg_partit}).
    }\label{fg_longdis}
\end{figure}

To investigate further the cause of growing residual load
imbalance, in Figure \ref{fg_longdis} we present the discrete
workload array (\S~\ref{sec_reptar}) at timestep 568.  A perfectly
load-balanced partitioning state would correspond to a target
partitioning state with all boundaries lined up in one vertical
column.  (In that way, a fraction $1/80$ of the work would be
assigned to each of the 80 processes.)  However, this is
impossible because processor boundaries cannot occur in white
sections (where by definition there are no processor boundaries)
and every column contains some white space.  Instead, the (nearly)
optimal solution is found using the method described in \S
\ref{sec_disreptar} and used to define the target partitioning
state corresponding to the black numbers giving the process
boundary for each process.

Figure \ref{fg_longdis} shows that the workloads of processes
48--53 are hard to adjust by repartitioning since most of the
cells of the workload array in their vicinity are white.  This
occurs because these processes have a small number of HC cells in
very high density regions requiring a significant amount of CPU
time to complete their PP force calculations.  The resulting
uneven workload assignment causes the systematic increase in the
measured instantaneous workloads for these processes after
timestep 470 in Figure \ref{fg_long_lbal} and therefore an
increase in the residual load imbalance in Figure
\ref{fg_imballong}.  We would not be able to carry out the P$^3$M
calculation much further before a single HC cell begins to take
more time than the local regions of other processes, leading to
severe load imbalance (eq.~\ref{eq_resimbalcon}).  Even with the
strong variation in workload present after timestep 500, Figure
\ref{fg_imballong} shows that our algorithm manages to achieve an
instantaneous load imbalance almost as small as the optimal
(residual) imbalance. We expect even better performance when (in a
later paper) adaptive mesh refinement is used to alleviate the PP
pair summation workload.

\subsection{P$^3$M Simulation of $\Lambda$CDM with Repartitioning:
Local Regions, Timing, and Memory Usage} \label{sec_long_other}

We continue discussing the same long $800^3$ P$^3$M simulation as
in the preceding section but now focus on aspects of code
performance other than load balance.

\begin{figure}[t]
  \begin{center}
    \includegraphics[scale=0.34]{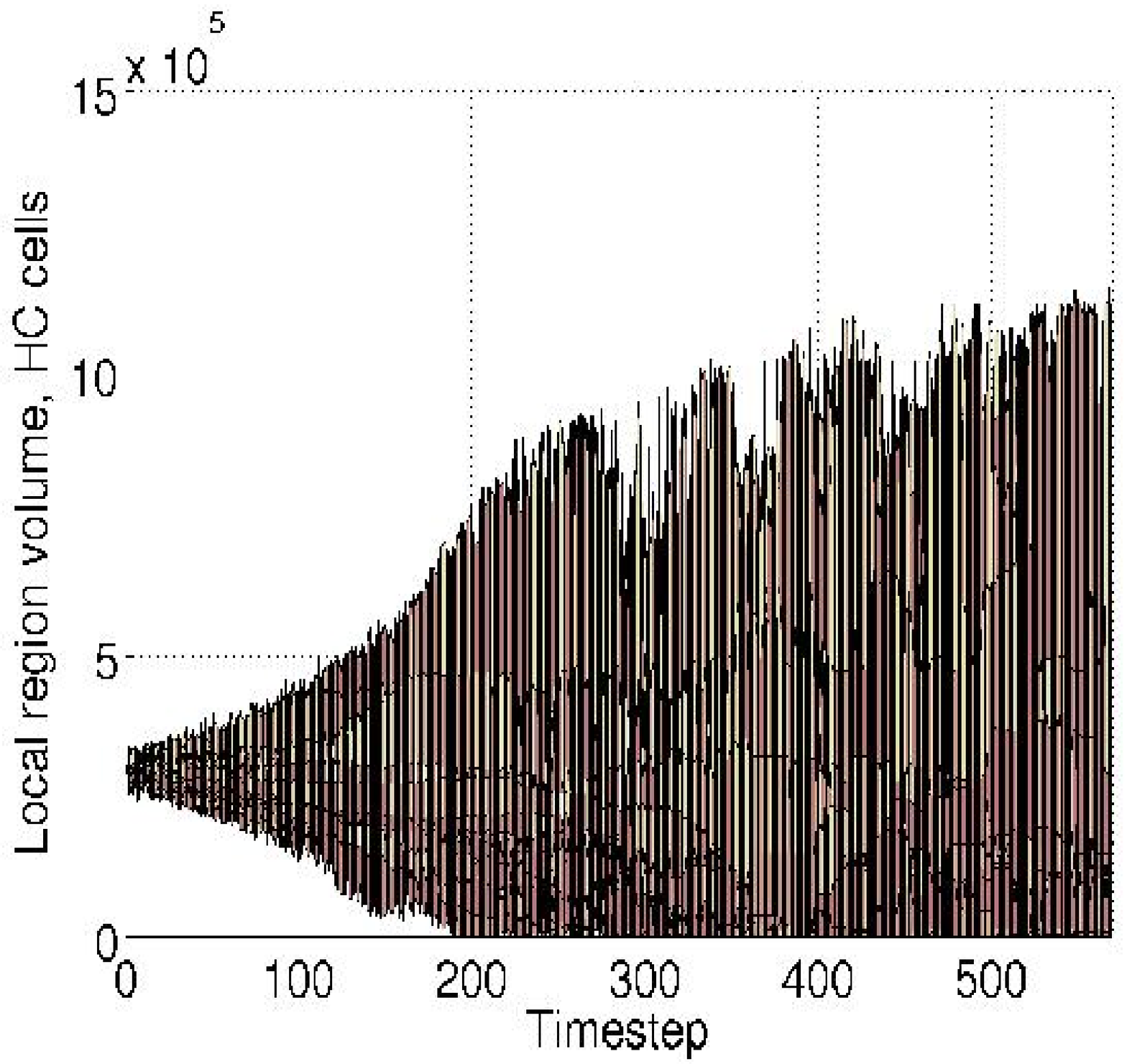}
    \hspace{0.2cm}
    \includegraphics[scale=0.34]{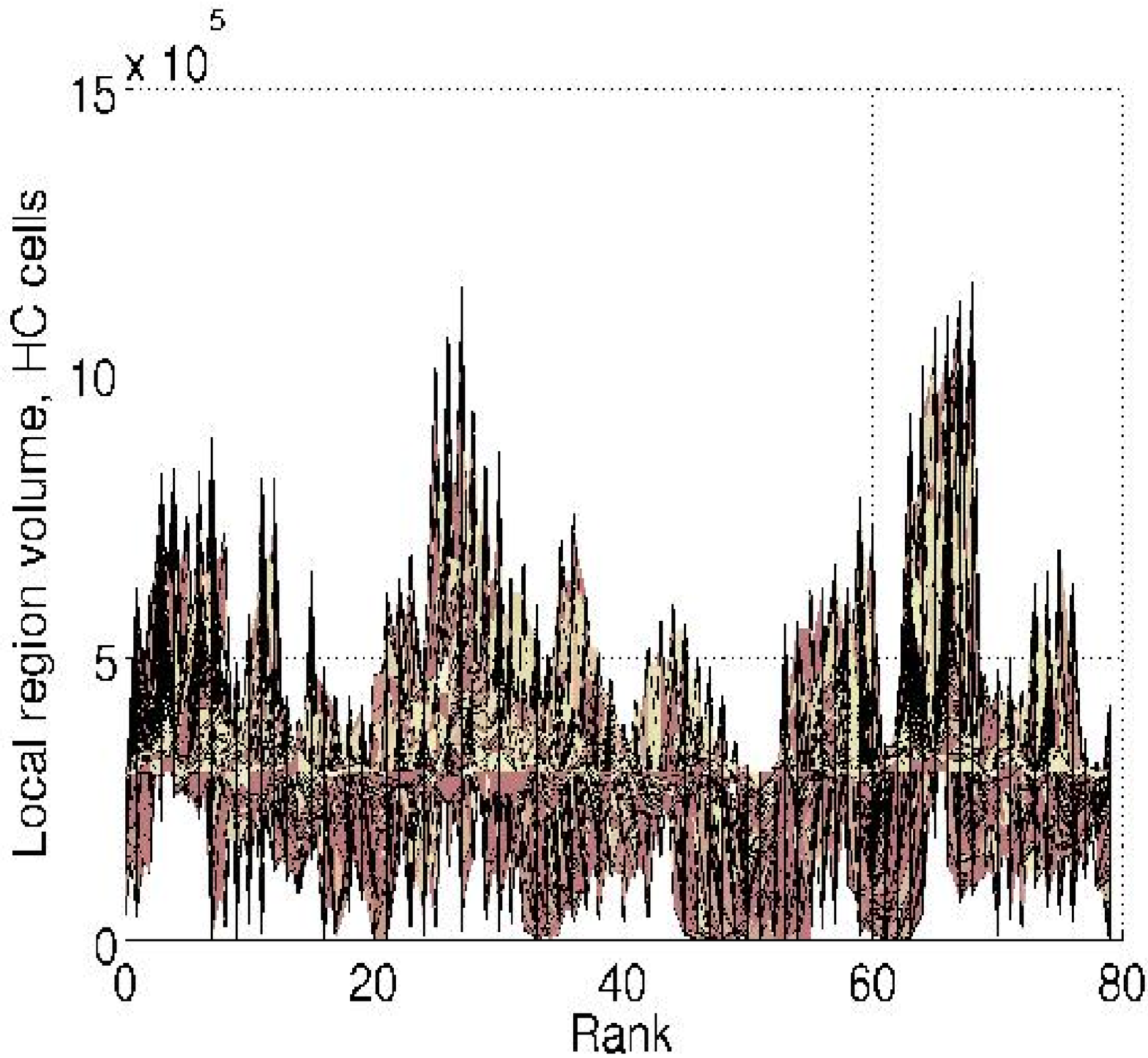}
  \end{center}
  \caption{
    Volume of HC local regions, as a function of timestep and
    process rank.  In the left panel, the full range of process
    ranks is shown for each timestep.  In the right panel, the
    full range of timesteps is shown for each process rank.
    The average volume of local regions is $3\times10^5$ cells.
  }\label{fg_long_rn}
\end{figure}

In Figure \ref{fg_long_rn}  we plot the volume $\dorn^i$ of HC
local regions as a function of timestep and process number. The
local region volume is large when the workload per HC cell is
small (i.e., in low-density regions) and is small when the
workload is high (i.e., in dense particle clusters).  At the
beginning of the simulation, all $287^3$ HC cells are uniformly
divided between the processes.  As clustering grows, a huge range
of volumes develops as the local regions adjust to follow the
change in their workload.  Because of the compactness property of
the Hilbert curve, particle clusters or voids tend to occupy
adjacent processes.

At timestep 568 (the end of the run), the smallest local regions
belong to processes $i=20$, 34, 48, 49 and 54, with $\dorn^i = 7$,
282, 4, 29, and 4 cells, respectively (see also their workload
structure in Fig.~\ref{fg_longdis}). Because the run is well load
balanced, the small number of local cells per process implies that
the workloads of those cells greatly exceed the average. Indeed,
the average value of workload per cell in the whole simulation box
is~$W_{\rm tot}/(\donh^0\donh^1\donh^2) = 4.23\times
10^{-8}\,W_{\rm tot}$. A process with only $4$ local cells has (on
average) workload $W_{\rm tot}/(4\donpc)=3\times 10^{-3}W_{\rm
tot}$, which is five orders of magnitude higher than the average
workload of a cell in the simulation volume. Such a high cell
workload results from the huge local number density of particles
for these cells leading to a heavy PP-force calculation load
(eq.~\ref{workload_p3m}). Indeed, process 48 holds $1.9\times10^5$
particles or $8.8\times 10^3$ times the average. Also, by an
unfortunate coincidence, two of the five most heavily loaded
processes (34 and 54) ended up on the same compute node, leading
to heavy demands on memory and (apparently) causing the spikes
seen in Figure \ref{fg_long_lbal}.

\begin{figure}[t]
  \begin{center}
    \includegraphics[scale=0.3]{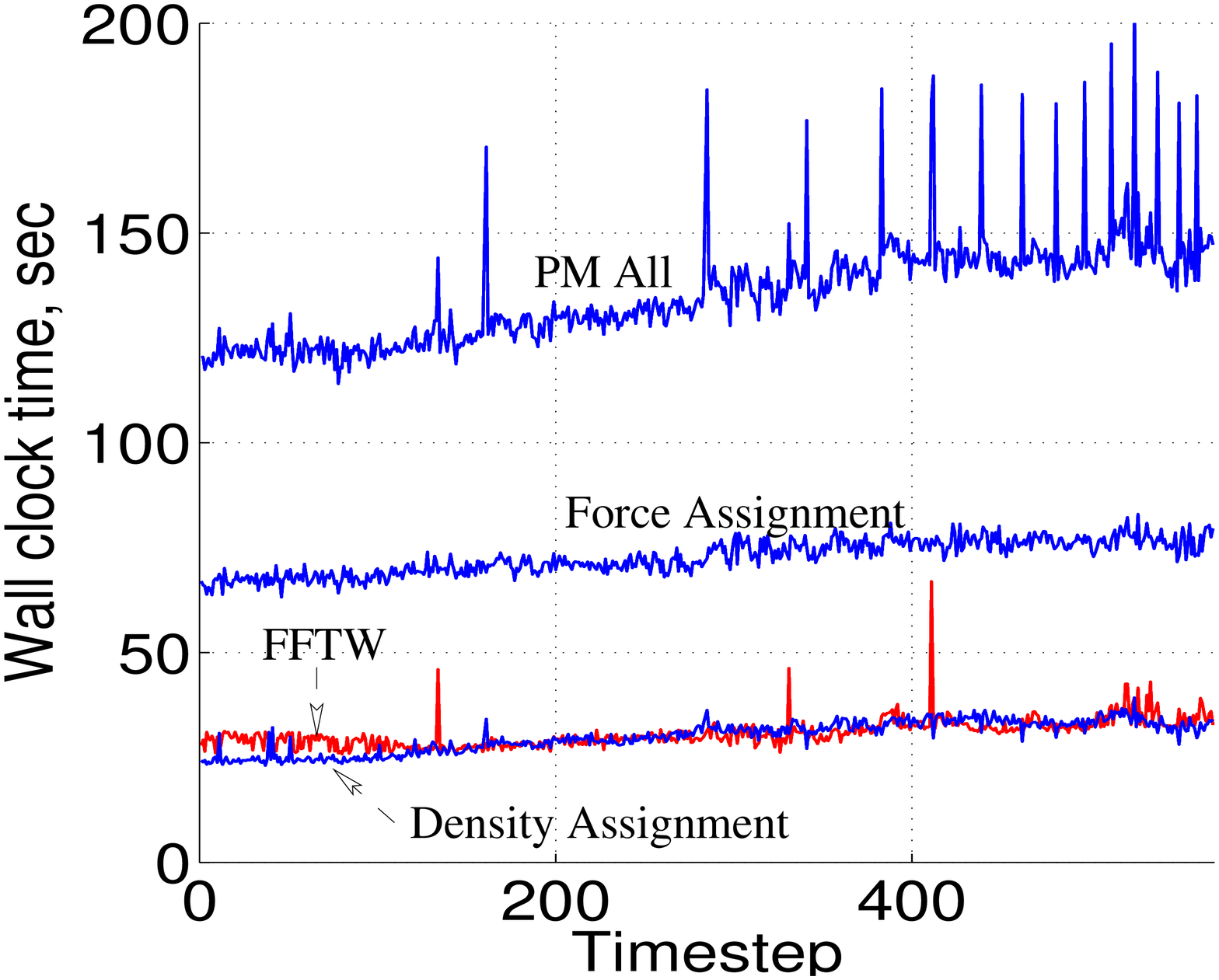}
    \hspace{0.5cm}
    \includegraphics[scale=0.3]{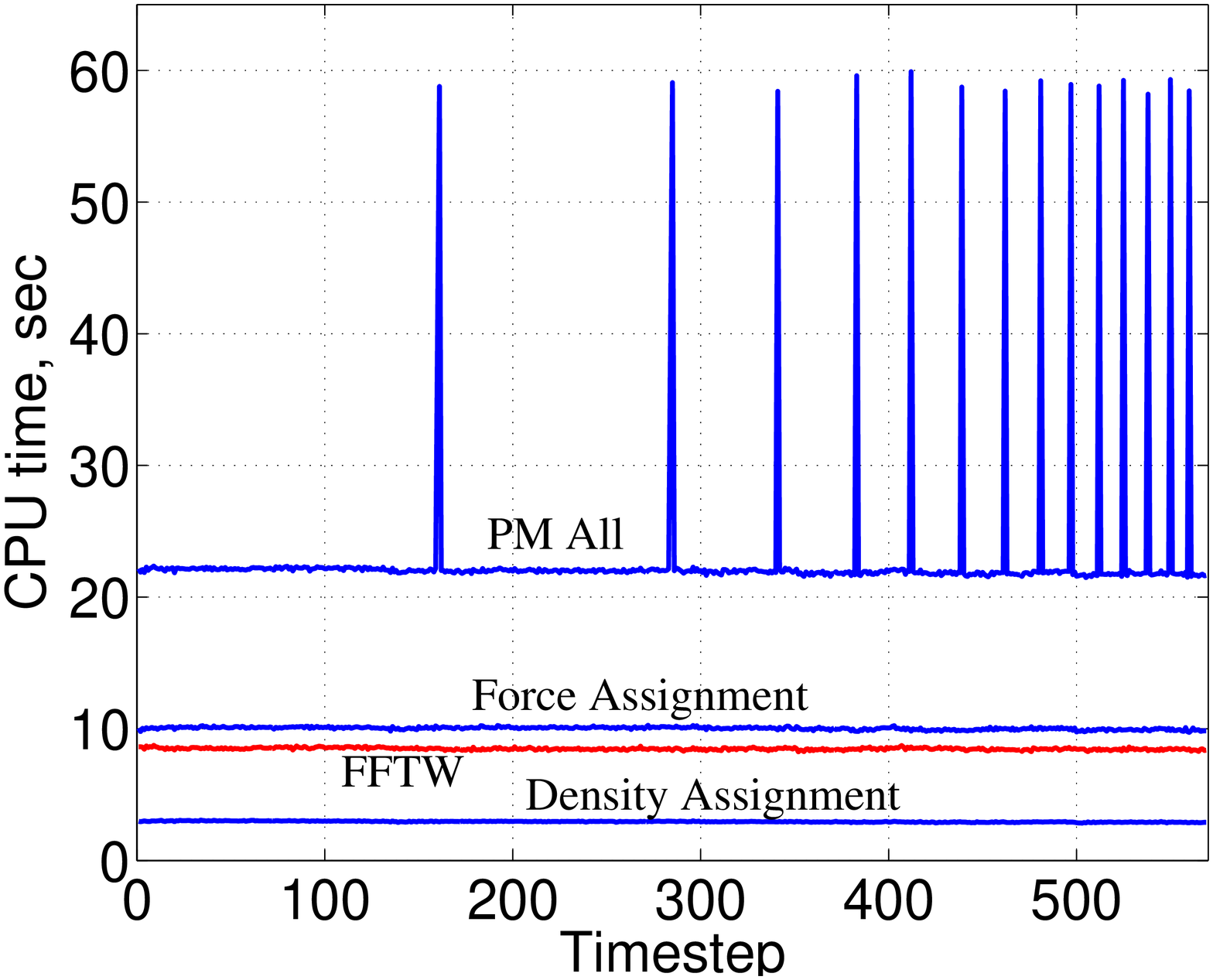}
  \end{center}
    \caption{Structure of the wall clock time (left) and CPU time
      (right) of the PM force computation.  The times are averaged
      over processes.  The 40-second spikes are due to recomputation
      of the PM Green's function when the code is restarted every 24
      hours. Unlike the wall clock time, the CPU time does not increase
      because it does not include time spent waiting for processes
      to finish.
    }\label{fg_pmt}
\end{figure}

Next we present a detailed analysis of the timing structure of the
force calculation.

In Figure \ref{fg_pmt} we present the structure of the wall clock
and CPU time of the PM force computation.  We see that the wall
clock time on average exceeds the CPU time by a factor of 7 during
the run, which means that during the PM force calculation
processes spend 85\% of their time waiting for interprocessor
communication requests to clear. This is a big fraction that can
be reduced significantly by the use of non-blocking requests for
PM density and force grid sends and receives between the local HC
and FFTW slabs (see \S~\ref{sec_hcpm}).  However, except at the
early stages of clustering (the initial timestep was 4 minutes,
growing to 2 hours), the PM time is a small fraction of the
total timestep.

\clearpage

\begin{figure}[t]
  \begin{center}
    \includegraphics[scale=0.3]{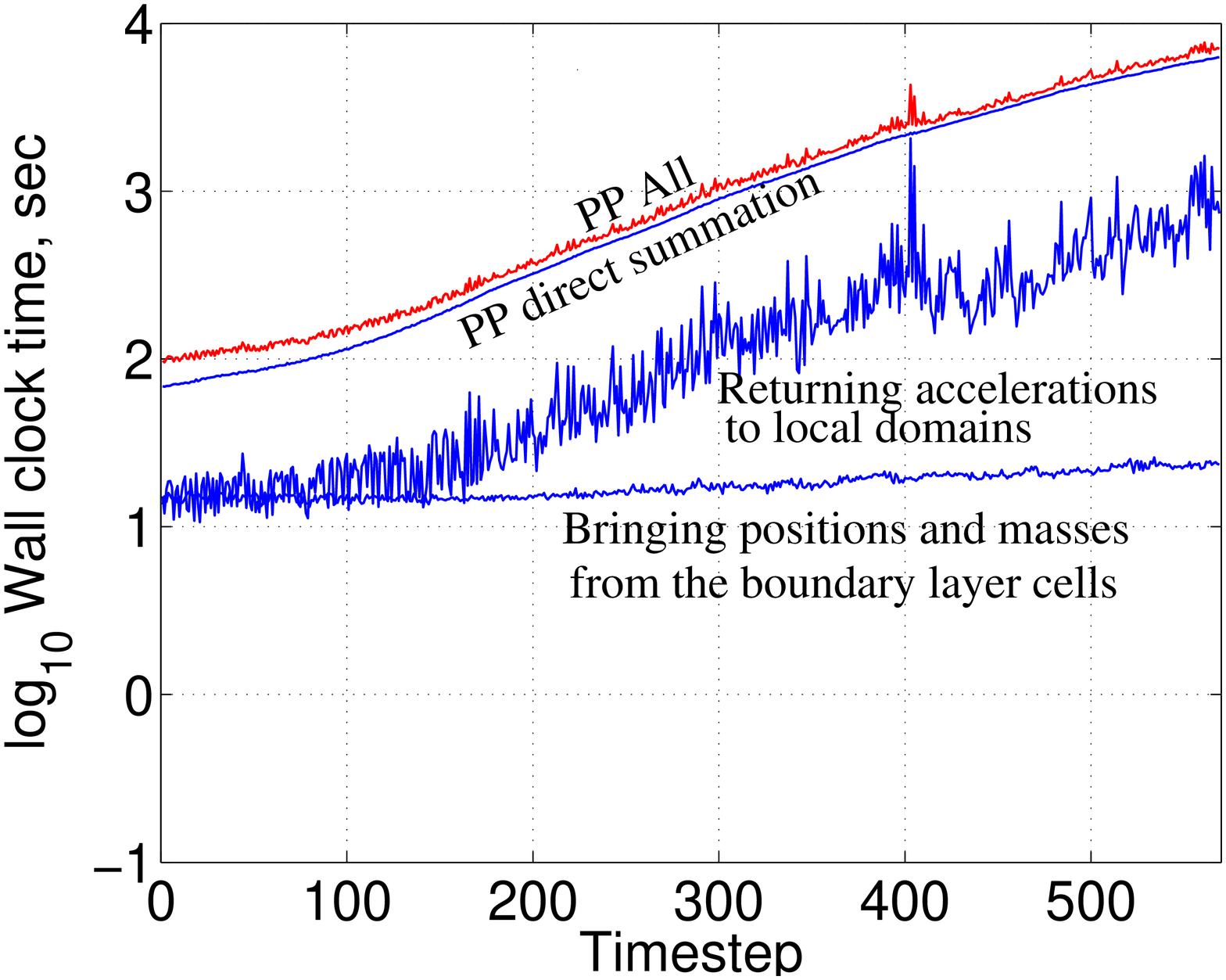}
    \hspace{0.5cm}
    \includegraphics[scale=0.3]{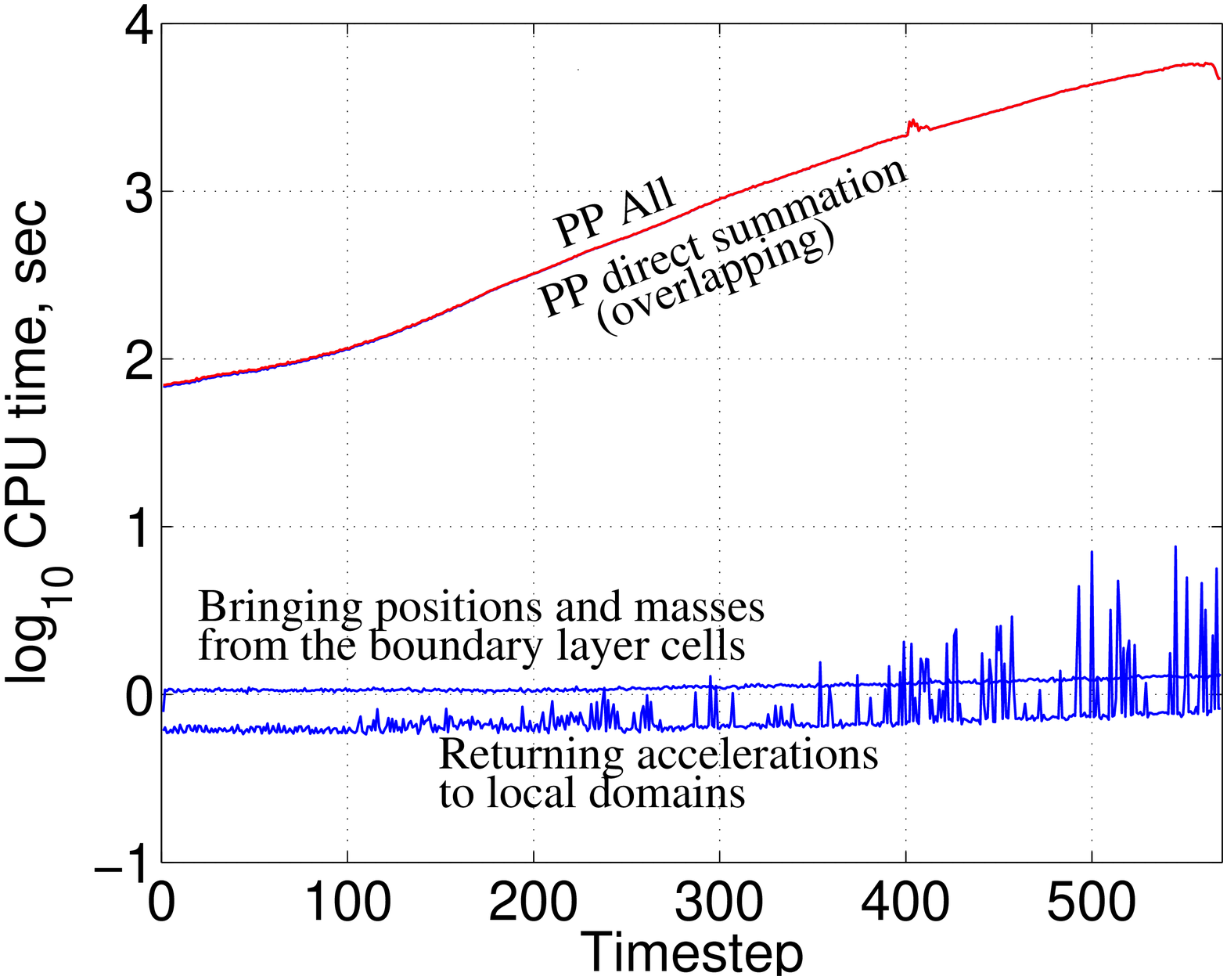}
  \end{center}
    \caption{Structure of the wall clock time (left) and CPU time
      (right) of the PP force computation.  The times are averaged
      over processes.
    }\label{fg_ppt}
\end{figure}

In Figure \ref{fg_ppt} we present the wall clock and CPU timing
for the parallel PP force computation.  Three tasks are required
for a PP step (\S~\ref{sec_hcpp}).  First, particle positions and
masses must be brought from boundary layer cells on other
processes. Second, pairwise gravitational accelerations are
computed by direct summation. Finally, accelerations are returned
across the domain boundaries as needed.  The cost of pair
summation dominates the other tasks and has essentially equal wall
clock and CPU times (hence involves almost no waiting).  For the
communication tasks (sending and receiving particle positions,
masses, and accelerations) the wall clock time greatly exceeds CPU
time because of blocking communication and the resulting waiting
time.

Figure \ref{fg_ppt} shows that the waiting time during PP is
dominated by the return of accelerations after their computation.
The wall clock time of this task is about 10\% that of the
pairwise computation. This is because the load imbalance in the
code arises almost entirely during the pairwise force computation.
Figure \ref{fg_imballong} shows that the average instantaneous
load imbalance is approximately 12\% during the whole run. From
equation (\ref{eq_imbaltar}), we see this means that one of the
processes requires about 12\% more time to complete its pair
summation than the others (since the total CPU time is dominated
by pair summation). Other processes cannot get all their
accelerations returned until this process finishes computing them,
which explains the order of magnitude difference in the direct
summation and return of acceleration wall clock times.

The P$^3$M force calculation accounts for nearly all the time of
each timestep.  Some time is spent by repartitioning every
timestep.  Because repartitioning may result in the exchange of
many HC cells and particles between processes, we might expect it
to take a significant amount of time.  In fact, the total wall
clock time spent on load balancing (analyzing workloads, finding
the target partitioning state, and exchanging data between
processes) takes on average less than 20 seconds per timestep, or
less than the CPU time of the PM force calculation.  Occasionally
the repartitioning time spikes up to nearly 80 seconds but it does
not grow steadily with clustering.  The load balancing time is
generally less than 8\% of the total wall clock time per timestep.
Less than one percent of the wall clock time is spent updating
particle positions and velocities and exchanging the particles
between processes as a result of their motion.

\begin{figure}[t]
  \begin{center}
    \includegraphics[scale=0.35]{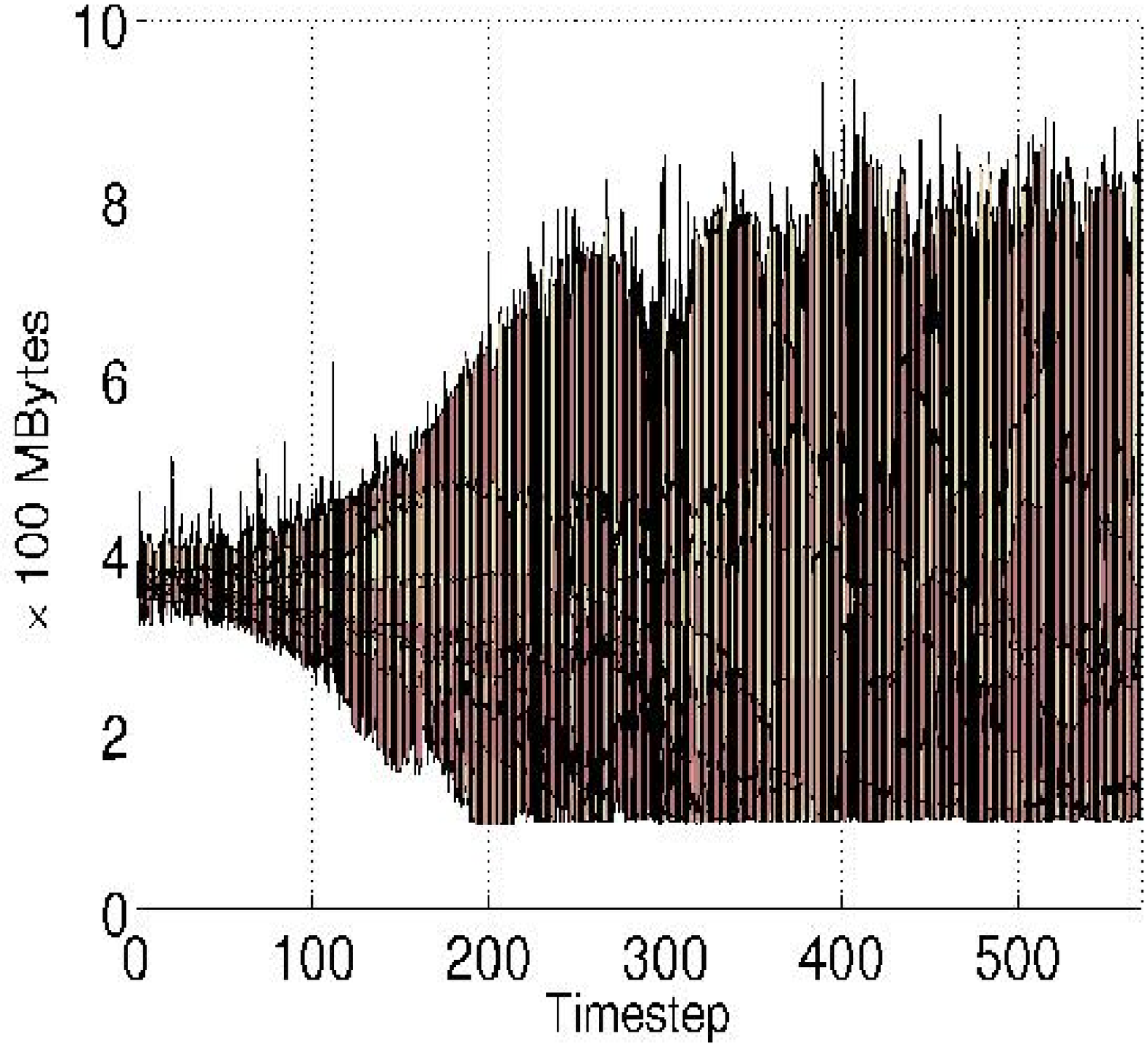}
    \hspace{0.3cm}
    \includegraphics[scale=0.35]{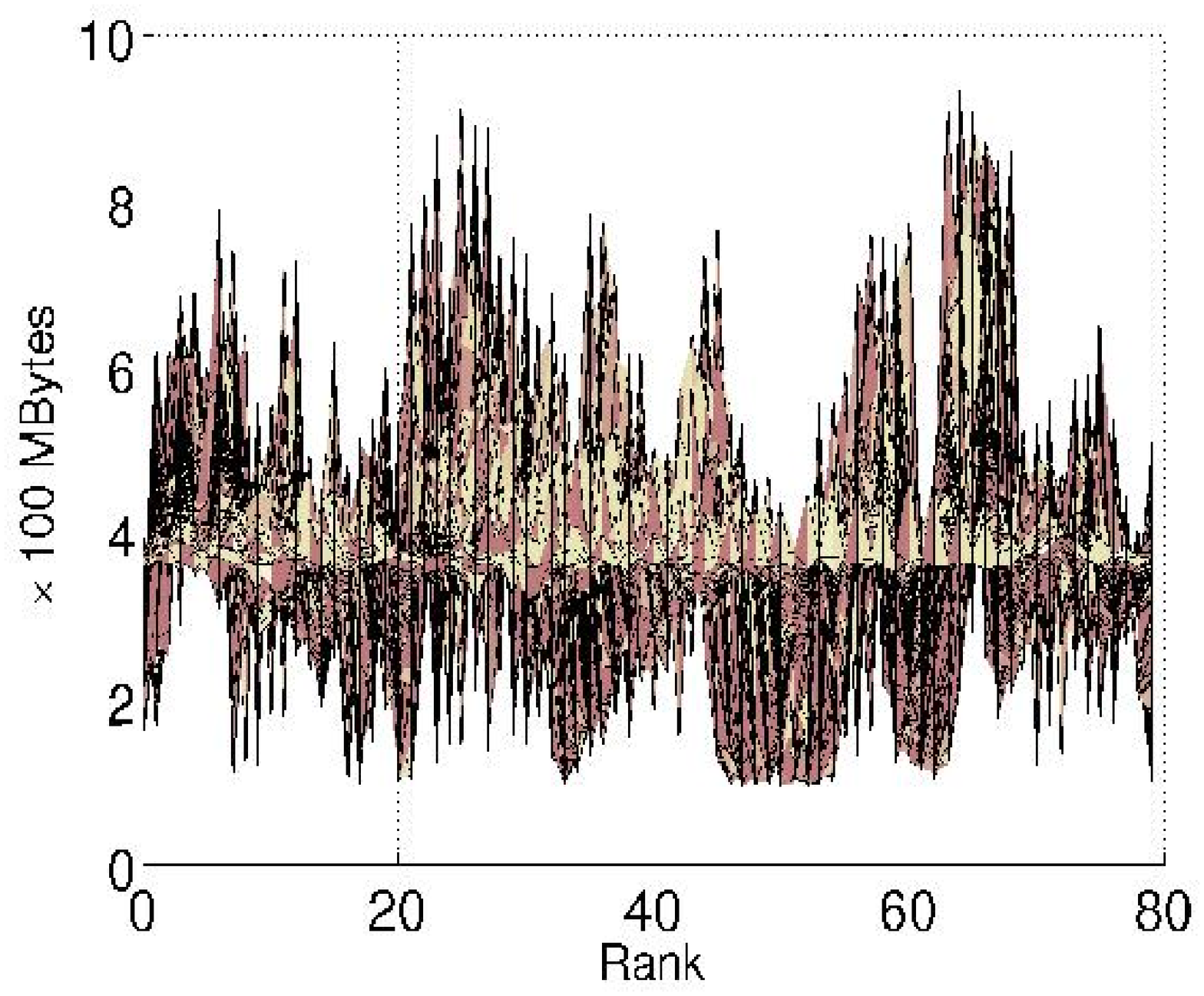}
  \end{center}
  \caption{ Maximum total memory allocated by any node, as a
    function of timestep (left panel) and process rank (right panel).
  }\label{fg_long_mem}
\end{figure}

Next we analyze memory usage during the run. In \S \ref{sec_llmem}
we estimated the memory requirements. We now compare these
estimates with measured memory usage.

In Figure \ref{fg_long_mem} we present the maximum amount of total
memory allocated by a given process during any timestep within the
run. In Linux, a memory request in excess of about 1.4 GB on any
process will crash the run (see \S~\ref{sec_mem}). As expected, at
the beginning of the run when particles are nearly uniformly
distributed, each process requires approximately the same amount
of memory. Using the data from Table \ref{tb_parmem}, we estimate
the initial maximum memory usage to be $M_{\rm P}+M_{\rm HC}+
M_{\rm G}+M_{\rm PM}+M_{\rm FFT}=
(11+0.23+0.5+2+1)\times800^3/\donpc \times4\,\hbox{bytes} = 359.6$
MB, compared with the measured value of  366.5 MB.  Most of the
difference comes from the table of HC entries (3.4 MB) plus slight
variations in the HC mesh and particle storage among processes.

At the end of the run the domains of each process have changed
substantially. All processes require at least $M_{\rm G}+M_{\rm
PM}+M_{\rm FFT} = (0.5+2+1)\times n^0n^1n^2\times 4/\donpc=85.4$
MB, compared with the measured minimum value of 98.3 MB. The
maximum amount of memory varies substantially and is hard to
control at the final timesteps. We limit the maximum memory using
the techniques mentioned in \S \ref{sec_llmem}. During the last
timestep 569, the maximum memory usage was reached on process 27.
During this timestep process 27 changed the volume of its domain
from $1.23\times10^6$ HC cells to $0.94\times10^6$ cells, which
reduced the number of particles in its local region from
$17.7\times10^6$ to $14.2\times10^6$. The number of PP boundary
layer particles received by this process (see \S~\ref{sec_hcpp})
during the same timestep was $1.1\times 10^6$. The measurement of
857.8 MB compares with the predicted value (before repartitioning)
$M_{\rm P}+M_{\rm HC} +M_{\rm G}+M_{\rm PM} +M_{\rm FFT} =
(778.5+24.6+12.8+51.2+25.6)\times 10^6 = 851.3$ MB.  Again the
table of HC entries (3.4 MB) makes up most of the difference.

The reader will notice the similarity between Figures
\ref{fg_long_rn} and \ref{fg_long_mem}.  The maximum memory usage
tracks the volume of HC local regions because the most variable
memory element is the number of particles, which correlates
strongly with the number of HC cells.  Strikingly, the CPU time
for the PP pair summation does not correlate well with the number
of particles or the maximum memory usage. At the last timestep,
process 27 (which used 857.7 MB) took 6037 seconds of CPU time for
the pair summation while process 48 (which used 98.1 MB) took 7152
seconds.  This is a measure of the success of load balancing,
which attempts to equalize CPU time rather than memory usage
across all processes.

\subsection{Parallel Scalability}\label{sec_sclb}

A key test of any parallel code is its scalability as the problem
size and/or number of CPUs increase.  For a fixed problem size, if
the wall clock time scales inversely with the number of CPUs, then
one may use more CPUs to realize the proverb ``many hands make
light work.''  The ideal inverse scaling is readily achievable
with so-called ``embarrassingly parallel'' codes that require
little or no communication, but high efficiency is much more
difficult to achieve for algorithms as complex as P$^3$M.  Even if
a code does not scale perfectly with a fixed problem size, it may
scale well when the problem size is increased, enabling one to
make effective use of supercomputers with hundreds or thousands of
processes to perform very large simulations.

\begin{table}[t]
\begin{center}
  \bigskip
  \begin{tabular}{|c|c|c|c|c|c|}\hline
     Run  &$\dong$ & Nodes & Proc./node &$\donpc$ &(Wall Clock Time)$\times$Nodes  \\\hline
    {3a}  & $288^3$ & 1     &  4    & 4    & \ 79.2h  \\  
    {3b}  & $288^3$ & 2     &  4    & 8    & \ 82.8h  \\  
    {3c}  & $288^3$ & 3     &  4    & 12   & \ 87.0h  \\  
    {3d}  & $288^3$ & 4     &  4    & 16   & \ 91.4h \\  
    {3e}  & $288^3$ & 8     &  4    & 32   & 108.1h \\  
    {3f}  & $288^3$ & 10    &  4    & 40   & 104.9h \\  
    {3g}  & $288^3$ & 12    &  4    & 48   & 111.8h \\  
    {3h}  & $288^3$ & 16    &  4    & 64   & 126.0h \\  
    {3i}  & $288^3$ & 20    &  4    & 80   & 152.3h \\  
    {4a}  & $288^3$ & 10    &  2    & 20   & 148.6h \\  
    {4b}  & $288^3$ & 20    &  2    & 40   & 170.2h \\  
    \hline
    {5a}  & $384^3$ & 5     &  4    & 20   & 421.2h \\  
    {5b}  & $384^3$ & 10    &  4    & 40   & 470.7h \\  
    {5c}  & $384^3$ & 14    &  4    & 56   & 503.5h \\  
    {5d}  & $384^3$ & 20    &  4    & 80   & 589.1h \\  
    \hline
  \end{tabular}
\end{center}
  \caption{Scalability Runs.
  }\label{tb_sclb}
\end{table}

We tested the scalability of \parhc\ using two problem sizes
($288^3$ and $384^3$ $\Lambda$CDM in a 200 Mpc box with Plummer
softening length $\epsilon=0.1$ Mpc, evolved to redshift zero,
taking 634 and 657 timesteps, respectively) and a range of numbers
of computing nodes as shown in Table \ref{tb_sclb}. Each computing
node has two CPUs. The runs have either two processes per node
(one per CPU, Runs 4a,b) or four processes per node (two per CPU,
using Intel hyperthreading). For perfect scalability, the times in
the last column would be equal for simulations of the same grid
size $\dong$.

From Table \ref{tb_sclb} we may draw several conclusions.  First,
\parhc\ does not scale perfectly like an embarrassingly parallel
application.  On the other hand, increasing the number of
processes up to 80 leads to a steadily decreasing wall clock time.
Comparing Runs 3a and 3g, we see that for up to 48 processes, the
wall clock time scales as $\donpc^{-0.86}$. Hyperthreading also
gives a significant speedup.  Comparing Runs 3f and 4b, which have
the same total number of processes but different numbers of
compute nodes, we see that hyperthreading improves the code
performance by a factor 1.62.  We also see that the code scales
reasonably well as the problem size is increased. Comparing Runs
3f and 5b, the wall clock time is proportional to $N^{1.74}$ where
$N$ is the number of particles.  When the wall clock time is
dominated by PP pair summation, we expect scaling as $N^2$.

\begin{figure}[t]
  \begin{center}
    \includegraphics[scale=0.4]{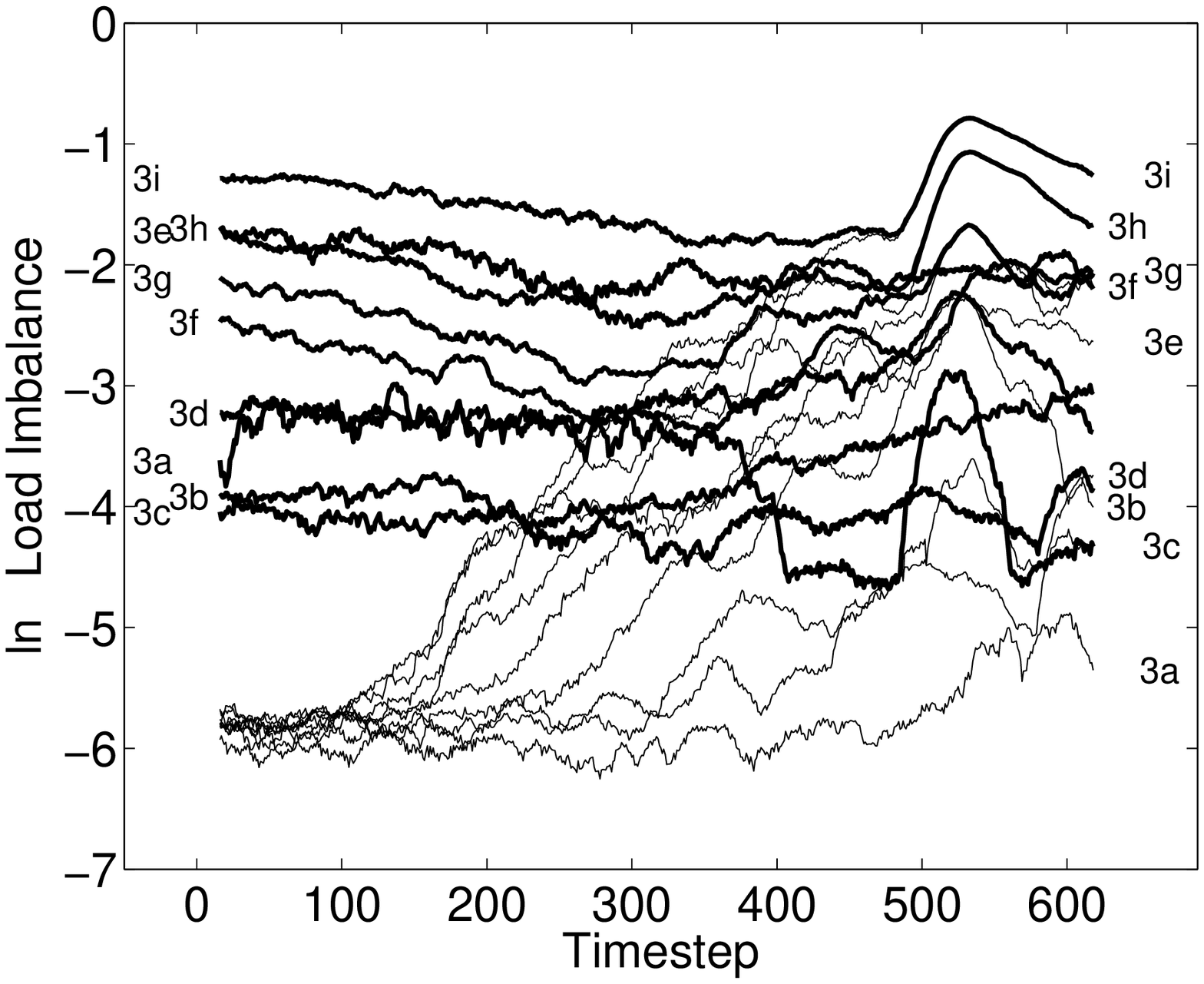}
    \hspace{0.5cm}
    \includegraphics[scale=0.4]{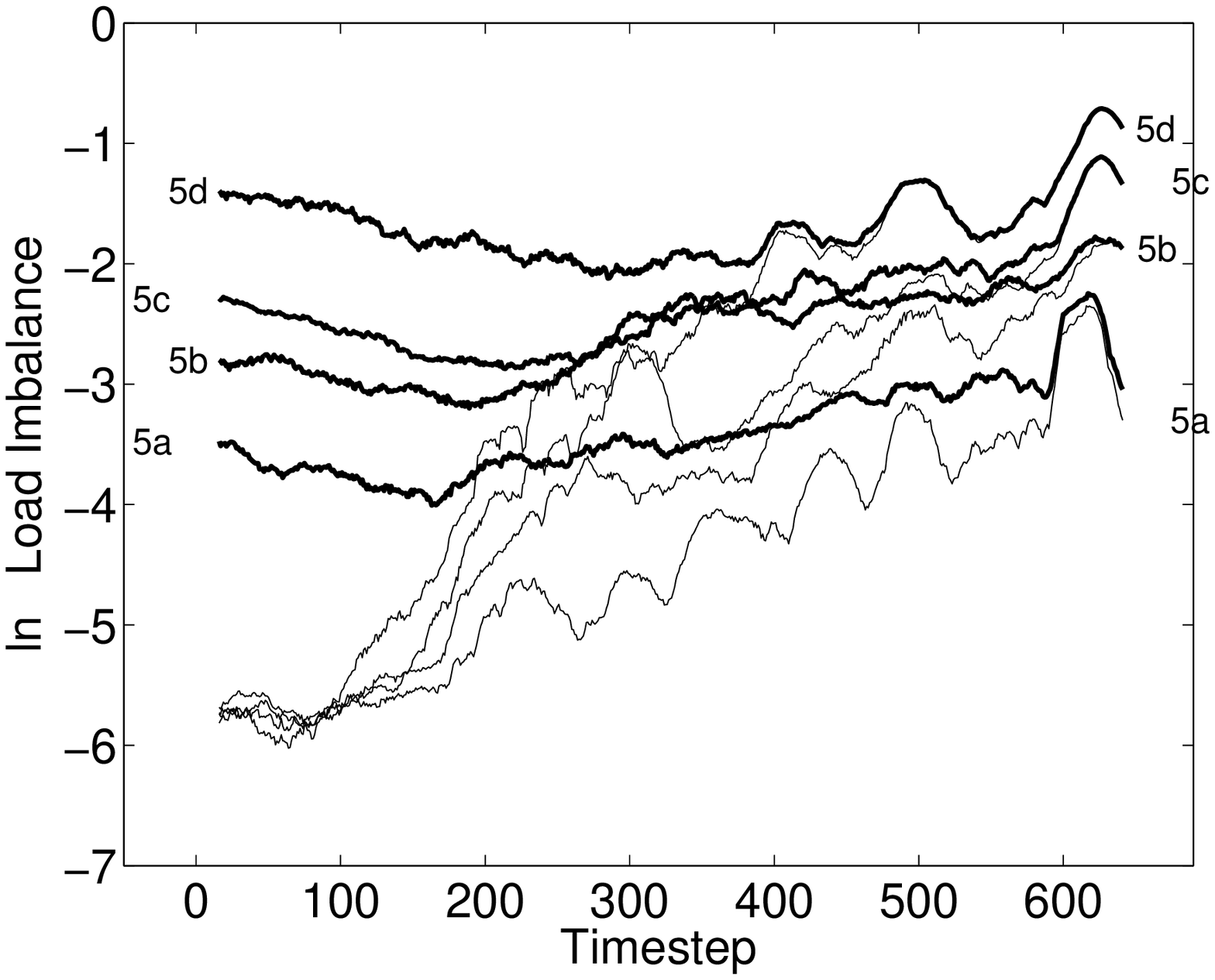}
  \end{center}
  \caption{ Instantaneous (heavy lines) and residual (thin
    lines) load imbalance as a function of timestep for Runs 3
    ($\dong=288^3$, left) and 5 ($\dong=384^3$, right).  The
    individual runs are labelled.
  }\label{fg_sclb-imbal}
\end{figure}

The most significant deviations from perfect scalability arise
with the largest numbers of processes, in particular Runs 3h, 3i,
and 5d.  These arise from load imbalance, as shown in Figure
\ref{fg_sclb-imbal}.  A significant increase in load imbalance
shows up after timestep 500 in Runs 3 and timestep 600 in Runs 4
due to the formation of a dense dark matter clump.  When the
number of processes is sufficiently large, this leads to one or a
few HC cells beginning to take as much time for PP pairwise
summation as the average time for the other processes. According
to equation (\ref{eq_resimbalcon}), the result is a growing
residual load imbalance.  Scalability breaks down beyond a certain
number of processes, given by equation (\ref{eq_npclim}). Once the
performance saturates, the instantaneous and residual load
imbalance match because it is no longer possible to improve the
load balancing by rearrangement of the partitioning.

Although the performance of \parhc\ is limited by the PP pair
summation and not by the PM force computation, it is worth
recalling that, because the current code uses blocking sends and
receives to pass data between the particle and grid structures,
the PM time also scales imperfectly.  When we implement adaptive
P$^3$M, the PP time will decrease significantly so that the PM
time becomes a significant fraction of the total wall clock time.
To improve the parallel scaling, it will be important to implement
non-blocking communication for the PM particle/grid messages.
\section{Conclusions}\label{sec_concl}

Parallelizing a gravitational N-body code involves considerably
more work than simply computing different sections of an array on
different processors.  The extreme clustering that develops as a
result of gravitational instability creates significant
challenges.  A successful parallelization strategy requires
careful consideration of CPU load balancing, memory management,
communication cost, and scalability.

The first decision that must be made in parallelizing any
algorithm is how to divide up the problem to run on multiple
processes.  In the present context this means choosing a method of
domain decomposition.  Because P$^3$M is a hybrid algorithm
combining elements of three-dimensional rectangular meshes and
one-dimensional particle lists, we chose a hybrid method of domain
decomposition.  A regular mesh, distributed among the processes by
a simple slab domain decomposition, is used to obtain the PM force
from the mesh density.  A one-dimensional structure --- the
Hilbert curve -- is introduced to handle the distribution of
particles across the processes and to load balance the work done
on particles.

Implementing Hilbert curve domain decomposition in a particle code
is the major innovation of our work.  To take full advantage of it
we had to employ a number of advanced techniques.  First, in \S
\ref{sec_loadbal} we devised a discrete algorithm to find the
nearly optimal partitioning of the Hilbert curve so as to achieve
load balance, the desirable state in which all processors have the
same amount of work to do. This is a much greater challenge in a
hybrid code than in a purely mesh-based code such as a
hydrodynamic solver or a gridless particle code such as the tree
code.  We then made the domain decomposition dynamic by
repartitioning the Hilbert curve every timestep, allowing us to
dynamically maintain approximate load balance even when the
particle clustering became strong.

In \S \ref{sec_voids} we presented a fast method for finding the
position of a cell along the Hilbert curve given its
three-dimensional location.  This procedure allows us to access
arbitrary cells in a general irregular domain by a lookup table
much faster than using the special-purpose Hilbert curve function
of \cite{doug}.

In \S \ref{sec_compress} we introduced run-length encoding to
greatly reduce the communication cost for transferring information
between the particle and mesh structures required during the PM
force computation.

In \S \ref{sec_adv} we optimized the particle distribution within
each process so as to improve the cache performance critical for
efficient pair summation in the PP force calculation.

By choosing the domain decomposition method appropriate for each
data structure, and by implementing these additional innovations,
we achieved good load balance and scalability even under extreme
clustering.  The techniques we introduced for effective
parallelization should be applicable to a broad range of other
computational problems in astrophysics including smooth-particle
hydrodynamics and radiative transfer.

Tests of our algorithm in \S \ref{sec_test} showed that we
achieved our goals of scalability and load balance, with two
caveats mentioned at the end.

In Figure \ref{fg_wall} we demonstrated the importance of using a
dynamic three-dimensional domain decomposition method instead of a
static one-dimensional slab decomposition.  The latter method is
unable to handle extreme spatial inhomogeneity.

Next, we performed a long $800^3$ $\Lambda$CDM simulation
(performed on only 20 dual-processor computing nodes) to
thoroughly test the load balancing algorithm.  The average load
imbalance for this simulation run with 80 processes was only 12\%,
meaning that 12\% of the total wall clock time of all the CPUs was
wasted. While not perfect, this is very good performance for the
P$^3$M algorithm.  The largest cause of load imbalance over most
of the simulation was our inability to predict the total CPU time
of the next timestep on each process because of variations in
cache memory usage.

Finally, we tested the limits of scalability by performing the set
of runs in Table \ref{tb_sclb}.  For up to 48 processes the code
performed with very good parallel speedup --- the wall clock time
scaled as $\donpc^{-0.86}$ for $\donpc$ processes, as compared
with $\donpc^{-1}$ for perfect scalability.

Our tests revealed two limitations to scalability that will be
addressed in a later paper presenting an adaptive P$^3$M
algorithm.  First, the current code uses blocking communication
for sending data between the particle and grid structures in the
PM force calculation.  In other words, some processes sit idle
waiting for others to complete their communications requests.
This inefficiency, while small when PP forces are expensive to
compute, will become more important when adaptive mesh refinement
reduces the PP cost. The solution is to restructure the
communication to work with non-blocking sends and receives.

Finally, we observed our code to become inefficient when a handful
of Hilbert curve cells (out of millions in the entire simulation)
begin to dominate the computation of PP forces.  Because a
non-adaptive code does not allow refinement of one cell, a single
process must handle these extremely clustered cells even if the
other processes have to wait idly while it finishes.  The solution
to this problem is simply to use adaptive refinement.  In a later
paper we present an algorithm for scalable adaptive P$^3$M
building upon the techniques introduced in the current paper.

Once this paper is accepted for publication, the simulation codes
presented here will be made publicly available at {\tt
http://antares.mit.edu/}.

\acknowledgments

A. Shirokov would like to thank Paul Shapiro and Mike Warren for
useful discussions and Serhii Zhak for helpful comments on
hardware issues.  This work was supported by NSF grant
AST-0407050.
\appendix

\section{Code Overview and Variables}\label{sec_overview}

\begin{figure}[t]
  \begin{center}
    \includegraphics[scale=0.55]{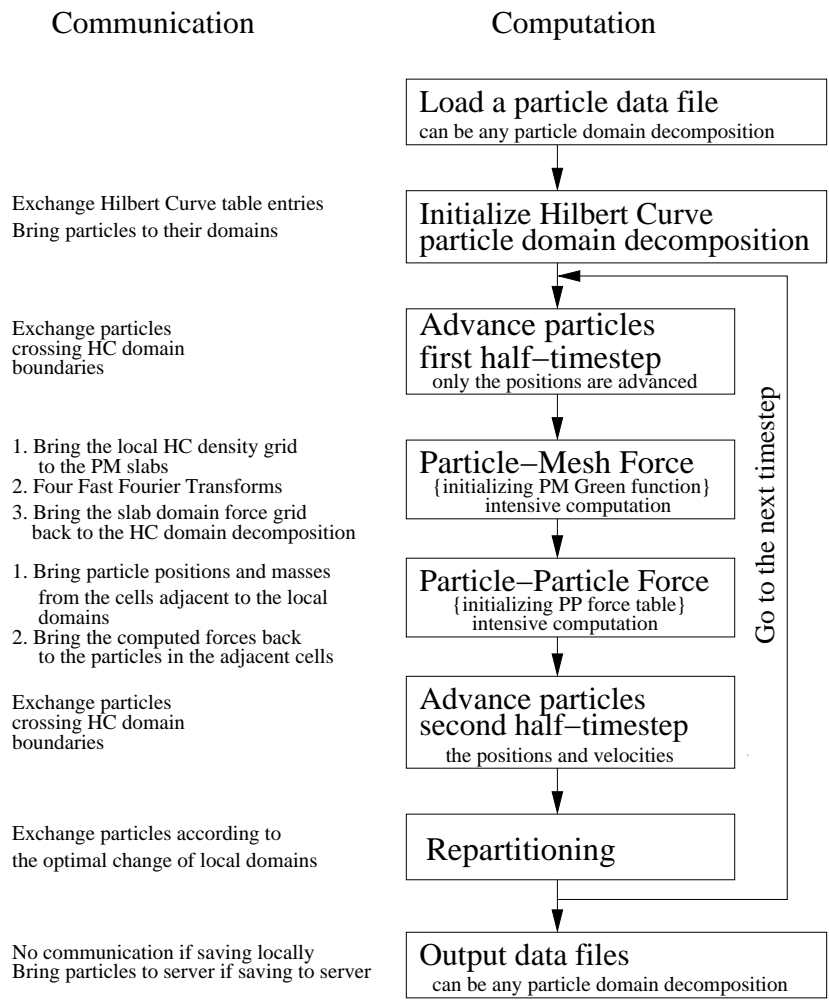}
  \end{center}
    \caption{Block diagram of the parallel P$^3$M code \parhc.
    }\label{fg_block}
\end{figure}

Figure \ref{fg_block} presents a block diagram of our parallel
Hilbert curve domain decomposition code \parhc.  The code may run
on any number of processes $\donpc$ (this is not restricted to
being a power of 2). The code is written in ansi C with MPI calls.
Excluding FFTW, it consists of about 33,000 lines of code. This
Appendix gives an overview of the code guiding the reader to the
relevant parts of the main paper.

The code begins by loading particle data from one or more files.
At the beginning of a simulation, these files contain the initial
conditions.  A simulation may also be started using particle data
that have already been evolved.  The particle data may be either
in one file on the cluster server or they may be in multiple
files, one stored on each cluster compute node.

The next step is to initialize the Hilbert curve for domain
decomposition based on the particle distribution, as described in
\S \ref{sec_hci}.  The \parhc\ code stores particle data (e.g.
positions and other variables as described in \S \ref{ser_pa})
differently than mesh data (e.g. density). Mesh-based data are
stored on a regular PM mesh which is divided by planes into a set
of thick slabs, one for each parallel process. Particle data are
organized into larger cells called Hilbert curve (HC) cells.
(These cells have a size just slightly larger than the cutoff
radius for the particle-particle or PP short-range force.) The
cells are then connected like beads on a necklace by a closed
one-dimensional curve called a Hilbert curve. The Hilbert curve
initialization step computes and stores the information needed to
determine the location of every bead on the necklace, that is, it
associates a one-dimensional address with each HC cell.

Once the Hilbert curve is initialized, the Hilbert curve is cut
into a series of segments, each segment (called a HC local region)
containing a set of HC cells and their associated particles.  Each
parallel process owns one of the local regions.  The particles are
thus sent from the process on which they were initially loaded to
the process where they belong.  When restarting a run on the same
nodes, the particles are already on the correct processes.  When
starting a new simulation, the partitions are set with equal
spacing along the Hilbert Curve and the particles are sent to the
appropriate processes.

This method of assigning particles to processes based on their
position along a one-dimensional curve of discrete segments is
called Hilbert curve domain decomposition and it is explained in
\S \ref{sec_par}.  The organization of particles within a process
is described in \S \ref{sec_layout}.

After these initialization steps the code integrates the equations
of motion given in \S \ref{sec_eom} using a leapfrog scheme
presented in \S \ref{sec_leapfrog}.  First the positions are
advanced one-half timestep, and if they cross HC local region
boundaries they are moved to the correct process.

Next, gravitational forces are computed. Most of the work done by
the code is spent computing forces.  The interparticle forces are
split into a long-range particle-mesh part computed on the mesh
and interpolated to the particles, plus a short-range
particle-particle correction, as described in \S\S \ref{sec_pm},
\ref{sec_pp}, and \ref{sec_hcforce}.  Most of the communication
between processes occurs during these steps.  If the particle-mesh
Green's function has not yet been computed, it is computed just
before the first PM calculation.  The Green's function is
essentially the discrete Fourier transform of $r^{-2}$, modified
by an anti-aliasing filter to reduce anisotropy on scales of the
PM mesh spacing.  After the particle-mesh forces are computed,
they are incremented by the particle-particle forces (the most
time-consuming part of P$^3$M).  After the forces are computed,
velocities and then positions are advanced to the end of the
timestep.  Once more, particles that cross HC local region
boundaries are transferred to the correct process.

After the particles have moved, the cuts along the Hilbert curve
are moved so as to change the segment lengths and thereby change
the domain decomposition.  This step is called repartitioning. Its
purpose is to ensure that, as much as possible, each process takes
the same amount of time to perform its work as every other
process, so that processes do not sit idle waiting for others to
finish their work.  (Certain operations, like the FFT, must be
globally synchronized.)  When this ideal situation is met, the
code is said to be load balanced. Repartitioning is performed
every timestep to optimize load balance, as explained in \S
\ref{sec_loadbal}.

At the end of the integration step, the code generally loops back
to advance another step. Periodically the code also outputs the
particle data, usually writing in parallel to local hard drives
attached to each compute node.

Table \ref{tb_cvars} presents a list of frequently used symbols
and variables in the code.

\begin{table}
\begin{tabular}{|@{ } > {\PBS\raggedright}p{0.12\textwidth} @{ } > {\PBS\raggedright}p{0.15\textwidth}  @{} > {\PBS}p{0.70\textwidth}@{  }|}
\hline
    Notation & Code variables & Description\\
\hline
      \multicolumn{3}{|l|}{\large \bf Serial and parallel codes}\\
      $L^0, L^1, L^2$          &                        & The simulation box size in comoving Mpc, $L^i \equiv n^i \Delta x$. \\
      $\donp$                  &                        & The total number of particles in the simulation volume\\
      $\dodx$                  &{\tt dx}                & The PM mesh spacing, same in all dimensions\\
      $\tilde\epsilon$         & {\tt epsilon}          & Plummer softening length, in units of $\dodx$\\
      $\eta_t$                 & {\tt etat}             & Time integration parameter, usually $\eta_t = 0.05$\\
      $\dormax$                & {\tt cr.max}           & PP-force length, in units of $\dodx$, typically $2.78$\\
      $\don^0, \don^1, \don^2$         &{\tt n0..n2}    & The size of the simulation box, in units of PM cells\\
      $\donc^0, \donc^1, \donc^2$ &{\tt ncm0..ncm2}     & The size of the simulation box in chaining mesh cells\\
      $\dong$                  & {\tt ngrid}            & $ = n^0 n^1 n^2$, the total number of PM grid points \\
      $\dodxc^i$           &{\tt cr.len0..cr.len2}  &\quad\quad\quad Chaining-mesh grid spacing along three dimensions\\
                       &\dopa, \dopaf           & Starting and finishing pointers of particle array~${\tt [pa, pa\_f)}$\\
                               &\dopafa                 & Pointer to the end of the preallocated particle array,
                                                          equals {\tt pa\_f} in the serial code. In the parallel
                                                          code~${\tt pa\_fa} \ge {\tt pa\_f}$.\\
      \multicolumn{3}{|l|}{\large \bf Parallel codes only}\\
                               & \dosloc[i]            & Starting index of FFTW slab $i$ for the FFT plan\\
                               & \donloc[i]            & Thickness of FFTW slab $i$. The whole slab on process\newline
                                                          $i$ has size $\don^0\don^1\donloc[i]$, where $i = 0\ldots\donpc-1$\\
      $\donh^0, \donh^1, \donh^2$ & {\tt hc\_{n0} \ldots hc\_{n2}} & The size of the simulation box in Hilbert curve (HC) mesh cells, per dimension\\
      $\donhc$                 &                        & $=\donh^0\donh^1\donh^2$, the total number of HC mesh cells.\\
      $\donp(i)$               &                        & The number of particles local to process $i$\\
      $\donpc$                 &{\tt wk.nproc}          & Number of worker processes, those containing particle data\\
      ${\bf \doc}$             &                         & Coordinates of a cell in the Hilbert curve mesh\\
      $\doh$                   &                         & A Hilbert curve index\\
      $\dor$                   &                         & A raw Hilbert curve index\\
      $\domhctoi({\bf \doc})$  &\dohctoi                 & Mapping between the HC index $h$ and the cell's coordinates\\
      $\domhitoc(h)$           &\dohitoc                 & The inverse of the above mapping\\
      $\domhctor({\bf \doc})$  &                         & Mapping between the HC raw index $h$ and the cell's coordinates\\
      $\dom$                   &                         & HC order: the number of cells in the HC mesh is $2^{md}$, $d=3$\\
      $\dodxh^i$               &                         & HC mesh spacing along dimension $i$, in units of $\dodx$\\
      $\dohclr^i$              &                         & HC local region of process $i \in [0, \donpc)$\\
                       &{\tt hc\_stg}            & 3-d ragged array with gaps of the cells of the $\dohclr^i$\\
      $\doK$                   &                         & The number of entries of the HC into the simulation volume\\
      $\doheb^k, \dohen^k$     &                         & The HC index of the $k$-th entry of the curve
                                                           into the simulation volume $k\in[0,K)$, and the number of the HC cells
                               that follow contiguously inside the simulation volume along the curve\\
      $\dohb(i), \dohn(i)$       &                         & The HC index of the bottom partition and number of cells on process $i$\\
      $\dorb(i), \dorn(i)$       &                         & Same, with the raw index\\
   \hline
\end{tabular}
\caption{Frequently used variables.} \label{tb_cvars}
\end{table}

\section{Moore's Hilbert Curve Implementation Functions}\label{sec_moore}

Working with a Hilbert (space-filling) curve requires a mapping
from HC index $\doh$ to HC cell position $\doc$ and vice versa.
\cite{doug} implemented C functions that accomplish these
mappings.

The most straightforward implementation of the Hilbert curve is
too slow, since a Hilbert curve is defined recursively by its self
similarity. Moore's implementation is based on a much faster
non-recursive algorithm of \cite{b71}.

A one-to-one correspondence between a cell and the HC index is
given by the following functions of Moore's implementation:
\begin{equation}
  \begin{array}{l}
    \doh=\dohctoi(d,\dom,{\bf \doc})\equiv\domhctoi_d({\bf \doc})\\
    {\bf \doc}=\dohitoc(d,\dom,\doh)\equiv\domhitoc_d(\doh)\ .
  \end{array}
\end{equation}
The Hilbert curve index $\doh$ is of type {\tt long long unsigned}
and ${\bf c}$ a vector of three integer indices giving the spatial
coordinates of the cell.  These two functions are inverse to each
other. They are implemented for any spatial dimension $d$. For
example for $d=2$ in Figure \ref{fg_virgo}, a function
$\domhitoc_2(0)$ will return the position of the curve's starting
point, and the function $\domhitoc_2(1)$ returns the position of
the next cell along the curve. We verified that the resulting
curve indeed provides a one-to-one mapping between the cell and
its HC index preserving space locality for all HC mesh sizes up to
$2048^3$.

Table \ref{tb_moore} shows the average measured CPU time to make
one call to the HC function \dohctoi\ on a 2.4 GHz Intel Xeon
processor. The time shown is compared with the average times to
make other simple arithmetic operations or memory references.  It
is surprising how fast the implementation is: it takes just two
minutes to make $512^3$ Hilbert curve function calls on a single
processor. However, in comparison with a simple arithmetic
operation or triple array dereferencing, it is very slow:
An average \dohctoi\ function call is about 120 times slower than
a  triple array dereferencing for the $512^3$ HC mesh;
the function call time increases linearly with the increase of the
Hilbert curve order $\dom$ as $(4.10+0.775m)\times10^{-7}$ sec.
We should therefore avoid using the HC implementation function
calls when it is possible to use memory dereferencing instead.  As
we discuss in \S\S \ref{sec_adv} and \ref{sec_hcpm} we
successfully avoid multiple calls to {\tt hilbert\_c2i}\ during
the force calculation and the particle advancement by proper
organization of memory usage.

\begin{table}[t]
  \begin{center}
    \begin{tabular}{|@{ } > {\PBS\raggedright}p{0.65\textwidth} | @{  } > {\PBS}p{0.25\textwidth}@{}|}
    \hline
      Operation within triple {\tt for} loop  & CPU time per call, $10^{-9}$ sec \\
      \hline
      nothing (bare triple {\tt for} loop)              &   7.75\\\hline
      inline multiplication (innermost integer index squared)   &   12.8 \\\hline
      arithmetic function call (innermost integer index squared)  &   18.57 \\\hline
      triple array dereferencing                      &   16.29 \\\hline
      \dohctoi\ function call ($m=9$)           &  1056.\\\hline
      \dohitoc\ function call ($m=9$)           &  920. \\\hline
    \end{tabular}
    \caption{ CPU time averaged for $512^3$ calls ($m=9$ bits per dimension)}
  \label{tb_moore}
  \end{center}
\end{table}
\def\newa{{NewA}}


\begin{thebibliography}{99}

\bibitem[Arnold (1978)]{a78} Arnold, V. I. 1978, Mathematical Methods of
  Classical Mechanics (New York: Springer-Verlag)

\bibitem[Bennett et al.\ (2003)]{b03}  Bennett, C.L., et al. 2003, \apjs, 148, 1

\bibitem[Bertschinger (1991)]{b91} Bertschinger, E. 1991, in After the First
  Three Minutes, ed. S. Holt, V. Trimble, \& C. Bennett (New York: AIP), 297

\bibitem[Bertschinger (1995)]{b95} Bertschinger, E. 1995, COSMICS software
  release (astro-ph/9506070)  

\bibitem[Bertschinger (1996)]{b93} Bertschinger, E. 1996, in
  Cosmology and Large Scale Structure, proc. Les Houches Summer School,
  Session LX, ed. R. Schaeffer, J. Silk, M. Spiro, and J. Zinn-Justin
  (Amsterdam: Elsevier Science), 273

\bibitem[Bertschinger (1998)]{b98} Bertschinger, E. 1998, \araa, 36, 599

\bibitem[Bertschinger \& Gelb (1991)]{bg91} Bertschinger, E. \&
  Gelb, J.M. 1991, Comp.\ in Phys., 5, 164

\bibitem[Binney \& Tremaine (1994)]{bt94} Binney, J. \& Tremaine, S. 1994,
  Galactic Dynamics (Princeton: Princeton University Press)

\bibitem[Bode \& Ostriker (2003)]{bo03} Bode, P. \& Ostriker, J. 2003,
  \apjs, 145, 1


\bibitem[Butz (1971)]{b71} Butz, A.R. 1971, IEEE Trans.\ Comp., 20, 424

\bibitem[Couchman (1991)]{c91} Couchman, H.M.P. 1991, \apj, 368, L23


\bibitem[Dav\'e, Dubinski, \& Hernquist (1997)]{ddh97} Dav\'e, R.,
  Dubinski, D.R. \& Hernquist, L. 1997, \newa, 2, 277 

\bibitem[Dubinski et al.\ (2004)]{dkp04} Dubinski, J., Kim, J., Park, C.,
  \& Humble, R. 2004, \newa, 9, 111  

\bibitem[Efstathiou \& Eastwood (1981)]{ee81} Efstathiou, G. \&
  Eastwood, J.W. 1981, \mnras, 194, 503

\bibitem[Efstathiou et al.\ (1985)]{edfw85} Efstathiou, G., Davis, M.,
  Frenk, C.S., \& White, S.D.M. 1985, \apjs, 57, 241

\bibitem[Ferrell \& Bertschinger (1994)]{fb94} Ferrell, R. \&
  Bertschinger, E. 1994, Int.\ J.\ Mod.\ Phys.\ C, 5, 933

\bibitem[Ferrell \& Bertschinger (1995)]{fb95} Ferrell, R. \& Bertschinger, E.
  1995, in proc.\ Soc.\ Comp.\ Sim.\ Multiconference (astro-ph/9503042)

\bibitem[Frigo \& Johnson (2003)]{fftw} Frigo, M. \& Johnson, S.,
  Fast Fourier Transform implementation at http://www.fftw.org/

\bibitem[Fryxell et al.\ (2000)]{flash} Fryxell, B.\ et al. 2000,
  \apjs, 131, 273; http://flash.uchicago.edu/

\bibitem[Gelb \& Bertschinger (1994)]{gb94} Gelb, J.M. \& Bertschinger, E.
  1994, \apj, 436, 467  




\bibitem[Hockney \& Eastwood (1988)]{he88} Hockney, R.W. \& Eastwood, J.W.
  1988, Computer Simulation Using Particles (Bristol: Adam Hilger)


\bibitem[MacFarland et al.\ (1998)]{mcpp98} MacFarland, T., Couchman, H.M.P.,
  Pearce, F.R., \& Pichlmeier, J. 1998, \newa, 3, 687  

\bibitem[Merz, Pen, \& Trac (2004)]{mpt04} Merz, H., Pen, U.-L., \&
  Trac, H. 2004, submitted to \mnras\ (preprint astro-ph/0402443)

\bibitem[Moore (1994)]{doug} Moore, D. 1994, Hilbert curve
  implementation at\hfil\break
  http://www.caam.rice.edu/\%7Edougm/twiddle/Hilbert/


\bibitem[Pilkington \& Baden (1996)]{pb94} Pilkington, J. \& Baden, S. 1996,
  IEEE Trans.\ Par.\ Dist.\ Systems, 7, 288 

\bibitem[Plummer (1911)]{p11} Plummer, H. C. 1911, \mnras, 71, 460  

\bibitem[Quinn et al.\ (1997)]{qks97} Quinn, T., Katz, N., Stadel, J. \&
  Lake, G. 1997, preprint (astro-ph/9710043)  

\bibitem[Ruth (1983)]{r83} Ruth, R. D. 1983, IEEE Trans.\ Nucl.\ Sci., 30, 2669

\bibitem[Salmon \& Warren (1994)]{sw94} Salmon, J. \& Warren, M. 1994,
  J.\ Comp.\ Phys., 111, 136

\bibitem[Saha \& Tremaine (1992)]{st92} Saha, P. \& Tremaine, S. 1992, \aj,
  104, 1633  

\bibitem[Springel, Yoshida, \& White (2001)]{syw01} Springel, V., Yoshida, N.,
  \& White, S.D.M 2001, \newa, 6, 79


\bibitem[Waldsley, Stadel, \& Quinn (2004)]{wsq04} Waldsley, J.W.,
  Stadel, J., \& Quinn, T. 2004, \newa, 9, 137

\bibitem[Yoshida (1990)]{y90} Yoshida, H. 1990, Phys.\ Lett., A150, 262





\end{thebibliography}
\end{document}